\documentclass[a4paper,fleqn,usenatbib]{mnras}

\usepackage{amssymb}
\usepackage{amsmath}
\usepackage{graphicx}
\usepackage{natbib}
\usepackage{hyperref}
\usepackage{adjustbox}
\usepackage{pdflscape}
\usepackage{longtable}
\usepackage{multicol}
\usepackage{multirow}
\usepackage{rotating}
\usepackage{float}
\usepackage{subcaption}

\usepackage{url}
\usepackage{rotating}
\usepackage{graphicx}
\usepackage{color}
\usepackage{txfonts}
\usepackage{booktabs}

\DeclareRobustCommand{\ION}[2]{%
\relax\ifmmode
\ifx\testbx\f@series
{\mathbf{#1\,\mathsc{#2}}}\else
{\mathrm{#1\,\mathsc{#2}}}\fi
\else\textup{#1\,{\mdseries\textsc{#2}}}%
\fi}

\newcommand{\HII}{\ION{H}{ii}}
\newcommand{\Hii}{\ION{H}{ii}}
\newcommand{\hii}{\ION{H}{ii}}

\newcommand{\nii}{[\ION{N}{ii}]}

\newcommand{\oiii}{[\ION{O}{iii}]}
\newcommand{\sii}{[\ION{S}{ii}]}

\newcommand{\Ha}{$\rm{H}\alpha$}
\newcommand{\Hb}{$\rm{H}\beta$}

\newcommand{\pyHII}{{\sc pyHIIextractor}}

\newcommand{\lkf}{$f_{\rm leak}$}
\newcommand{\frec}{$f_{\rm rec}$}
\newcommand{\DF}{$\Delta_{\rm F}$}
\newcommand{\rat}{$\frac{r_{\rm out}}{r_{\rm in}}$}
\newcommand{\rEW}{$\frac{\Delta \rm EW(H\alpha)}{\rm EW(H\alpha)}$}

\usepackage{xcolor}

\newcommand{\orcid}[1]{\href{https://orcid.org/#1}{\includegraphics[width=16pt]{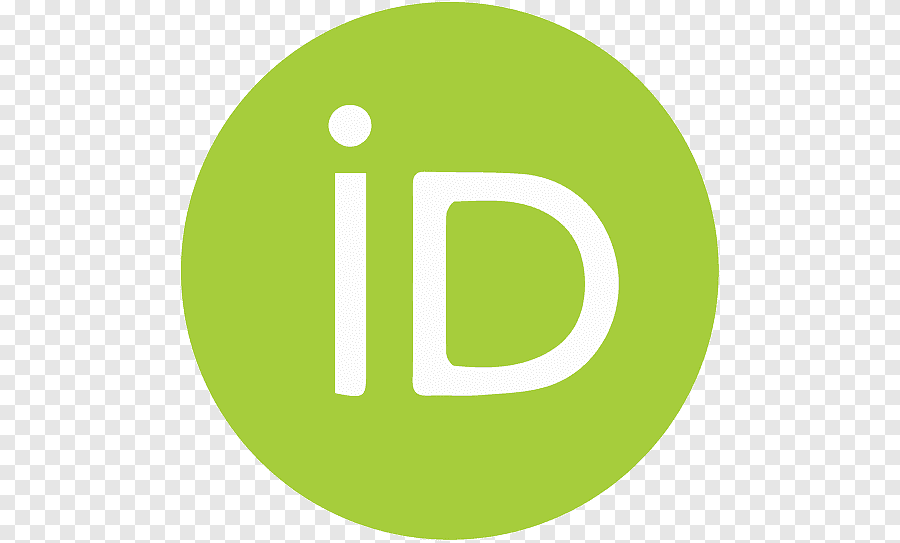}}}

\title[pyHIIextractor]{pyHIIextractor: A tool to detect and extract physical properties of \Hii\ regions from Integral Field Spectroscopic data}

\author[A.Z.Lugo-Aranda]{
A.Z. Lugo-Aranda$^{1}$,
S.F. S\'anchez$^{1}$, 
C. Espinosa-Ponce$^{1}$,
C. L\'opez-Cob\'a$^{2}$\orcid{0000-0003-1045-0702},\newauthor
L. Galbany$^{3,4}$,
J.K. Barrera-Ballesteros$^{1}$,
L. Sánchez-Menguiano$^{5,6}$,
J.P. Anderson$^{7}$\orcid{0000-0003-0227-3451}
\\
$^1$Instituto de Astronom\'ia, Universidad Nacional Aut\'onoma de  M\'exico, A.~P. 70-264, C.P. 04510, M\'exico, D.F., Mexico.\\
$^{2}$Institute of Astronomy and Astrophysics, Academia Sinica, No. 1, Section 4, Roosevelt Road, Taipei 10617, Taiwan. ORCID:0000-0003-1045-0702\\
$^{3}$Institute of Space Sciences (ICE, CSIC), Campus UAB, Carrer de Can Magrans, s/n, E-08193 Barcelona, Spain.\\
$^{4}$Institut d’Estudis Espacials de Catalunya (IEEC), E-08034 Barcelona, Spain.\\
$^{5}$Dpto. de F\'isica Te\'orica y del Cosmos, Campus de Fuentenueva, Edificio Mecenas, Universidad de Granada, E-18071 Granada, Spain.\\
$^{6}$European Southern Observatory, Karl-Schwarzschild-Str. 2, Garching bei M{\"u}nchen, D-85748, Germany.\\
$^{7}$European Southern Observatory, Alonso de C\'ordova 3107, Casilla 19, Santiago, Chile.
ORCID:0000-0003-0227-3451\\
}

\date{Accepted XXX. Received YYY; in original form ZZZ}

\pubyear{2021}

\begin{document}
\label{firstpage}
\pagerange{\pageref{firstpage}--\pageref{lastpage}}
\maketitle

\begin{abstract}

We present a new code named \pyHII, which detects and extracts the main features (positions and radii) of clumpy ionized regions, i.e. candidate \hii\ regions, using \Ha\ emission line images. Our code is optimized to be used on the dataproducts provided by the {\sc Pipe3D} pipeline (or dataproducts with such a format), applied to high spatial resolution Integral Field Spectroscopy data (like that provided by the AMUSING++ compilation, using MUSE). The code provides the properties of both the underlying stellar population and the emission lines for each detected \hii\ candidate. Furthermore, the code delivers a novel estimation of the diffuse ionized gas (DIG) component, independent of its physical properties, which enables a decontamination of the properties of the \hii\ regions from the DIG. Using simulated data, mimicking the expected observations of spiral galaxies, we characterise \pyHII\ and its ability to extract the main properties of the \hii\ regions (and the DIG), including the line fluxes, ratios and equivalent widths. Finally, we compare our code with other such tools adopted in the literature, which have been developed or used for similar purposes: {\sc pyhiiexplorer}, {\sc SourceExtractor}, {\sc HIIphot}, and {\sc astrodendro}. We conclude that \pyHII\ exceeds the performance of previous tools in aspects such as the number of recovered regions and the distribution of sizes and fluxes (an improvement that is especially noticeable for the faintest and smallest regions). \pyHII\ is therefore an optimals tool to detect candidate \hii\ regions, offering an accurate estimation of their properties and a good decontamination of the DIG component.

%We conclude that \pyHII\ is an appropriate tool to recover a large fraction of candidates to \Hii\ regions, offering an accurate estimation their properties and a good decontamination of the DIG component.

%The program was compared with simulations of spiral-type galaxies, we varied arm´s number, leaking, and the b/a ratio to characterize the ability of the program to recover \Hii\ regions number, flux per \Hii\ region, regions size, fluxes of the emission lines of \nii, \oiii, \Ha, \Hb\ and EW(\Ha) for decontaminated \Hii\ regions and DIG.

\end{abstract}

\begin{keywords}

ISM: \hii\ regions -- techniques: spectroscopic -- ISM: general -- galaxies: star formation.

\end{keywords}

\section{Introduction}
\label{sec:intro}

One of the most frequent methods to explore the properties of the interstellar medium in the optical regime is by studying the emission lines produced by the ionized gas. These lines can be produced by different physical processes such as: a) young and massive OB stars that ionize the surrounding nebula \citep[i.e. \Hii\ regions,][]{strom39}, b) low and high speed shocks \citep[e.g.][]{heckman90, dopita1995, kehrig:2012, lopezsanchezmoiseev2017}, c) old hot evolved stars \citep[post-AGBs, HOLMES,][]{stasinskavale2008, flor11}, d) supernova remnants \citep[e.g.][]{woodhilljoung2010, barneswoodhill2014}, e) leaking of photons from ionized nebulae \citep[e.g.][]{zuritabeckmanrozas2002, rela12}, f) ionization due to turbulent dissipation \citep{binette2009}, and g) heating by cosmic rays or dust grains \citep{reynoldscox1992}. The last six components are usually observed as diffuse ionized gas \citep[DIG, e.g.][]{valeasaricoutofernandes2019} covering the optical extension of galaxies. In general, all these ionizing processes can be observed simultaneously in the same galaxy, although their relative importance depends on the morphological type of the galaxy and/or region within it \citep{ARAA, sanchez20}. 

%This study is focused on the detection of \hii\ regions in emission line images (e.g., \Ha\ maps), their segregation and decontamination by the DIG component, and the extraction of their main properties. For this reason is important to understand the main properties of those two components. 

As outlined above, \hii\ regions are produced by the ionization due to young massive stars. Those stars are short-lived, and therefore are good tracers of both the star formation and the chemical composition of the ISM.
%of star-formation, on one hand, and of the chemical composition of the ISM, on the other. 
For these reasons \hii\ regions have been broadly used as tracers of the evolution of galaxies \citep{sanchezproperties2013}. The general properties of these nebulae have been extensively described in the literature. Their typical densities are in the range from 10 to 10$^{3}$ particles per cm$^{-3}$, gas temperatures vary from 5$\times$10$^{3}$ to 15$\times$10$^{3}$ K and the typical observed masses span from 10$^{2}$ to 10$^{4}$ M$_\odot$ \citep{osterbrock89, 2006oster, 2007kwok, peim67}. An archetypal idealized \hii\ region is a spherical structure of ionized gas, supported either by gravity or pressure, surrounding the ionizing stars, recently formed from the same gas (and therefore, sharing the same chemical composition). Their physical extension depends on the size of the ionizing cluster (or single star), ranging between a few pc to half a kpc \citep{hunt2009}. At a spatial resolution of $\sim$100 pc (a typical value of the observations used in this study), extragalactic giant \hii\ regions are observed as clumpy/peaky structures in emission line images. Given their importance and utility to explore the evolution of galaxies, robust methods are required that maximise \hii\ detection while preserving their intrinsic properties.

%Given their importance, are key to explore the evolution of galaxies, and for that reason it is required to provide with a method that maximizes their detection preserving their properties.

%At scales such as $\sim$100 pc, 
The advent of new high-spatial resolution Integral Field Spectroscopy (IFS) and Fourier Transform Spectrometer (FTS) data, such as those provided by MUSE \citep[Multi Unit Spectroscopic Explorer][]{bacon01} and SITELLE \citep{drissenmartinrousseau2019}, motivates the need to develop automated procedures to efficiently detect candidate \hii\ regions. 
%clumpy regions (candidates to \hii\ regions)
%, in particular when new IFS data deal with high-resolution images and it is expected to have hundreds or thousands of ionized regions on a single image. 
There are several codes in the literature that have been used to achieve this goal; such as {\sc pyhiiexplorer}, {\sc SourceExtractor}, {\sc HIIphot}, and {\sc astrodendro}
\citep{espi20, bert96,thilk00, astrodendro}. However, none of these codes are completely optimal for treating high-resolution data: {\sc pyhiiexplorer} is only valid for low spatial resolution data; {\sc SourceExtractor} was designed for other goals, and therefore it is adapted to segregate extended sources; {\sc HIIphot} was written in IDL and not maintained (to our knowledge); and {\sc astrodendro} was designed to segregate structures of different spatial scales. Therefore, a new code is required to process high spatially resolved IFS data, as none of the above extract and characterise \hii\ regions in an optimal manner.

The decontamination of pollutants such as the DIG in the derivation of candidate \hii\ region properties is another aspect not tackled by previous tools. The DIG has been described as a warm ionized gas phase ($\sim$10$^{4}$K) with low density ($\sim$10$^{-1}$ particles per cm$^{-3}$), with different possible ionizing sources (as described above). It is found in the neighborhood of the \hii\ regions and inter-arm regions in face-on spirals, at great distances above the plane of the galaxy in edge-on discs, or in spheroidal structures such as bulges \citep{kreckelblanc2016, levy2019apj, gomes16a, haffner2009}. It is ubiquitously observed in galaxies of all morphological type, representing up to $\sim$60\% of the total \Ha\ emission even in star-forming galaxies \citep{dettmar1990, oey07, haffner2009, lacerda18}. By definition, the DIG is observed as a smooth, low surface brightness component without significant structures in emission line images of similar resolution to those considered here \citep{haffner2009, censushii}.
%As indicated before the main contribution to the DIG may be leaked photons from \hii\ regions, ionization by HOLMES and shocks ionization \citet{ARAA}. Additional contributors, like turbulent dissipation \citep{binette2009}, and heating by cosmic rays or dust grains \citep{reynoldscox1992} could present a significant contribution too.  
Depending on the relative contribution of each component, the expected properties of the DIG may present strong variations galaxy by galaxy (and within a galaxy). 
%This is reflected in a significant variation of the observed properties of the associated ionized gas (e.g. flux intensities, EW's and line ratios). 

Due to this variety in the physical origin and observed properties very different approaches have been adopted to select and study the DIG in different galaxies. For instance, ($i$) \citet{flor11} explored the extra-planar ionized regions above the thin disk in 
late-type galaxies; ($ii$) \citet{blanclu2019, erroz19} selected as diffuse those regions with high [\ion{S}{ii}] $\lambda$6717/\Ha\ line ratios; ($iii$) \citet{zhangyan2017} adopted an upper limit to the surface brightness of the \Ha\ intensity ($\Sigma_{H\alpha}$); ($iv$) \citet{blancheiderman2009} and \citet{kaplanjogee2016} used a combination of the { [\ion{S}{ii}] $\lambda$6717}/\Ha\ line ratio and $\Sigma_{H\alpha}$; ($v$) \citet{lacerda18} defined the DIG as those regions with low equivalent width in \Ha\ (EW(\Ha)), below 3\AA. However, all these procedures have flaws because they have to make certain assumptions on the dominant ionizing source among, which is often unknown. 

%In this study we propose a new procedure to build-up a successful DIG model that will be used to decontaminate the \hii\ regions.

Following the above, \pyHII\ seeks to meet the goal of obtaining as much information as possible (underlying stellar populations and emission lines) from extragalactic \hii\ regions with high resolution data (from AMUSING++), decontaminated from to the contribution by DIG. This aim is met by building a successful and novel DIG model independent of their intrinsic properties.

The paper is organized as follows: Section \ref{sec:data} presents the data and its derivation; Section \ref{sec:pyHII} describes the algorithm; Section \ref{sec:results} shows the main results derived from simulations and comparisons of \pyHII\ with other codes; finally, Section \ref{sec:summary} highlights our main conclusions.

\section{Data}
\label{sec:data}
%\Com{I think we need to add just a note on the data you are using (MUSE), and maybe the descriptions of the dataproducts ({\sc PIPE3D}), that you will use for the code}

We foresee \pyHII\ as a general and flexible code to be used for the detection of \hii\ regions in a wide range of emission line maps (we recommend use \Ha\ map) generated either using direct observations (i.e., the classical narrow-band images) or through the use of Fabry-Perot (FP), IFS or FTS observations. However, we specifically test it using data from the All-weather MUse Supernova Integral field Nearby Galaxies ++ compilation \citep[AMUSING++,][] {galbanyandersonrosales2016, lopezcobasanchez2020}. These observations correspond to a collection of galaxies in the nearby Universe (z$\sim$0.015, DL$\sim$100 Mpc) observed with MUSE \citep{baconaccardoadjali2010, baconconseilmary2017}. Therefore, the code is optimized for the detection of \hii\ regions in data sampled with a seeing-limited $\sim$1\arcsec resolution (pixel size $\sim$0.2$\arcsec$), that at the projected distance corresponds to $\sim$300 pc. These are the data that we will use for the examples throughout this article, and those that we will attempt to mimic with our simulations. In any case, as indicated before, its use is not limited to this particular dataset. Indeed, we have tested it on other data, such as the emission line maps provided by the CALIFA \citep{califa} and MaNGA \citep{manga} IFS surveys, and direct narrow-band images of galaxies at the same cosmological distance observed with the William Herschel Telescope (Sánchez-Menguiano, priv. comm.). Generally speaking, an emission line map is required of a transition that traces the ionization associated with young stars, and we adopt H$\alpha$ for most cases (given that it is generally the brightest emission line). 

In the case of IFS data - such as that provided by AMUSING++ - emission lines are not directly accessible. They are a by-product of a particular analysis of the spectroscopic data. \pyHII\ is optimized for dataproducts exploration with the same format as the dataproducts provided by the {\sc Pipe3D} pipeline \citep{pipe3d_ii}, although this is not a strong limitation or restriction of the code that can be easily adapted to be used with the dataproducts provided by other tools.

%Due to the data used in this work and by the optimization of pyHIIextractor, below, we will explain obtaining and structure of the dataproducts. 
Since \pyHII\ is primarily developed and optimized to work with the dataproducts provided by {\sc Pipe3D} \citep{pipe3d_ii} extracted from IFS data, we briefly describe that code here. {\sc Pipe3D} performs a spectral fitting decomposition of the stellar continuum based on a combination of synthetic stellar population (SSP) spectra, convolved and shifted using a Gaussian kernel to account for the line-of-sight velocity distribution, and attenuated by an extinction law. In this process, the algorithm fits ($i$) the light-fraction (weight) of each SSP in the V-band, ($ii$) the stellar systemic velocity and velocity dispersion, and ($iii$) the dust-extinction at the V-band to recover the best spectral model of the stellar component. This procedure is applied to each IFS datacube, performing a spatial binning to ensure an optimal signal-to-noise ratio to provide a reliable spectral model. Once the best stellar model is obtained for each spectrum in each cube, this model is subtracted from the original spectra creating a cube with the ionized emission line component (plus noise and residuals of the stellar model subtraction). Then, a weighted-moment analysis is performed on each spectrum of each datacube to derive the main properties of a set of pre-defined emission lines. Finally, the full set of derived parameters (dataproducts) are rearranged in the original spatial distribution of the data creating a set of maps that are packed in datacubes in which each slice (map in the $z$-axis) corresponds to a physical parameter. This way, the final dataproducts are arranged in a minimum of four datacubes: SSP (that contains maps of the average stellar properties, such as the age or the metallicity), SFH (that contains maps of the fraction of light that each SSP contributes to the observed spectrum), { FLUX\_ELINES (that contains maps of the emission line properties, such as flux, EW, velocity and velocity dispersion)} and INDICES (that contains maps of the spatial distribution of classical stellar indices, such as the lick indices). These products are the input data used by \pyHII\ discussed in this study. { In the Appendix \ref{sec:info_pipe3d} we expand the information about these dataproducts. For more details see \citep{pipe3d_ii}.} 

%\Com{Now include the explanation of the {\sc PIPE3D} dataproducts.}

%%%%%%%%%%%%%%%%%%%%%%%%%%%%%%%%%%%%%%%%%%%%%%%%%%%%%%%%%%%%%%%%%%%%%%%%%%%%%%%%%%%%%%%%%%%%%%%%%%%%%%%%%%
\begin{figure*}
    \minipage{0.99\textwidth} 
    \includegraphics[width=\linewidth]{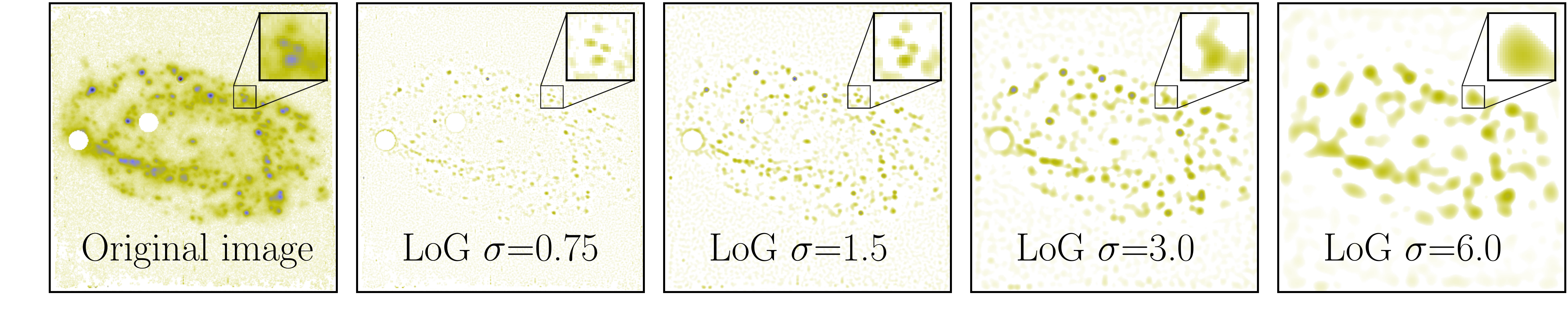}
    \endminipage
    \caption{Observed \Ha\ image of the galaxy NGC 2906 extracted from the MUSE datacube (first panel), together with the images resulting from applying the Laplacian of Gaussian procedure based on the convolution of the \Ha\ map with Gaussian functions of different sigma values (arbitrarily chosen, second to fifth panels), indicated in each label.}
    \label{fig:LoG}
\end{figure*}
%%%%%%%%%%%%%%%%%%%%%%%%%%%%%%%%%%%%%%%%%%%%%%%%%%%%%%%%%%%%%%%%%%%%%%%%%%%%%%%%%%%%%%%%%%%%%%%%%%%%%%%%%%
%\Ha\ image of the galaxy NGC 2906 extracted from MUSE observations after being used in \pyHII\ as input parameter (left panel), together the original image convolved with a Laplacian of Gaussian (2nd to 5th panel), using different widths ($\sigma$), indicated in each label.

\section{Description of the algorithm}
\label{sec:pyHII}

\pyHII\, is an evolution of previous attempts from our group to detect \Hii\ regions using IFS data. In particular, it inherits many concepts and ideas from {\sc HIIexplorer}\footnote{\url{http://www.caha.es/sanchez/HII_explorer/}}: a set of tools designed to detect \Hii\ regions in low spatial-resolution emission line images, extracted primarily from IFS data. Once each region has been detected and segregated, the package includes a tool to extract the individual spectra storing the results in a row-stacked spectra (RSS) format \citep{sanchez06a}. Originally coded in {\sc Perl} \citep{sanchez12b}, {\sc HIIexplorer} was fully transcribed to {\sc Python} \citep{espi20}\footnote{\url{https://github.com/cespinosa/pyHIIexplorerV2}}, adding the ability to extract the spectroscopic properties of both the stellar populations and the emission lines derived by other fitting tools (e.g. {\sc Pipe3D}), stored in the same World Coordinate System (WCS) of the original data. The software provides a set of tables with the spectroscopic properties of the detected \Hii\ regions. Different versions of this code were used to explore the \Hii\ regions in different surveys and compilations such as the PPAK IFS Nearby Galaxy Survey (PINGS) \citet{rosales-ortega10}, Calar Alto Legacy Integral Field Area survey (CALIFA) \citet{sanchez12b, sanchez15}, and AMUSING++ \citet{censushii, laura18}. So far, {\sc pyhiiexplorer} ({\sc Python} version of {\sc HIIexplorer}) has documented the largest catalogue of spectroscopic properties of \Hii\ regions, extracted from the CALIFA datacubes \citep{espi20}\footnote{\url{http://ifs.astroscu.unam.mx/CALIFA/HII_regions/}}. 

Despite its capabilities and extensive use, the {\sc pyhiiexplorer} code provides only a coarse spatial segregation of the \Hii\ regions. Following a scheme similar to the one applied by other tools \citep[e.g.][SourceExtractor]{bert96}, it generates a segregation map with a running identification number above zero for each region and zero for the areas outside any region. Therefore, a certain location in the original emission line map either belongs to one region or not. This procedure is sufficiently good for the coarse spatial resolution provided by IFS surveys such as CALIFA or MaNGA, both with a Full-Width at Half-Maximum (FWHM) of the Point Spread Function (PSF) $\sim$2.5$\arcsec$,  corresponding to a physical resolution of $\sim$0.8 kpc and $\sim$2 kpc on average, respectively \citep[e.g.][]{ARAA}. However, {\sc pyhiiexplorer} presents certain limitations by design. The strongest one is that it is unable to properly segregate adjacent \Hii\ regions of relatively large difference in brightness. In general, it aggregates the faintest region to the segregation map of the brightest one. For the same reason it is unable to segregate faint \Hii\ regions on top of the tail/wings of a much brighter one. Another limitation is that it assigns very similar sizes to all the detected \Hii\ regions, near to the PSF FWHM, regardless of their brightness (luminosity). Finally, as a consequence of both limitations, it underestimates the area outside any region, limiting its ability to detect regions ionized by other sources \citep[e.g., DIG][]{espi20}. All these limitations are not critical when dealing with IFS data with spatial resolutions of PSF FWHM $\sim$2.5$\arcsec$, however for data of better spatial resolution, e.g. PSF FWHM$\sim$1$\arcsec$ of galaxies at a similar redshift range, this package does not allow to extract information of the observed \hii\ regions in an optimal way. This limits our ability to separate them from the diffuse ionization and among themselves. 

Following the above, we thus developed a new tool whose main objective is to detect and segregate clumpy regions in high-resolution data, which we call candidate \Hii\ regions, while additionally characterising and removing the DIG component.

\subsection{Initial detection of \Hii\ region candidates}
\label{subsec:initdetect}

Following the philosophy of previous codes (e.g. {\sc HIIexplorer}), the main parameter to detect \Hii\ regions in an emission line map is that they are clumpy ionized structures, centralized and peaked, clearly contrasted with the diffuse (background) ionization. This geometrical property is fundamental to distinguish the ionization by a cluster of young stars from other sources of ionization in galaxies at the explored spatial resolutions \citep[e.g.][]{ARAA,sanchez20}. 
%A local maximum or peak in an image is, by definition, a point at which the gradient present a maximum. Thus, to detect those peaks a practical procedure is to derive the two dimensional distribution of the gradient by applying the 
%$ \nabla^2  = \Sigma \frac{\partial^2} {\partial x^2} $
Different procedures have been adopted in the literature for the selection of this kind of clumpy structure. We will discuss later some particular examples of previously published codes that we have explored without a fully satisfactory result. Thus, we have developed a new procedure that adopts the {\sc blob\_log} algorithm included in the {\sc scikit-image} package\footnote{\url{https://scikit-image.org/}} as the basic procedure to detect \Hii\ regions. The use of this particular algorithm instead of other ones available (even in the same {\sc python} package) is justified based on the simulations that we will present later.

The {\sc blob\_log} algorithm detects clumps/peaky structures (i.e., blobs) in an image by looking for local maxima (peaks) and making a simultaneous estimation of the size of the detected structures (characterized by a radius, $r$). For doing so the algorithm looks for those points at which the two dimensional distribution of gradients present a peak. However, to estimate the size, instead of deriving the gradient in the original image, it is derived on a set of images resulting from its convolution with a Gaussian function covering a range of values for the dispersion ($\sigma$). It is found that the value of the gradient in the local peak is maximal when the size of the structure corresponds to $r=\sqrt{2}\sigma$ \citep{blob}\footnote{\url{https://en.wikipedia.org/wiki/Blob_detection}}.

In summary, {\sc blob\_log} applies the so-called Laplacian of a Gaussian (LoG) operator \cite[we choose that operator because it provides sharpness to outstanding regions; in our case peaked regions with low noise sensitivity,][]{LOG} to the original image varying $\sigma$ (the dispersion of the Gaussian function) within a particular range defined by the user. For doing so, it implements the {\sc gaussian\_laplace} routine included in the {\sc scipy} {\sc python} module. This routine convolves the original image with a 2D-Gaussian function of a certain width (i.e., $\sigma$). This is:
\begin{equation}
 {\rm img}_{G} (x,y,\sigma) = G(x,y,\sigma) * {\rm img}(x,y) 
\end{equation}
\noindent where ${\rm img}_{G}(x,y,\sigma)$ is the convolved image by the Gaussian function, $img(x,y)$ is the original image and $G(x,y,\sigma)$ is the Gaussian function, defined as:
\begin{equation}
 G(x,y) = \frac{1}{2\pi\sigma^2} e^{-\frac{x^2+y^2}{2\sigma^2}}
\end{equation}
\noindent then, it applies a Laplacian operator
\begin{equation}
 \nabla^2 {\rm img}_G(x,y,\sigma) = \frac{\partial^2 {\rm img}_G}{\partial x^2} + \frac{\partial^2 {\rm img}_G}{\partial y^2}
\end{equation}
\noindent finally, it normalizes the resulting image by the width of the Gaussian function, and multiplies by $-$1, providing with the final LoG image,
\begin{equation}
 {\rm LoG} (x,y,\sigma) = (-1) \sigma^2 \nabla^2 {\rm img}_G(x,y,\sigma) 
\end{equation} 
\noindent this LoG corresponds to the two dimensional distribution of the gradient of 
${\rm img}_{G} (x,y,\sigma)$.

%in order to provide with images which intensities can be directly compared for different values of $\sigma$. 

This procedure is illustrated in Figure \ref{fig:LoG}, where the observed \Ha\ image extracted from the NGC 2906 MUSE datacube is shown (first panel), together with the LoG images resulting from convolving the \Ha\ map with LoG operators of different $\sigma$ values (arbitrarily chosen, from second to fifth panels).

%where it is shown an \Ha\ image extracted from the MUSE datacube of NGC 2906, and a images set generated by applying the LoG operator, using an arbitrary set of increasing values for $\sigma$.

Then, the {\sc blob\_log} re-arranges the set of ${\rm LoG} (x,y,\sigma)$ images derived when varying $\sigma$ within the given range in a three dimensional array (cube) in which each slice along the z-axis corresponds to one of these images, ordered in ascending values of $\sigma$. For the particular example in Fig. \ref{fig:LoG}, the resulting cube would comprise the four LoG images, ordered from left to right along the z-axis.
%that we will interpret later on as the covering the minimum and maximum expected sizes of the \Hii\ regions (which is assumed to be $\sqrt{2}\sigma$).
%The resulting set of LoG images are stored in a three-dimensional array (or cube) in which each slice corresponds to the LoG image for a given $\sigma$.
Then, the {\sc blob\_log} algorithm selects those coordinates (x,y,z) in this cube that verify to be local maxima in the three dimensional space (where z stands for different values of $\sigma$). For doing so, the value in this coordinate is compared with those values for all the 26 adjacent coordinates: i.e., the 8 adjacent pixels with the same value of $\sigma$ (i.e., [x$\pm$1,y$\pm$1], excluding [x,y]) and the 18 adjacent pixels with consecutive values of $\sigma$ (i.e., [x$\pm$1, y$\pm$1, z-1] and [x$\pm$1, y$\pm$1, z+1]). As indicated before, this way both the intensity peak in the original image and the size of the structure around this peak are found, given by the local maximum (given by $\sqrt{2}\sigma$).

Finally, only those regions that have peak intensities above a global threshold value (the global threshold value is a limit proposed by the user) are recorded. The output of this procedure is a set of positions and sizes that characterize the spatial distribution of clumpy ionized regions ({\it blobs}) in the original image.

The {\sc blob\_log} function implements the {\sc HIIblob} routine (as the basic algorithm to detect \hii\ regions). {\sc HIIblob} and in all subsequent steps, the code is written entirely by us and requires as input parameters an emission line image and a corresponding continuum image of the same object (with the same size and WCS). The continuum image is included to guarantee that there is an stellar continuum associated with the peak intensity in the emission line image, in order to exclude other sources of ionization not associated with a continuum source (e.g., shocks). If this continuum image is not available it is possible to use the emission line image as this second input parameter. The routine also requires a flux intensity threshold above which the peak intensity of the blobs are selected ($min\_peak$), and a threshold below which the original image is masked ($val\_mask$). It is recommended that the first value is selected as the 1$\sigma$ detection limit of the emission line image ($min\_peak\sim \sigma_{\rm eline}$), and the second one is selected as the 1$\sigma$ detection limit for the provided continuum image ($val\_mask\sim\sigma_{\rm cont}$), based on simulations. If they are not provided the algorithm makes an educated guess based on a statistical analysis of the input emission line image (in essence those values below zero in the emission line image are used to trace the noise). Then, it masks the original image for values below the second threshold, and runs the {\sc blob\_log} algorithm over the resulting image. In this first iteration {\it blobs} are detected using a very low threshold corresponding to 1.5$\sigma_{\rm eline}$, temporally storing the location of the peak intensities and the size of the regions. Again, those values were explored and tuned based on extensive exploration of different IFS datasets and the simulations that we will describe later on. 

\subsection{Initial estimation of the DIG}
\label{subsec:digmapinit}

One of the main goals of \pyHII, in addition to detecting and characterising \hii\ region candidates, is to build a model of the DIG across the optical extension of the studied galaxy and use this to decontaminate the \hii\ regions from the DIG. As discussed previously, the DIG is a smooth component of the ISM that was first observed in between the \Hii\ regions of the Galactic disk \citep{reynolds1971}. On the contrary to other previously mentioned explorations, we try to be as general as possible, making the minimum number of assumptions to select the diffuse gas (and remove it from the detected \hii\ regions). We therefore use the only general property of the DIG that is intrinsic to its definition and does not depend on the nature of the ionization (since this is still under discussion): this ionization is {\it diffuse}, i.e., smooth, without clear clumpy or peaked structures, ubiquitous in galaxies, and not evidently associated with other specific ionizing source (e.g., not linked to the presence of a strong AGN or a \Hii\ region). 

Simultaneously with the detection of the \Hii\ regions, \pyHII\ builds a diffuse ionized gas model image. To do so, it carries out a Delanuay triangulation \citep{dela}. First, we group the \Hii\ regions in triplets of the three nearest regions. Then, for each triplet the point located at the maximum distance among them is found. This set of points that traces the locations in the map at the maximum distance to any \Hii\ region, is stored as the tracer of the diffuse component. Then, a cleaning to discard repeated points is performed, removing those points too close to any adjacent \Hii\ regions (i.e., at distances lower than the size of the region). Subsequently, the flux intensity at the location of each of these points tracing the diffuse gas is estimated (co-adding the values within an aperture corresponding to the FWHM of the image). Finally, a DIG map is created by interpolating these values to recover the shape of the original image.
This initial DIG map is then subtracted from the original emission line map to create an image decontaminated by this component.

%The full procedure could be iterated if required. 

\subsection{Final detection of candidate \Hii\ regions}

The detection of the \hii\ candidates is carried out iteratively three times, varying the value of the flux intensity detection threshold in each iteration (each iteration is decontaminated by contribution due to diffuse gas). The three different thresholds used are obtained by multiplying the global threshold by factors: 1.5, 2 and 5 (factors values were chosen based on the results of the simulations described below). The detection of the candidate \Hii\ regions is done as described in the following procedure. As indicated before, a first exploration is performed adopting a very low detection threshold of just 1.5$\sigma_{\rm eline}$ above the noise level of the emission line map. This first iteration provides the user with a large number of candidates, that are used to estimate the DIG as outlined in the previous section. Then, we subtract the derived DIG map from the original emission line image and we proceed to a second selection of candidates but using a threshold of 2$\sigma_{\rm eline}$ (without using the candidate list from the previous iteration). Once the new list of candidates is generated, the flux intensity of each of them on the emission line corresponding to the original map used by user, is estimated as described at the end of this section. Using this flux intensity and the size (estimated in the detection as explained in the Sec. \ref{subsec:initdetect}) for each candidate, we generate a model image of all candidate \hii\ regions, assuming that they are perfectly round regions following a 2D Gaussian function (valid as a first order approximation). Then, this model is subtracted from the DIG-decontaminated emission line image, deriving a residual map. This residual map is used in the third detection loop with a very high detection threshold (with a value of 5$\sigma_{\rm eline}$) that allows us to uncover weak regions near bright ones, specifically, to avoid spurious detections from imperfect subtraction of the bright \hii\ region candidates detected in the second iteration. Finally, the lists of candidates provided by the second and third loop are joined together. Those regions that are located at a distance below 0.5 the size of one of them are removed since we consider them part of the same region. { The selection of the value 0.5 of the size is based on the results of the simulations and the tests performed on real data. We seek for the best compromise between minimizing the value of $\chi^2$ and maximizing the recovery of candidates to \hii\ regions, with the aim of avoiding false positives given by the over-fragmentation of the same ionized region}. After that, we obtain the final list of blobs, i.e., candidates defined by their position and size. 

Once the radii and the positions of the detected candidate to \hii\ regions have been obtained, we extract the fluxes. \pyHII\ includes options for different kind of extractions, including (i) pure aperture extraction of the co-added fluxes, (ii) average of the considered quantity, or (iii) weighted average or sum. By default, the fluxes are extracted using a weighted sum with the weight following a Gaussian distribution to the fourth power, with the width proportional to the size of each \hii\ region. This weight was adopted after experimenting with the simulations that we will describe below (Sec. \ref{sec:ref_sim}), with the ultimate goals of (i) recovering the most representative fluxes of the considered regions and (ii) minimizing the possible contamination from adjacent/nearby blobs. After this procedure, the program provides a catalog comprising the parameters of each blob (position and size) together with the corresponding flux.

%The extraction of the fluxes depends on two parameters: "kind" and "we". "Kind" has three options: that the extraction of the flow within the \Hii\ region is the sum of the values (k = 0), the average (k = 1) or the variance (k = 2). On the other hand, the "we" parameter also has 3 options: no weight (we = 0), a Gaussian type weight (we = 1) and a Gaussian quadration weight (we = 2). Regarding the component of \Hii\ regions, our code returns arrangements of positions, radii, flows, and an image of the final \Hii\ regions detected after the three iterations and their respective cleanings.... to webpage

In addition to this final catalog of \hii\ regions, our program can provide a segmentation map for the candidates, following the same scheme adopted by other tools such as {\sc SourceExtractor} or {\sc HIIexplorer}. The \pyHII\ segmentation map is constructed by assigning to each area corresponding to each candidate \hii\ region all pixels located at a distance to its center lower than the size of the candidate \hii\ region provided by the catalog.
%\pyHII\ segmentation map is obtained by assigned to a particular region those pixels at a distance lower than the size of the region to the center of each considered region. 
Finally, a mask map of the areas outside of the candidate \Hii\ regions is provided, as those pixels/spaxels are not associated with any region. We should recall that these maps (segmentation and mask) are incomplete representations of the real shape and distribution ionized regions since they could truncate their intensity distributions.

{ In future versions of \pyHII\ and for data of higher spectral resolution, we do not rule out the use of an additional information such as the velocity and velocity dispersion to detect and characterize the ionized regions \citep[following ][]{dagos21,law21}}

%Another optional parameter of the program is the return of the segmentation map of the \Hii regions and the masking map of the same \Hii regions, to save both maps, it is necessary for the user to choose to maps$\_$seg = 1. To obtain the segmentation map and masking map, the create$\_$HII$\_$seg and create$\_$mask$\_$HII$\_$seg functions uses radii, HII region positions and image size as inputs. For the case of the segmentation map create$\_$HII$\_$seg function assigns an index to each detected \Hii region, while the rest map has only zeros. On the other hand for the case of masking map the create$\_$mask$\_$HII$\_$seg function assigns a zero index to each detected HII region, while the rest map has only ones....to webpage

%%%%%%%%%%%%%%%%%%%%%%%%%%%%%%%%%%%%%%%%%%%%%%%%%%%%%%%%%%%%%%%%
\begin{figure*}
    \minipage{0.99\textwidth}
    \includegraphics[width=\linewidth]{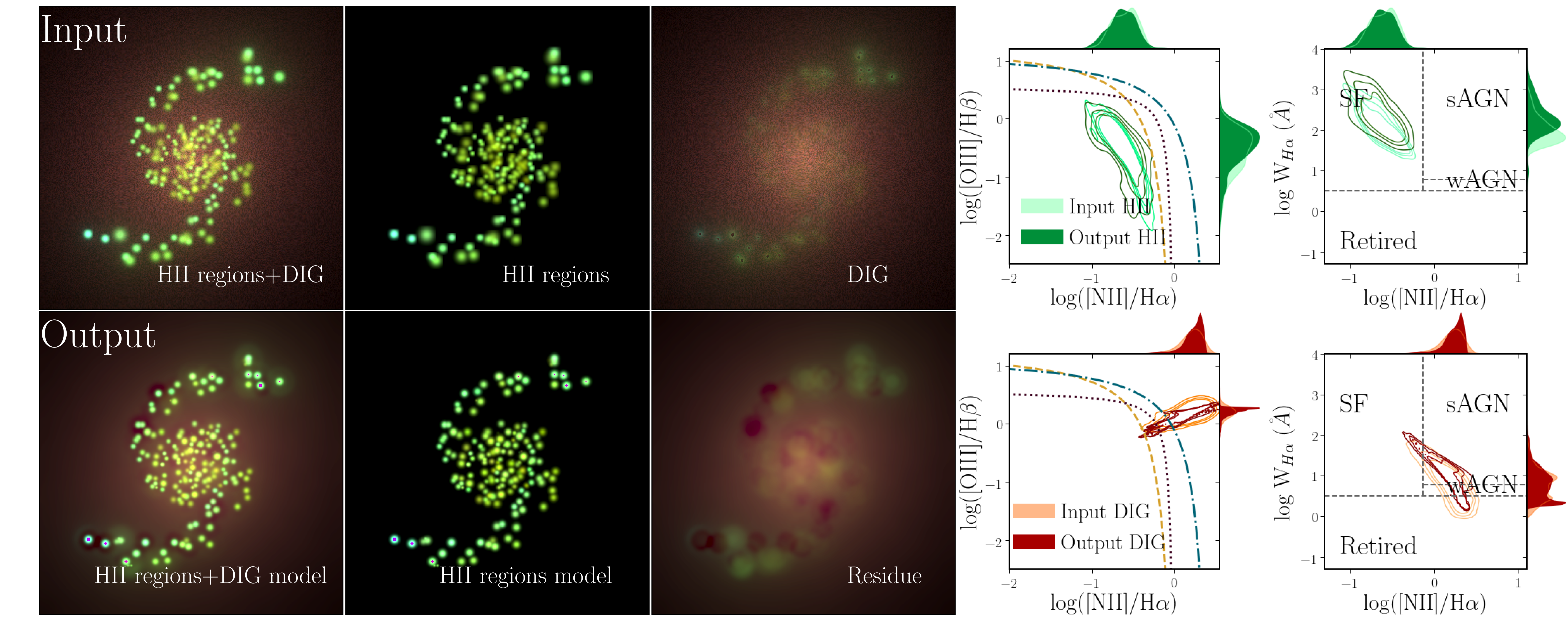}
    \endminipage
    \caption{Comparison between the simulated emission line images (top-left panels) and the corresponding ones recovered by \pyHII\ (bottom-left panels), assuming the galaxy is face-on (i.e., an inclination of 0$^{\circ}$). Each of the left-side panels corresponds to an RGB image where red corresponds to the \nii\ flux intensity, green to the \Ha\ one, and blue to the \oiii\ one. Left-most panels show the images for the complete simulation, once the three components considered here are co-added: \hii\ regions, DIG due to HOLMES and DIG due to leakage. Middle panels show the image comprising only the emission by the \hii\ regions; and finally, right-most of the left-size panels show the images including the emission by the two DIG components. In addition, the panels on the right hand side show two sets of BPT \citep{baldwin81} and WHAN \citep{cid11} diagrams, showing the distributions for the simulated (input) and recovered (output) \hii\ regions (top panels) and the DIG (bottom panels). In each diagnostic diagram the distributions are shown as density contours enclosing 15\%, 50\% and 99\% of the regions, respectively. In each BPT diagram, yellow dashed, blue dot-dashed and black-dotted lines represent \citet{kauffmanheckmantremonti2003}, \citet{kewley+2001} and \citet{espi20} demarcation lines, respectively. Each WHAN diagram shows the regions assigned to different ionizations defined by \citet{cid11}: star formation (SF), strong and weak AGN (sAGN, wAGN), and retired sources.}
    \label{fig:simula}
\end{figure*}
%%%%%%%%%%%%%%%%%%%%%%%%%%%%%%%%%%%%%%%%%%%%%%%%%%%%%%%%%%%%%%%%

\subsection{Final DIG evaluation}
\label{subsec:digmapfin}

Once the final catalogue of candidate \hii\ regions or blobs is derived (with the corresponding flux intensities) the code provides an optional re-evaluation of the diffuse gas. This re-evaluation makes use of both this catalog of blobs and the original distribution of points sampling the diffuse gas (described in Subsec. \ref{subsec:digmapinit}). First, a model image is created as described above (i.e., assuming Gaussian functions located at the position of the blobs, with a $\sigma$ corresponding to $\sqrt{2}$ of the size, and integrated flux corresponding to the extracted by the procedure described before). A preliminary estimation of the diffuse gas is created by subtracting blobs models from the original emission line image. Then, for each pixel of the original image the flux associated with the diffuse gas is derived { by co-adding the nearby points within a distance of one FWHM (i.e., in an aperture of radius=FWHM). For instance, for a MUSE data with a FWHM$\sim$1$\arcsec$ and a pixel of 0.2$\arcsec$, it corresponds to ~80 pixels, while, for a CALIFA data with a FWHM$\sim$2.5$\arcsec$ and a pixel of 1$\arcsec$, it will correspond to $\sim$20 pixels.} This co-adding is done following a weighting system that enhances positively the values near point tracers of the diffuse gas (the points farther away from a triplet of candidate \hii\ regions, as described above) and negatively the values near candidate \hii\ regions. The distribution of weights follows Gaussian functions centred on the point tracers of the diffuse gas and candidate \hii\ regions, with the former weighing always 3/2 more than the latter. We found that this estimation of the DIG reproduces better the simulated data that we will describe in the forthcoming sections.

\subsection{Adopted range of sizes for the candidate \Hii\ regions}
\label{subsec:adoptedrange}

As indicated above, the algorithm scans a range of possible sizes for the \hii\ regions, by defining a minimum and maximum value for the $\sigma$ parameter. For the {\sc blob\_log} routine this range is somehow arbitrary (i.e., defined by the user). This is not an optimal solution to make an automatic detection of \hii\ regions in images of different nature (sampling, physical and projected resolutions, or depths). Therefore, in our analysis, we define this range of parameters based on an exploration of the data themselves. For the minimum size, we adopt the scale of the pixel. { Strictly speaking a non-}clumpy structure should be smaller than the resolution element, that is usually 2 to 3 times the pixel-scale for well sampled data. However, the pixel-scale is certainly an absolute minimum size in any data set (even for under-sampled data). Finally, to determine the maximum size the procedure is repeated several times, using a wide range of values for this parameter, covering values from the minimum size indicated above to a maximum size corresponding to 2 kpc projected at the distance (and pixel-scale) of the image. For each maximum size we repeat the full exploration described before, deriving for each adopted set of $\sigma_{min}$ and $\sigma_{max}$, a model image of the \hii\ regions plus the DIG emission. This model is then compared to the original image, deriving a $\chi^2$ value. Based on this likelihood the final range of minimum and maximum sizes are selected as that which minimizes the $\chi^2$ value.

\subsection{Extraction of the properties}

One of the purposes of \pyHII\ is not only to detect candidate \hii\ regions in emission line maps, but also to extract the information of the stellar populations and emission lines (as already explained in Sec. \ref{sec:data} in the case where the user uses the dataproducts derived by {\sc PIPE3D}). For each of the properties delivered by this pipeline our code provides an estimation associated with each detected \hii\ region candidate and with the diffuse gas.
For this purpose, the data products of {\sc PIPE3D} are required as input (i.e., the four cubes of FLUX\_ELINES, SFH, SSP, and Lick-Indices), together with the catalog of \hii\ regions and the point tracers of the diffuse gas.

The extraction procedure is slightly different for additive/extensive quantities (like the emission line fluxes), than for non-additive/intensive ones (like the velocity). For the first, the extraction performs an estimation of the contribution of the DIG and a decontamination following the procedure described in the previous sections. For the second no decontamination is performed. Finally, for relative quantities, such as the EW of the emission lines, a two-step procedure is required, since first the decontaminated flux is derived and then the decontaminated EW is obtained following the equation:
\begin{equation}\label{eq:sp}
 EW^{d} = \frac{f^{d}}{f} \times EW 
\end{equation}
\noindent where EW$^{d}$ and EW are the decontaminated and original equivalent widths, and $f^{d}$ and $f$ are the decontaminated and original fluxes, respectively.

{ Regarding the error maps contained in dataproducts of Pipe3D, these are propagated through the extraction process according to the PSF FWHM. For more information on these error maps were estimated for each property, we recommend that the reader refer to section 3 (“Accuracy of the derived parameters”) in \cite{pipe3d_ii} and section 5 (“Accuracy of the fitting code”) in \cite{lacerda2022}}.

The final product of the extraction is a set of tables (stored in ecsv format), each one corresponding to each of the analyzed dataproduct cubes (FLUX\_ELINES, SFH, SSP, Lick-Indices), comprising an ID number for each candidate \hii\ region, their location, and the corresponding set of extracted properties. Finally, a set of models for each dataproduct provided by {\sc Pipe3D} are stored in the same original format, describing the 2D distribution for the candidate and the DIG, separately.

%THESIS AND WEBPAGE! Due to memory storage logistics, we strongly recommend the user to use mainly the tables which are in astropy format.
%\Com{Description of the original HIIexplorer, its version 2.0 (Carlos), and the new version: }
%\begin{itemize}
%    \item { HII detection}
%    \item { Diffuse construction}
%    \item { Extraction procedure}
%    \item { simulations and validation}
%    \item { Comparison with previous codes: SExtractor, PSFEx? photHII (look at Sanchez et al. 2012), HIIexplorer v2.0}
%\end{itemize}
%\subsection{HII regions selection}
%\label{sec:sel}
%\Com{We use the EW-f\_young procedure to select from the clumpy
%regions the HII regions.}
%\subsection{Spiral }
%Once the code was finalized, we characterized it with simulations of spiral-type galaxies.
%Once the code is finished, we continue with its characterization by simulations, use and comparison with other codes.
% and 90$^{\circ}$ degrees. 
% For both cases, we use A=0.5, B=1, C=5, FWHM=1.0, spaxel scale=0.2 and a leaking flux intensity$\sim$60\% of flux integrated. In each panel, the first three images (from left to right) correspond to: RGB images, 

\section{Results}
\label{sec:results}

The main goal of \pyHII\ is to derive the properties of emission lines for the candidate \hii\ regions with a reliable estimation (and decontamination) from the DIG. To characterize the behavior of the code and determine how well this goal is achieved, we perform a set of simulations focused on the recovery of the emission line fluxes (see Sec. \ref{sec:sim} and \ref{sec:ref_sim}). We simulate spiral-like galaxies since those are the objects where the vast majority of \hii\ regions are found in the Universe. However, we deliver the code adopted for these simulations together with the \pyHII\ code to allow any further, more tuned, exploration. Finally in section \ref{sec:comp_other_codes}, we compare the performance of our code with others adopted in the literature with a similar goal, both using real and simulated data. 
 
\subsection{Description of the simulations}
\label{sec:sim}

Each simulation comprises four main ingredients: 1) the structure of the spiral arms (Sec. \ref{sec:arms}); 2) the distribution of \Hii\ regions (Sec. \ref{sec:sim_Hii}); 3) a DIG component due to HOLMES (Sec. \ref{sec:sim_holmes}), and 4) a DIG component due to leaking of photons from the \Hii\ regions themselves (Sec. \ref{sec:sim_leak}). Other possible components of the DIG, such as shocks due to outflows, have not been considered so far. { The simulation of a more realistic spiral galaxy could be generated using N-body and hydrodynamical simulations by a radiative transfer code, e.g. SUNRISE \citep{sunrise} or SKIRT \citep{skirt}. Such simulations have been explored, for instace, to determine the reliability of the products derived by Pipe3D \citep{ibarra19}. However, the current mock-spiral simulation is well enough for the purposes of this study, as we have a better control on the input parameters.} 

Below we describe how we build each component.
%The following function describes the radial path of any spiral arm. It allows to generate a bar automatically, following a continuous and fixed relation to each arm of arbitrary winding sweep
\subsubsection{Spirals arms structure}
\label{sec:arms}

We adopt the prescriptions proposed by \cite{ring09} to describe the structure of the spiral arms (number and location) following the same procedures described by \cite{sanchez12b}. The location of a single spiral-arm with an arbitrary winding sweep is described by the equation in polar coordinates:
\begin{equation}
 {r (\theta) = \frac{A}{\log(B\:{\rm tan} (\frac{\theta}{2C})}} \end{equation}
\noindent where A is a scale parameter, C controls the tightness of the spiral arms (much like the Hubble scheme), and B controls the bar-to-arm size. Together B and C determine the spiral pitch (and pitch angle). In summary an increase (decrease) of each parameter results in: (A) an increase (decrease) in the size of the galaxy scale; (B) a larger (smaller) arm sweep and a smaller (larger) relative bar; and (C) a tighter (looser) winding. The previous equation describes the location of one single spiral-arm in the physical plane of the galaxy disk. To describe the full observed spiral structure, for a galaxy with $N$ arms, more copies of the considered arm rotated by an angle of 360$^\circ$/$N$ have to be added. In addition, we need to project the structure on the plane of the sky taking into account the inclination and position angle of the galaxy.

%To describe the full observed spiral structure, for a galaxy with more arms, it is required to project it at the observed plane (taking into account the inclination and position angle) and to add more copies of the considered arm, rotated by an angle of 360$^\circ$/$N$, with $N$ being the number of arms.

%The typical parameters that describe the arms structure were taken from the analysis by \citet{sanchez12b}. Those were derived based on an interactive fitting algorithm for a sample of spiral galaxies at a similar redshift range. The fitting was based on two main assumptions: ($i$) spiral arms trace the strong emission of H$\alpha$ and ($ii$) \Hii\ regions are most frequently found in the spiral arms vicinity. 

Throughout all simulations we adopted the following set of values: A=8$\arcsec$, B=1, and C=5$^\circ$, that should trace an average spiral galaxy \citep{sanchez12b}. Ad hoc simulations trying to reproduce a particular pattern would require tuning those parameters or fitting them following the procedure outlined in \citet{sanchez12b}. Finally, we projected this distribution into the plane of the galaxy assuming a certain position angle and inclination, and considering that all the \hii\ regions are preferentially located in the disk of the galaxy (i.e., any vertical distance is purely random).

\subsubsection{Distribution and properties of \hii\ regions}
\label{sec:sim_Hii}

%We chose to simulate a late-type galaxy because they are those that have a high number of \Hii\ regions. 

The code that generates the simulation requires a number of \hii\ regions per spiral arm ($N_{\rm HII}$, { by default in the code assumes a value of $N_{\rm HII}$=100}). Based on that number it populates each spiral arm following a random distribution with equal probability along the $r(\theta)$ location traced by the equation described in the previous section. In this way, due to the winding and number of spiral arms the actual density of \hii\ regions is larger in the central regions than in the outer ones mimicking the known pattern of a spiral galaxy. A tangential random shift that increases with the galactocentric distance is included to provide a more realistic simulation of the actual distribution of \hii\ regions. Finally, the current simulation populates the galaxy with \Hii\ regions without considering the presence of a bulge (generally depopulated of those regions). { It is important to note that for face-on galaxies the possible overcrowding of \hii\ regions introduced when including a new spiral arm in these simulations could have a similar impact as the inclination.}

%It is important for the user to consider that when simulating face-on galaxies the overcrowding of \hii\ regions may be the dominant impact factor in comparison with other parameters such as the inclination when using the simulator.

%To generate the simulated datacubes that
To cover the goals of the current simulation we need to set not only the location of the \hii\ regions across the galaxy, but their sizes, emission line intensities, and line ratios (for a set of emission lines), { trying to mimic realistic regions}. In this particular simulation we generate (i) the flux intensities of \Ha, \Hb, \oiii$\lambda$5007 and \nii$\lambda$6583, (ii) the equivalent width of \Ha, i.e., EW(\Ha), and (iii) the size of the regions. 
%For doing so, for each region, we simulate the \oiii$\lambda$5007/\Hb, \nii $\lambda$6583/H$\alpha$ deriving for each region emission lines fluxes for the emission lines of \Ha, \Hb, [OIII]5007, [NII]6583, and additionally EW(\Ha).
Thus, we first generate a continuum emission distribution, assuming an arbitrary global bright surface density and a radial exponential decay, with disk scale-length of $h=$10$\arcsec$. This scale-length is of the order of the average value found for the disk galaxies in the CALIFA \citep{sanchez16} and AMUSING++ \citep{lopezcobasanchez2020} samples. Then, the EW(\Ha), and the line ratios are generated for each \hii\ region following the radial distribution of those parameters described for a sample of nearby galaxies in \citet{sanchez12b}, their Table 8. { This is an empirical approach, but a broader range of parameters (and physically different \HII regions) could be implemented based on the results from photoinization models \citep[e.g.][]{3MdB,war}.} { The two line ratios used throughout this work are N2 and O3:}

\begin{equation}
 N2 = [NII]\lambda 6583/H\alpha 
\end{equation}

\begin{equation}
 O3 =[OIII]\lambda 5007/H\beta 
\end{equation}

A random value is added to each parameter following the reported standard deviations in the same study. Once the EW(Ha) is estimated, together with the line rations and the continuum intensity, the flux density is derived for each emission line for each simulated \hii\ region. Finally, we estimate the angular sizes of the \hii\ regions. This is achieved by deriving a relation between the radius of an \hii\ region and its observed \Ha\ flux intensity based on the relation between the Strömgren radius and the \Ha\ luminosity. We assume that the observed objects are located at a typical redshift of z$\sim$0.015 (i.e., the average redshift of the CALIFA and AMUSING++ samples). This relation depends on the ionization parameter, which we roughly estimate based on the N2 ratio and the electron density, which we fixed to $n_e=100$ cm$^{-3}$. The main aim of this exercise is to provide simulated \Hii\ regions of reasonable sizes, and in particular sizes that present a strong correlation with the \Ha\ flux. It is known that the real size of an \Hii\ region does not follow such a simple prescription, as recently highlighted by \cite{barnesgloverkreckel2021}. Finally, we include the effects of dust attenuation on the observed emission line fluxes. Like in the case of the line ratios, we adopted the average radial gradient in A$_V$ described by \cite{sanchez12b} for a sample of nearby galaxies, taking into account a random value that reproduces the reported standard deviation around the mean distribution. For implementing the dust effects we use the \citet{cardelli:1989} extinction-law with an R$_{V}$ = 3.1, and the theoretical value for the un-obscured line ratio for case B recombination of \Ha/\Hb = 2.86 \citep{osterbrock89}. { 
This treatment of the dust is clearly an oversimplification (although it is broadly adopted). In essence it does not adopt any radiative transfer procedure to consider the effect of the dust on light when crossing the galaxy. Therefore, for highly inclined galaxies the effect of dust is underestimated. However, for those galaxies almost all existing procedures to explore \hii\ regions present difficulties, and in general they are usually excluded in this kind of explorations \citep[e.g.][]{espinosaponce2022}. }

%Although our model of correction due to dust is simple, and therefore in our current simulations the effect of inclination due to dust in spiral galaxies is underestimated, in general, the codes present limitations in the study of highly inclined galaxies, for example, in the recovery of the properties of the underlying \hii\ regions. This is a well-known problem whose solution is based on the omission of the use of those highly inclined galaxies, 

%%%%%%%%%%%%%%%%%%%%%%%%%%%%%%%%%%%%%%%%%%%%%%%%%%%%%%%%%%%%%%%%
\begin{figure*}
    \minipage{0.99\textwidth}
    \includegraphics[width=\linewidth]{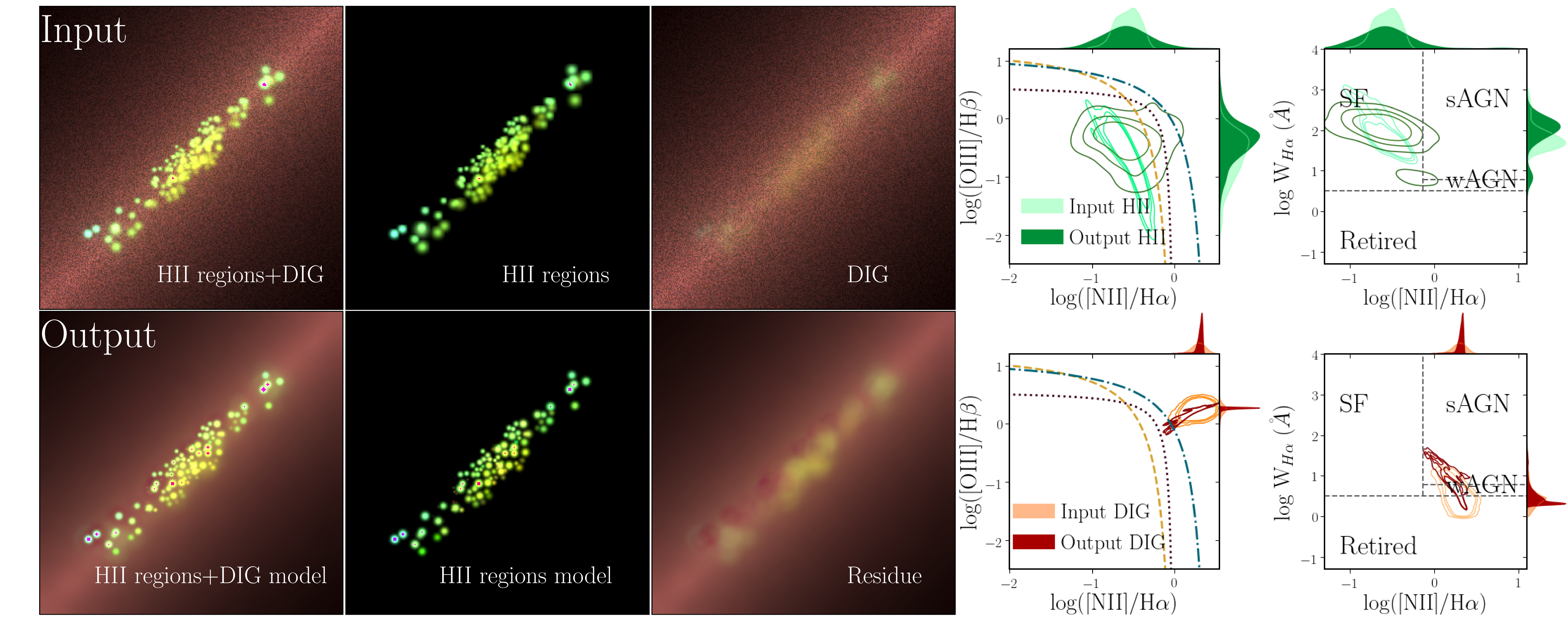}
    \endminipage
    \caption{Same figure as Fig. \ref{fig:simula}, but for an edge-on galaxy (i.e., inclination of 90$^{\circ}$).}
    \label{fig:simula2}
\end{figure*}
%%%%%%%%%%%%%%%%%%%%%%%%%%%%%%%%%%%%%%%%%%%%%%%%%%%%%%%%%%%%%%%%

It is important to highlight that the simulated distribution of \hii\ regions follows by construction some of their typical properties: (i) they are found at the well-known loci of these regions in the BPT \citep{baldwin81} and WHAN \citep{cid-fernandes10} diagrams (as we will see in the Sec. \ref{sec:ref_sim}); (ii) they present similar flux and size distributions as real regions, with a larger number of faint/small regions and a lower number of bright/large ones; and (iii) they are more frequent along the spiral arms than in the inter-arm regions. Based on this distribution we create a set of emission line maps, one for each of the considered transitions, mimicking the field-of-view (FoV) and spaxel(pixel)-scale of a MUSE observation. Thus, we create a set of empty images of 300$\times$300 pixels, assuming a pixel-scale of 0.2$\arcsec$/pixel (i.e., a FoV of 60$\arcsec$ $\times$60$\arcsec$). Then, for each emission line map, we run a loop on the \HII\ regions adding in each step a 2D Gaussian function with the flux intensity and size of the region at the considered location as included in the simulated distribution. In this procedure, it is assumed that the galaxy is well centered in the simulated image (i.e., the center of the galaxy is located in the pixel [150,150]). Thus, the simulation provides both an input catalog of the \HII\ regions and their main properties and a set of emission lines maps that corresponds to those regions. In addition, a continuum and EW(H$\alpha$) map is generated using the same projection, based on the distributions described previously.

%We estimate the emission lines ratio, based on a rough estimation of the ionization parameter $\log_{10}$ U=-1.4$\log_{10}$(\nii/\Ha $\times$2.86)-3.26 as a measurement of the strength of the ionization radiation \citep{diaz00}, and using an empirical relation with the radius; this parameter and the flux intensity in \Ha\ is described such as radius=0.0322$\sqrt{H\alpha} \times$(-U-2.8)/2. This relation is based on the relation between the Strömgren radius and the \Ha\ luminosity, under the assumption that the observed objects are at around z$\sim$0.015 (i.g., the typical distance of surveys as CALIFA and AMUSING++). This relation can be easily adopted for other cosmological distances.

%Specifically to define H$\beta$ and H$\alpha$ lines, we use the theoretical value for the unobscured line ratio for case B recombination of \Ha/\Hb=2.86 and \Ha=$\log_{10}$[EW(\Ha)] $\times$ continuum, respectively. The ionization parameter estimated as $\log_{10}$ U=-1.4$\log_{10}$(\nii/\Ha $\times$2.86)-3.26 is a measurement of the strength of the ionization radiation \citep{diaz00} and is used to calculate the radius of the \Hii regions as radius=0.0322$\sqrt{H\alpha} \times$(-U-2.8)/2.

\subsubsection{DIG due to HOLMES}
\label{sec:sim_holmes}

We simulate the DIG due to the presence of old stellar populations (HOLMES) as a smooth distribution of the flux intensity of the considered emission lines, whose strength follows the continuum emission (the simulation of which was described above). This component is simulated by assuming a uniform EW(\Ha), on average, with values that range between 1-3 \AA\ across the extension of the simulated galaxy. We assume no dust attenuation for this component, so the \Ha/\Hb\ ratio corresponds to the canonical value of 2.86. This choice is justified based on the fact that low or no star-formation activity is usually associated with low gas content, that is connected to the dust attenuation via dust-to-gas relation \citep{barrerautomobolatto2020, barreraheckmansanchez2021}. Finally, the N2 and O3 line ratios are simulated to reproduce the expected distribution for this ionization along the low-ionization nuclear emission-line (LINER-like) regime of the BPT diagram. The actual adopted values follow a smooth gradient with the aim of reproducing the values reported for this kind of ionization in early-type galaxies (e.g., figures 4, 2 and 6 of \cite{sanchez20}, \cite{capettibaldi2011}, and \cite{annibalibressanrampazzo2010}, respectively). The result of this simulation is a set of emission line and EW(H$\alpha$) maps of the same size, pixel scale, and projection on the sky as the maps generated for the contribution of the \HII\ regions.

%A smooth gradient in both parameters is adopted to reproduce the inside-out trend reported in Sanchez (2020) \ComAL{?????????} for this ionization, with a random component aimed to cover the full regime of expected values, as represented in Fig. NN of Sanchez et al. 2021 \ComAL{???????}. 
%WEBPAGE ...This component is described with the create\_DIG function, where its input parameters are the coordinates of the center of the smooth distribution, the pitch angle, the parameter ab that describes the tilt of the galaxy, the scale height, the peak of the continuum distribution, and the size in pixels of the galaxy to simulate.
%?????Like in the case of the \Hii\ regions simulation, the final product of this simulation is a set of emission lines maps and the EW(\Ha) that are re-arranged in a similar, that mimics the FLUX\_ELINES extension of the {\sc PIPE3D} fits files.
%The return to the user of this function is a cube corresponding to the DIG only due to HOLMES where the order of the indices is as follows: \Hb, \oiii, \Ha, \nii, and EW(\Ha).

\subsubsection{DIG due to leaking from \hii\ regions}
\label{sec:sim_leak}

Finally, we simulate the leaking of photons from \Hii\ regions by assuming that its flux is proportional to the corresponding one in each region. This proportionality is modulated by a leaking factor (f$_{leak}$) with values between 0 (no leaking) and 1 (all photons escape). We acknowledge that none of those extreme values are physically realistic. Nevertheless, we consider that the leaked flux intensity decreases with the distance to the center of each \Hii\ region following an $r^{-2}$ decay, due to pure dilution effects. In this simulation, we do not consider the effects of differential dust coverage or scatter in the leaking \citep{zuritabeckmanrozas2002}.

In summary, for each \Hii\ region we simulate a leaking component by adding to each image a component that it is proportional to the flux intensity of the \Hii\ region, multiplied by the f$_{leak}$ factor, and following an $r^{-2}$ functional form. For practical reasons, we consider that the leaking factor is the same for all \Hii\ regions in each simulation. However, a more realistic approach would be to consider a different value for each region or even a value that varies spatially regarding the geometry of the region (or its dust attenuation). We will explore that possibility in future studies.

Like in the previous components, the result of this procedure, when iterated through all the simulated \Hii\ regions, is a set of maps of the flux intensity for each considered emission line, and for the EW(\Ha), with the same size, scale, and projection at the images generated in Sec. \ref{sec:sim_Hii} and Sec. \ref{sec:sim_holmes}. Finally, all the maps corresponding to the same emission line and different ionizing sources are summed and convolved with the corresponding observational PSF (by default a 1$\arcsec$/FWHM Gaussian function). Photon noise due to the astronomical targets or the sky can be easily added to generate even more realistic simulations (although for simplicity we have not considered it here). Then, the different maps are re-arranged and packed into a single datacube that mimics the format of the FLUX\_ELINES extension of the {\sc Pipe3D} dataproducts.

%\Com{Here I am: 13.10.2021}

\subsection{Results from the simulations}
\label{sec:ref_sim}

Using the procedures described above, we create different sets of simulations modifying the inclination of the galaxy, the number of spiral arms, and the leaking factor. So far, we have not considered the effects of the noise in these simulations. Thus, the detectability of the \HII\ regions depends only on the ability of separating them and the contrast with respect to other regions or the DIG. For each simulated dataset, we run our detection and extraction code, generating both (i) a catalog of \Hii\ regions comprising their locations, sizes, and flux intensities, and (ii) a model image for each emission line for the \Hii\ regions and the DIG. Figures \ref{fig:simula} and \ref{fig:simula2} illustrate qualitatively this procedure by showing two particular simulations of the same galaxy, a { Grand Design} spiral, face- and edge-on, respectively. In both cases, the default A, B, and C parameters were adopted for the spiral arms, the default FWHM for the PSF, and a leaking factor of 60\%. Each figure comprises a set of color images created using the flux intensity maps of \oiii\ (blue), \Ha\ (green), and \nii\ (red) for both the simulated dataset (including both the contribution of \hii\ regions and the DIG, and each component separately) and the corresponding images based on the components recovered by the code. We note that as expected the \Hii\ regions are observed as greenish clumpy structures since by construction they are simulated by simple Gaussian functions and with their emission dominated by \Ha. On the contrary, the DIG presents two distinct components, clearly observed in the right-most panel of the panels of the left, with a brownish smooth component that follows a clear radial decay (corresponding to the ionization by HOLMES), and some greenish smooth structures around the location of the \Hii\ regions (corresponding to the ionization by leaked photons). In addition, we include in Fig. \ref{fig:simula} and \ref{fig:simula2} a comparison of the emission line ratios and EW(\Ha) simulated and recovered by the code as distributed in the classical BPT \citep{baldwin81} and WHAN \citep{cid11} diagrams. It is worth noticing that as we have foreseen the distributions of these parameters mimic the real location of those ionizing components in the considered diagrams.

%we see that in the sum of components (left-most panel), the radii of the detected and recovered regions are more defined making their structure more marked, the leaking differs from the \hii\ regions as a smoother circular ionized region around each \hii\ region and the radial decline of the DIG due to older stellar populations is smoother and continuous, these observations visually indicate that the separation of the components to the first order are good

A qualitative comparison between the set of input and output RGB emission-line images for the two particular examples shown in Fig. \ref{fig:simula} and \ref{fig:simula2} suggest that our code has done a good job in modelling and replicating the observed distributions, to first order. A particularly good agreement is found between the input and output emission line maps for the \Hii\ regions component (middle-panels). The recovered flux intensities, sizes and line ratios are similar, since there is no appreciable change in the color of the \Hii\ regions. Furthermore, the structure of the simulated galaxy is well recovered without introducing fake \hii\ regions in inter-arm regions or grouping of the smallest \Hii\ regions into bigger ones. For the diffuse component (right-most panel), we recover the smooth decline of brownish emission associated with DIG due to HOLMES. However, the recovery of the leaking component seems less precise, extracting larger structures and with some evident failures (mostly for weak \Hii\ regions and in the areas where the HOLMES DIG seems to dominate). Finally, the code seems to recover well the sum of the two components, which is expected since the \Hii\ regions are well recovered and they are the dominant ionization source. 

As a parameter to gauge how well our code recovers the distribution of \Hii\ regions, we introduce the fraction of flux assigned to those regions by our code relative to their simulated flux. We consider that this parameter is a better proxy of the behavior of the code than the number of \Hii\ regions or their distribution of recovered fluxes since for regions below the size of the PSF there is an inherent mixing, merging them into larger and brighter regions. We cannot segregate those regions with the current data and analysis. However, it is worth knowing how well the averaged flux is recovered even in these circumstances. We highlight that this is not an intrinsic limitation of the code, but a limitation of the simulated dataset (that tries to mimic real data that would present similar problems).

In short, the code performs better for the face-on galaxy (82\% recovery) than for the edge-on system (65.5\%).
This flux that it is miss-assigned is distributed in the DIG components. Indeed, the DIG distribution in the case of the face-on galaxy match better the observed distribution, replicating the radial decrease for the brownish component and few residuals associated with weak \Hii\ regions, than the DIG distribution for the edge-on galaxy. In this case, the greatest residual is in the central part of the galaxy, showing a stronger leaking component that is absent in the simulated image.

With respect to BPT diagrams in general terms for both cases (edge-on and face-on galaxies) the line ratios for the decontaminated \hii\ regions are found in the star-forming area. In particular, the output density contours of the face-on galaxy are below the three demarcation lines tracing well the input density contours. When comparing the histograms of the N2 and O3 line ratios a very good match between the simulated and recovered distributions is seen. %Finally, the distribution for the EW(H$\alpha$) present a slight offset, with recovered values being slightly larger than simulated ones (although both of them are of the same order).
On the other hand, for the edge-on galaxy, the output density contours widen more, covering a broader dynamic range for N2 without reaching the lowest values for O3. Despite this effect, the average location of the input and output distributions are well reproduced, even for this extreme case. Regarding the DIG component, we see that in both cases the average locations of the recovered distributions match those of the simulated ones. However, the recovered distributions are narrower than the simulated ones (this is expected since our DIG derivation provides a smooth component, unable to reproduce the pixel by pixel variations introduced in the simulation). It is interesting that in the case of the face-on galaxy the output density contours are more extended attaining until the inter-demarcations lines region. This means that DIG due to photon leaking is somewhat recovered. On the other hand, for the edge-on galaxy, the dominant ionization in the DIG is located at the LINER-like region, being well recovered by our code. The dominant component to the DIG in this galaxy is due to HOLMES, and it seems to be well recovered, in particular in the extra-planar regions, where the contamination by \Hii\ regions and the leaking of photons is weak.
%This is particularly true for the extra-planar regions, where this ionization is clearly dominating.
%while that in the BPT for the DIG component correspondent to edge-on galaxy the output is only LINERs region (this means that DIG due to old stellar populations is the recovered in the majority and the dominant one for this second case). 
 
%\Com{Here I am: 13.10.2021}

%%%%%%%%%%%%%%%%%%%%%%%%%%%%%%%%%%%%%%%%%%%%%%%%%%%%%%%%%%%%%%%%%%%%%%%%%%%%
\begin{figure*}
    \minipage{0.99\textwidth}
    \centering
    \includegraphics[scale=0.5]{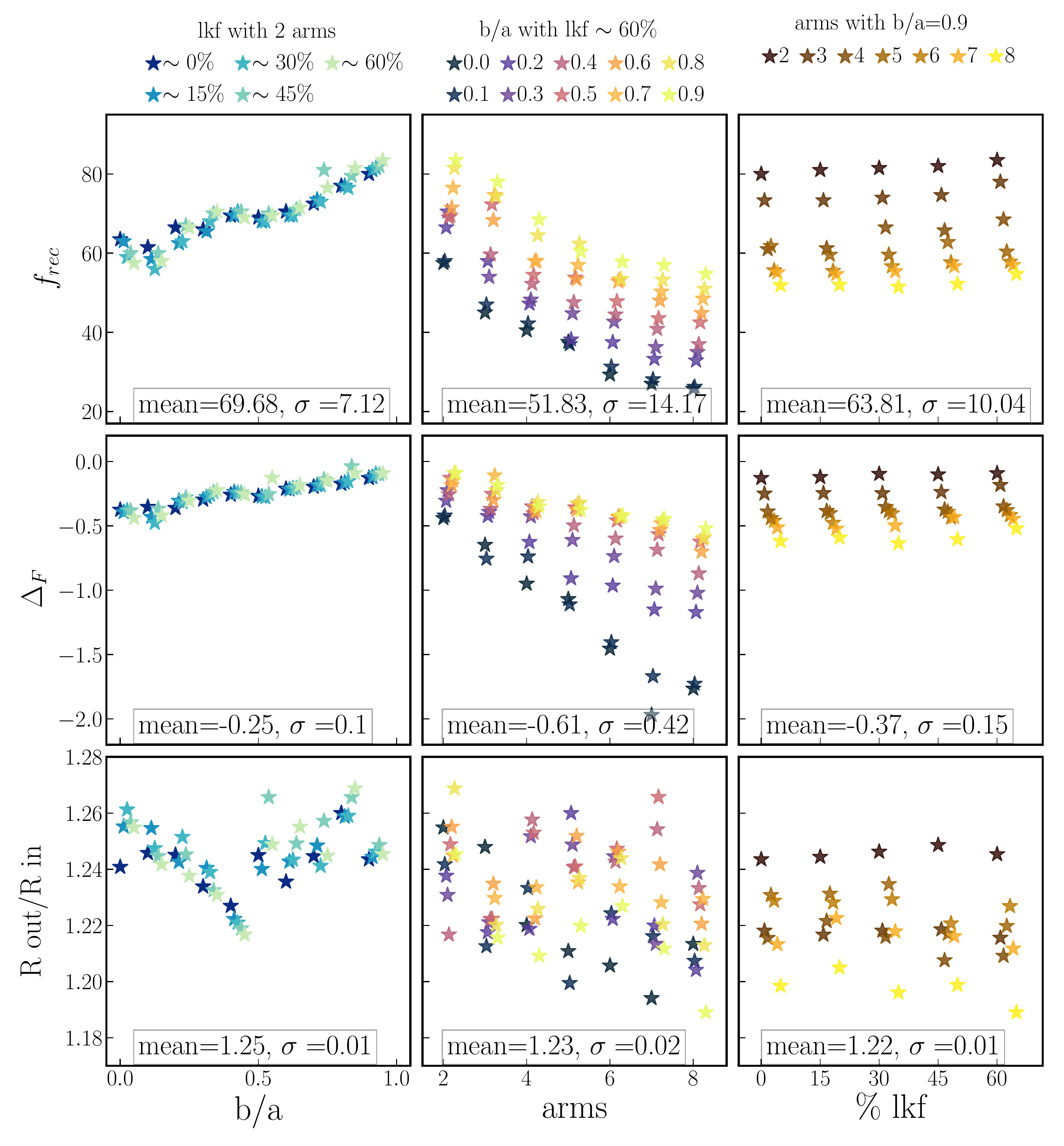}
    \endminipage
    \caption{Distribution of \frec\ (fraction of recovered regions, {\it upper panels}), \DF\ (relative fraction of recovered flux, {\it middle panels}) and \rat\ (ratio of recovered and simulated sizes, {\it bottom panels}) as a function of the different parameters varied in the three set of explored simulations: {\it left-panels}, varying $b/a$ (x-axis) and \lkf\ (color) for a fixed number of spiral arms; {\it central-panels}, varying the number of spiral arms (x-axis) and the $b/a$ ratio (color); and {\it right-panels}, varying \lkf\ (x-axis) and the number of spiral arms (color). In all cases, each consider parameter (\frec, \DF\ and \rat) is represented in the Y-axis, with each star corresponding to the average value for each individual simulation. In the values of all the panels a displacement is made in the X-axis for a better visualization.}
    \label{fig:rrr}
\end{figure*}
%%%%%%%%%%%%%%%%%%%%%%%%%%%%%%%%%%%%%%%%%%%%%%%%%%%%%%%%%%%%%%%%%%%%%%%%%%%%

Regarding the WHAN diagrams for the \hii\ regions, important differences are seen depending on the inclination (face-on vs edge-on). For the face-on simulation, all output density contours are located at the SF region, and in general terms, the density contours match with the simulated/input ones, however, there is a mild systematic overestimation of the derived values of EW(\Ha), that happens mostly at low N2 values. This is not the same for the case of the edge-on galaxy, where not all density contours are confined in the SF region, since there are a few regions that enter in the wAGN region of the WHAN diagram. In this case, the values of EW(\Ha) are slightly overestimated at high values of N2, while for low values of N2 they are mildly underestimated, although in general terms the density contours present a good match. 

In the WHAN diagrams of the DIG component for the face-on galaxy, the input/output density contours present a better match compared to the edge-on galaxy ones. In this second case, the values of EW(\Ha) are more clearly overestimated. At the same time, for the face-on galaxy, we can observe that the derived DIG contours extend towards the SF region while in the second case they are shifted towards the sAGNS and wAGN ones (mostly the first ones). In any case, there is a considerable agreement in the average values of the explored parameters for the DIG component in both cases. { Additionally, we estimate the contribution to the integrated flux by \hii\ regions and DIG, being $\sim$45\% and $\sim$65\% respectively, being similar to the estimated based on observations \citep[e.g.][]{rela12}}

%we tested the best recovery (face-on galaxy) and we derive that the percentage of \Ha\ emission from the map containing only candidate to \hii\ regions is $\sim$45\%, while DIG contributes $\sim$65\% of the total.}

%These two particular examples shows that qualitatively the procedure is able to recover well the input parameters...

%\Com{*** Describe what you see when comparing the input-output images for the two components together and separately. What seems to be recovered well and the possible different input/output that illustrate the errors. How the face- and edge-on images looks different or similar. Then jump to the BPT diagrams. How the input-ouput distributions looks like. In which regimes they are different (95\%?), the shape and patterns. Which component dominates (leaking in the DIG). How the effect of inclination blur/affects the results. 
%Here is the core of the article, Alejandra! Just describe what you see!}

Now that we have illustrated qualitatively how well \pyHII\ works for a simulation adopting a particular set of parameters, we perform a set of simulations varying different variables in order to determine the accuracy and precision of the recovered parameters depending on the variables of the simulated galaxy. In particular, we explore the number of recovered candidates, their radii and fluxes, as well as the line ratios described before (O3 and N2), and the EW(\Ha). These last three parameters will be explored for both components, i.e., the \hii\ regions and DIG. For doing so we vary in the simulations the inclination (characterized by the $b/a$ ratio), number of arms, and the percentage of photon leakage from the \hii\ regions (\lkf). The exploration was performed by fixing one of the three parameters (2 arms, \lkf $\sim$60\% and a $b/a$=0.9, respectively) and varying the other two covering a wide range of values. For all this set of test simulations, we use the same parameters to describe the spiral arms, the shape and resolution of the simulated image, and input parameters adopted before (i.e., A=8$\arcsec$, B=1, C=5, FWHM=1$\arcsec$, spaxel-scale=0.2$\arcsec$ and a max\_size=1.75). 

%This last value is determined by iterating to obtain minimum value $\chi^{2}$.

%We 
%We separate  structural parameters of \Hii\ regions, \hii\ regions line ratios and DIG line ratios. Line ratios are define as: N2=\nii $\lambda$6583/H$\alpha$ and O3=\oiii $\lambda$5007/\Hb, $\Delta$\Ha\ and $\Delta$EW(\Ha) are the difference output-input/output from their mean values and in particular case line ratios $\Delta$=output-input from their mean values.

%\Com{*** Now you need to describe all the range of parameters covered by the simulations. Then you need to describe all the different parameters that you have tested. The fraction of recovered Hii region, their relative flux, the radius, the line ratios, the EW of Ha. Then the parameters tested for the DIG (making clearly that we cannot distinguish between the two DIG components so far).

%*** Then you have to go panel by panel of each figure (from fig 3 to 10) and describe what you see. What is the range of variation of the explored parameter vs  b/a, # arms, \% \lkf\... the range of variation, the min-max values, if there are trends with the explored parameter.

%*** You cannot write the discussion and conclusions without all these results. This is the code of the article!}

\subsubsection{Structural parameters of the \hii\ regions}
\label{sec:sim_struct}

We explore three parameters that characterize how our code recovers the main structural properties of the \hii\ region and their distribution: (1) the fraction of regions recovered in comparison to the number of simulated ones (\frec); (2) the relative difference between the input and recovered \Ha\ flux, corrected by the fraction of recovered \hii\ regions, i.e. defined as \DF$=100\times\frac{F_{H\alpha,out}-F_{H\alpha}}{F_{H\alpha} f_{\rm rec}}$. The correction by $f_{\rm rec}$ is needed to compensate for the \Hii\ regions lost in the detection process. In other words, \DF provides with an estimation of how well the fluxes are recovered but only for the detected regions, without considering the non detections (a parameter that it is already coded \frec).
%This parameter determines how well the fluxes of the detected regions are recovered, and for this reason it is required to multiply by $f_{\rm rec}$.
%by including a correction for the fraction of recovered ones; 
(3) the ratio between the estimated (output) and simulated (input) radii (\rat). 

Figure \ref{fig:rrr} shows the distribution of these three parameters for three sets of simulations: (i) varying the $b/a$ ratio and \lkf\ for a fixed number of 2 spiral arms; (ii) varying the number of arms and the $b/a$ ratio and fixing \lkf\ to 60\%; and (iii) varying \lkf\ and the number of arms with a fixed value of $b/a$=0.9. For the fraction of recovered regions (top panels), we show that the total dynamic range is between 30\% and 80\% for all the simulations. The simulation that includes a variation of $b/a$ ratio and \lkf\ (left panel) is the one with the smallest scatter ($\sigma$=7.12), a more limited dynamic range with a mean=69.68 and a clear trend to better recovery rates at lower inclinations (i.e., larger $b/a$ ratios) and larger values of \lkf (i.e., $\sim$45\% and $\sim$60\%). The simulation that comprises a variation of the number of arms and the inclination (middle panel) covers the widest dynamical ranges of the explored parameter ($\sigma$=14.17), with a poorer recovering rate (mean=51.83) for more inclined galaxies and a higher number of spiral arms.
%The different values of b/a are those that also provide the greatest dispersion and it show a decreasing trend towards a greater number of spiral arms, with a better recovery at high b/a values. 
Finally, the simulation where both the \lkf\ and the number of arms are varied (right panel) covers a similar dynamical range and dispersion as the first one (with a mean=63.81 and $\sigma$=10.04). However, there is no clear trend observed of the rate of recovery as a function of the \lkf, being the variation fully dominated by the previously described trend with the number of spiral arms.
%with medium dispersion, the dynamic range is between 80\% and 55\% without any tendency at increase the value of \lkf\ but obtaining a greater recovery of candidates with arms number lower.
As a preliminary conclusion, we find that the recovery of candidate \hii\ regions is affected to a much lesser extent by the \lkf\ than by the $b/a$ ratio and arms number. The best recovery rates are found for face-on galaxies with few spiral arms (i.e., Grand Design spirals).

In the case of \DF, the relative recovery of the fluxes, the trends are similar to the ones found for the \frec, i.e., the fraction of recovered candidates. The dynamic range covered by the full set of simulations is between -1.7 dex and 0 dex. However, the most extreme cases, i.e., the lowest values of \frec, are found only for almost edge-on galaxies. The different simulations show, like the case of \frec, that there is a clear trend towards better recovered fluxes as the $b/a$ ratio is higher and the number of spiral arms is lower, with a limited effect of the \lkf. The concordance in the trends observed for \DF\ and \frec\ indicates that the inability to detect \Hii\ regions is correlated with the inability to recover the correct flux intensity for the detected regions.

%The variations of b/a ratio as function of \lkf\ (left panel), arms number as function of b/a (middle panel) and \lkf\ as function of arms number (right panel) have the particular dynamic ranges between 0 and -0.5, 0 and -1.5, and 0 and -0.7, respectively. 
%From this we can preliminary conclude that the percentage of candidates to \hii\ regions recovered as well as the relative flux intensity are similarly correlated with the parameters of \lkf\, arms number and b/a ratio. \%lfk does not affect the recovery of the relative fluxes like b/a ratio and arms number. Then, we recovery the best flux relative of candidates to \hii\ regions for face-on galaxies and with few spiral arms.

Finally, for the case of the recovery of the radii (bottom panels), the \rat\ parameter covers a range between 1.19 and 1.27 for the full set of simulations, despite the fact that the ultimate goal would be to recover a value as near as one as possible. None of the three panels (left, central, and right), that correspond to three different sets of simulation, show clear trends with any of the explored parameters ($b/a$, number of arms, or \lkf). However, the average \rat\ value is larger (mean=1.25) with a lower dispersion ($\sigma$=0.01) for the first simulation (when the number of spiral arms are fixed to two, and varying $b/a$ and \lkf). For the second simulation (when the \lkf\ is fixed) the dispersion is larger ($\sigma$=0.02) and the average \rat\ value is slightly smaller (mean=1.23). Finally, the lower average value (mean=1.22), the one more near to the optimal value, is found for the third simulation (fixed inclination: face-on galaxy). Curiously, the best \rat\ value is recovered for the simulation that includes the largest number of spiral arms, that it is somehow counter-intuitive because having a larger population of nearby \hii\ regions or overlapping regions, it should be more difficult to delimit their sizes. In summary, we know that our code introduced a systematic bias towards larger sizes of the \hii\ regions of about a 19-27\%.

%the range of dispersed values is less when varying b/a ratio as a function of \lkf\ (left panel), while the panel of more dispersed values is that of the variation of the arms number as a function of the b/a ratio (middle panel). Some values of the best recovery of radii is at high values of b/a (middle panel) and several spiral arms (right panel).

%Based on this, we know that "systematically" our code errs in the recovery of radii between 19\% and 27\%.

%%%%%%%%%%%%%%%%%%%%%%%%%%%%%%%%%%%%%%%%%%%%%%%%%%%%%%%%%%%%%%%%%%%%%%%
% Simulations Emission line HII regions
\begin{figure*}
    \minipage{0.99\textwidth}
    \centering
    \includegraphics[scale=0.5]{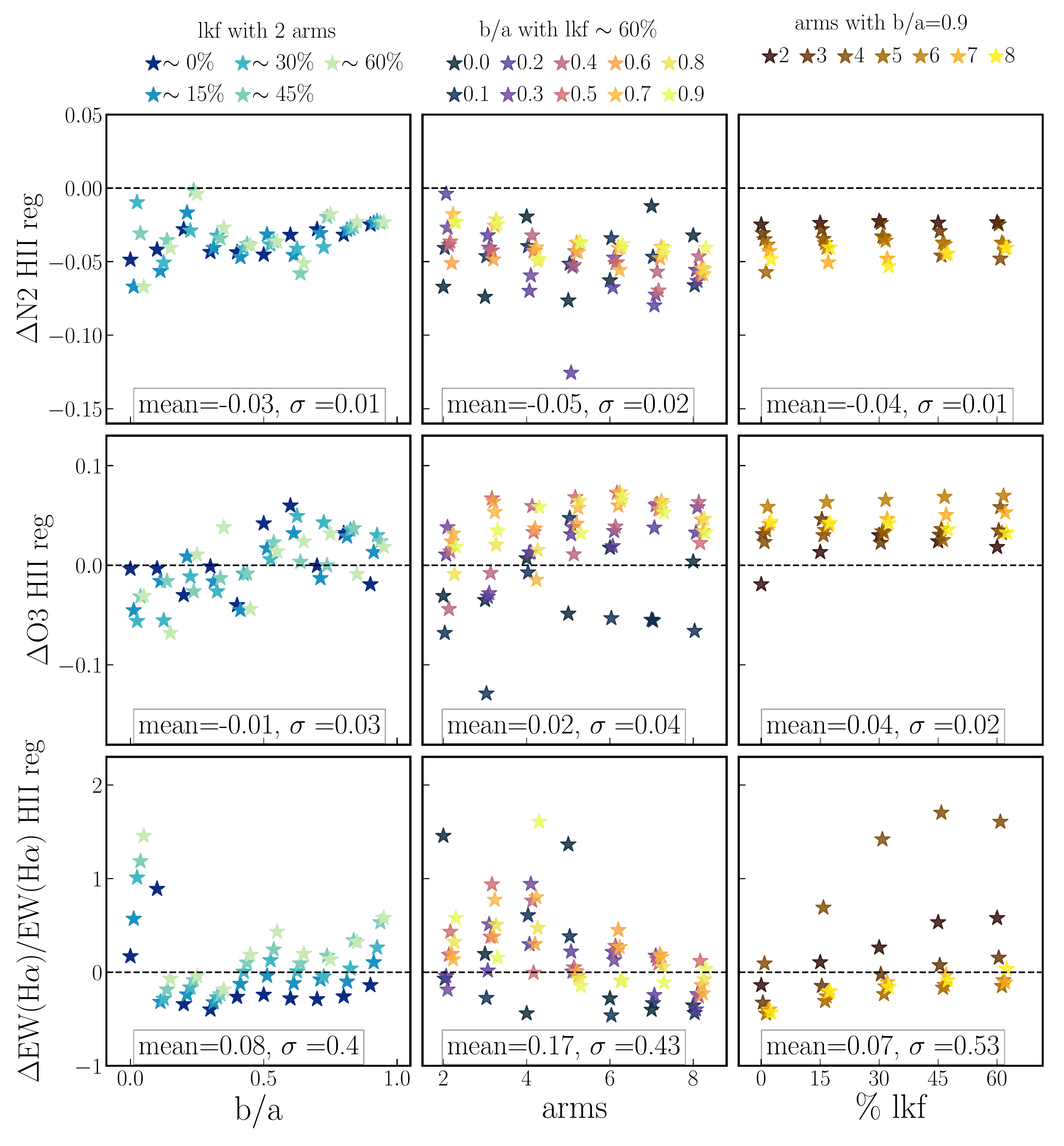}
    \endminipage
    \caption{Recovering of the line ratios of the \hii\ regions from the explored set of simulations. From top to bottom each panel shows the distribution of the average $\Delta$N2, $\Delta$O3, and \rEW, for each simulation as a function of one of the parameters varied in the simulation, with the color indicating the other varied parameter, following the same scheme presented in Fig. \ref{fig:rrr}.}
    \label{fig:hii}
\end{figure*}
%%%%%%%%%%%%%%%%%%%%%%%%%%%%%%%%%%%%%%%%%%%%%%%%%%%%%%%%%%%%%%%%%%%%%%%

\subsubsection{Line ratios of the \hii\ regions}
\label{sec:sim_rat}

As indicated above, we explore how well the code recovers the N2 and O3 line ratios and the EW(\Ha) for both the \hii\ regions and DIG. Figure \ref{fig:hii} shows the distribution of the average values of the difference between the recovered (output) and simulated (input) for the line ratios ($\Delta$N2 and $\Delta$O3, top and middle panels), and the relative difference for the EW(\Ha, bottom panels), i.e., \rEW. We note that ideal goal of our code would be to recover values for these parameters as similar as the simulated ones, i.e., with the represented differences as near to zero as possible (dashed lines in all panels of Fig. \ref{fig:hii}).
%\hii\ regions case is illustrated in figure \ref{fig:hii}, where shown the result for a set of simulations in variation with the same parameters as \ref{fig:rrr}. The optimal recovery value is zero and in all panels it is drawn as a gray dotted line. 

For $\Delta$N2 (top panels) we see that we recover values between -0.12 dex and 0.02 dex, with an average bias of $\sim$-0.05 dex for the full set of simulations. Thus, N2 is recovered with a bias corresponding to a value $\sim$5\% lower than the original one, with a maximum error of $\sim$7\%. The second set of simulations (middle panel), i.e., the one where the \lkf\ is set to a fixed value and both the number of spiral arms and the inclination is varied, is the one that presents the largest scatter and the mean furthest from the optimal value (mean=-0.05 and $\sigma$=0.02). On the other hand, the other two simulations, (i) the one adopting a fixed number of spiral arms (2), while varying the inclination and $f_{\rm leak}$ (left-panel), and (ii) the one adopting a fixed inclination (face-on galaxy), while varying the number of spiral arms and the $f_{\rm leak}$ (right-panel), offering best mean and scatter values, respectively. In this set of simulations, a mild decrease of $\Delta$N2 with the number of spiral arms is observed, without any significant trend with any of the explored parameters. Only in the case of the first simulation (left panel) a very weak trend is observed with the inclination towards a lower bias in the recovery of N2 for face-on galaxies with a low number of spiral arms (i.e., two, in our case).

%In average the code produces a bias in $\Delta$N2 of about -0.04 dex, 

% SFS 22.10.21

%, being the variation of arms number as a function of b/a ratio with a \lkf$\sim$\%60 (middle panel) the ones that has a greater scattering in their values and the variation of \lkf\ as function of arms number with a b/a=0.9 (right panel) the one with the least dispersion between its values. For the panel corresponding to the variation of b/a ratio as function of the \lkf\ value (left panel) for the lowest values of b/a ratio the N2 recovery is better with higher values of \lkf\, while for higher b/a values there is no significant difference. The general trend of this plot is increasing at high b/a values.
%For the next panel, corresponding to the variation of the arms number as function of b/a ratio (middle panel) there is no clear trend, however for the variation of \lkf\ as function of arms number (right panel) the values of a higher recovery is with few spiral arms.

%In general terms, we show that the $\Delta$N2 values tend to be underestimated in their recovery by our code, finding them up to 10\% less than their input value.

Regarding O3, we recover the original values within a range between -0.13 and 0.09 dex for the bulk of the simulations. The values are slightly overestimated, with an average bias of $\sim$3\%, and an error limited to 6\%, when this bias is considered positive. However, we find significant differences for the different simulations. Like in the case of N2 the first simulation shows a mild trend between $\Delta$O3 and the inclination, with O3 slightly underestimated for highly inclined galaxies, and slightly overestimated for face-on ones. However, the trend with the number of spiral arms, explored in the second (central panels) and third (right panels) set of simulations, is the opposite to that reported for $\Delta$N2. In this case, there is a mild increase of $\Delta$O3 with the number of spiral arms. No further trends with the explored parameters are found.

{ The N2 and O3 line ratios are frequently used to derive the oxygen abundance in the ISM. Therefore, we can estimate the effect of the inaccuracies in the recovery of both ratios on the derivation of this parameter. For doing so we adopted the calibrator proposed by \cite{marino13}, defined as:

\begin{equation}
    12+log(O/H)=8.545-0.202 \times O3N2 
\end{equation}

where O3N2 = log([OIII] $\lambda$5007/\Hb\ $\times$ \Ha/[NII] $\lambda$6583 (i.e., O3-N2). Assuming a typical offset for the line ratios ($\Delta$N2=-0.5 dex and $\Delta$O3=0.4 dex) and the highest standard deviation observed for those offsets ($\sigma$N2=0.02 dex and $\sigma$O3=0.04 dex), we estimate a $\Delta$O3N2$\approx$0.1 dex, and a  $\sigma$O3N2$\approx$0.06 dex. This is translated into a rather small, not significant, effect in the oxygen abundance ($\Delta$(O/H)$\approx$-0.02$\pm$0.01 dex) when addoting the calibrator indicated before.}% Therefore, with this we tested that the offset between the input and output candidates does not significantly affect the oxygen abundance.} 

%As to $\Delta$O3 (middle panels) the total dynamic range is between -0.13 and 0.9, and although no general trends are observed in any of the 3 panels corresponding to this parameter, in general terms we can say that O3 output values are overestimated with differences of up to 13\%.

Finally, for the EW(\Ha), we find the strongest differences between the input and output values, with \rEW\ ranging between -0.5 and 1.8. However, we find strong differences between the different simulations. Indeed, in most of the cases ($\sim$80\%), the relative difference is restricted to $\pm$0.5, i.e., the recovered EW(\Ha) is restricted to 50\%\ of the simulated value, with a clear average bias but with most \rEW\ values centered on zero. The best recovery is found for the first set of simulations, the one where the number of spiral arms is fixed to two (a low value). This simulation, and the third one (when the inclination is fixed to a face-on galaxy), uncover a weak positive trend of \rEW\ with the \lkf. Thus, the EW(\Ha) is slightly underestimated for low leaking factors and slightly overestimated for high values of \lkf. \rEW\ presents also a weak negative trend with the number of spiral arms, that is hinted at in the second simulation (the one where \lkf\ is kept fixed) and clearly appreciated in the third one. The worst results are found for highly inclined galaxies in general, which is somewhat expected since in these kinds of galaxies there is an overlapping of observed flux that corresponds to very different regions due to projection effects. On the other hand, the EW(\Ha) is better recovered when the \lkf\ is $\sim$30\%, for a mildly or low inclined galaxy (b/a$\sim$0.7-0.9), and 5-6 spiral arms. However, for the other simulations  except those for edge-one galaxies, the differences are subtle.

%SFS 23.10.21

% The case in which the recovery is better are in the variation b/a as function of \lkf\ (left panel). While in the panels corresponding to the variation of the arms number as function of b/a ratio (middle panel) and the variation of \lkf\ as function of arms number (right panel) do not show clear trends in the recovery of this parameter, but it can be overestimated by up to 200\%.

%From this we can preliminary conclude that EW(\Ha) it is a parameter of difficult recovery with an error of up to $\sim$200\%, although we show that its recovery increases at low values of \lkf\.

% SFS: Tentative different caption
%, as a function of one the parameters varied in the simulation (left-panels:$b/a$, central-panels:number of arms and right-panels:\lkf), with the color indicating the other varied parameter (left-panel:\lkf, central-panels:$b/a$ and right-panel:number of arms). As indicated in the text, the third involved parameter is kept set to a fixed value in each simulation (left-panel:number or arms, central-panels:

%%%%%%%%%%%%%%%%%%%%%%%%%%%%%%%%%%%%%%%%%%%%%%%%%%%%%%%%%%%%%%%%%%%%%%%
% Simulations Emission line DIG regions
\begin{figure*}
    \minipage{0.99\textwidth}
    \centering
    \includegraphics[scale=0.5]{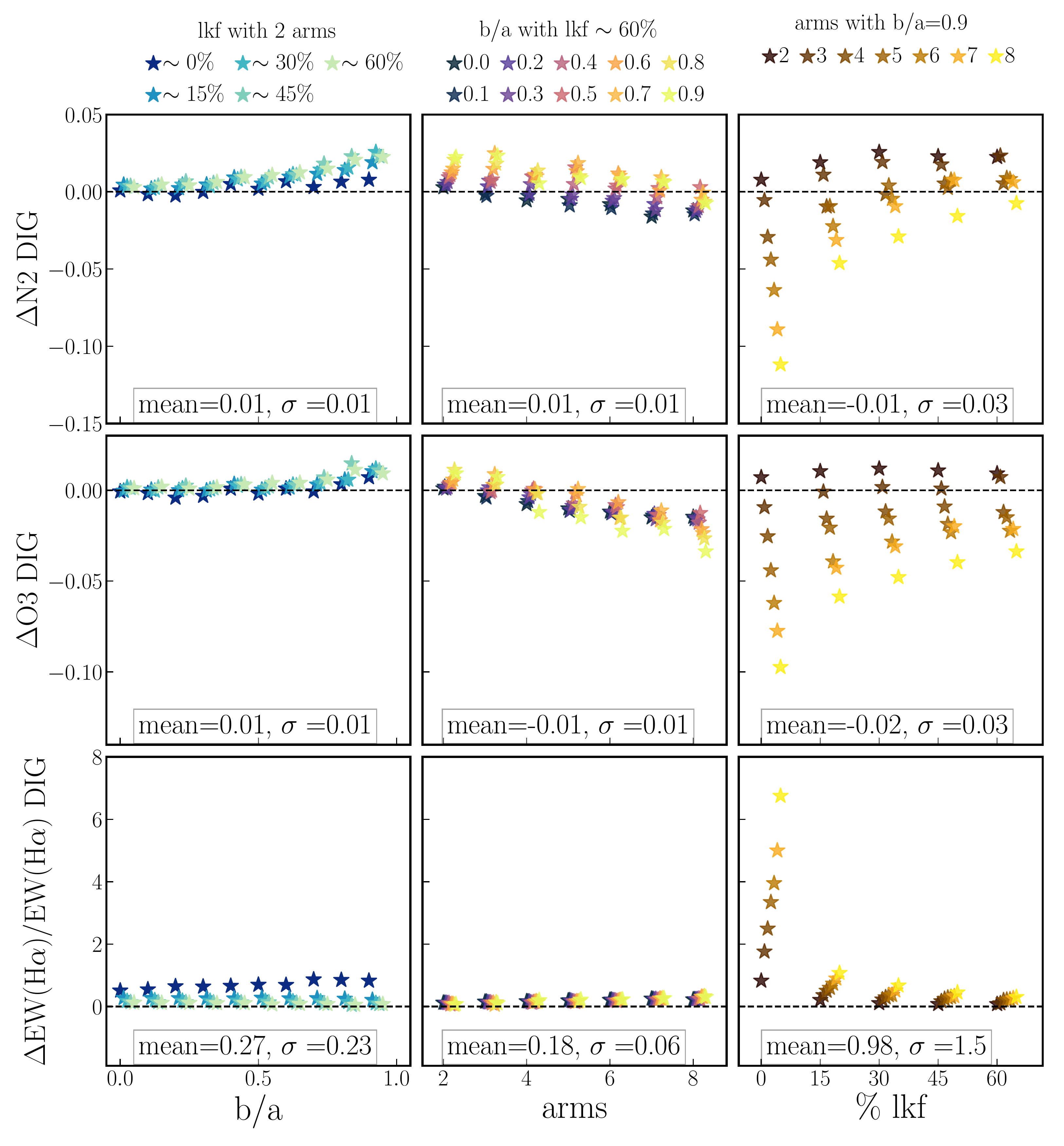}
    \endminipage
    \caption{Recovering the line ratios of the DIG component from the explored set of simulations. Panels, figures, colors, and symbols are all similar to those presented in Fig. \ref{fig:hii}, but for the diffuse emission.}
    \label{fig:dig}
\end{figure*}
%%%%%%%%%%%%%%%%%%%%%%%%%%%%%%%%%%%%%%%%%%%%%%%%%%%%%%%%%%%%%%%%%%%%%%%

\subsubsection{Line ratios of the DIG component}
\label{sec:sim_rat_dig}
 
Even though the main aim of \pyHII\ is the estimation of the properties of the \hii\ regions, since the code derives an estimation of the DIG (Sec. \ref{subsec:digmapinit} and \ref{subsec:digmapfin}), we use the simulations to determine how well the explored line ratios and EW(\Ha) are recovered. Like in the case of the \hii\ regions, for this purpose we investigate the behavior of the $\Delta$N2, $\Delta$O3, and \rEW\ parameters as a function of the varying parameters in each of the three sets of simulations. 

Figure \ref{fig:dig} represents the same distributions shown in Fig. \ref{fig:hii}, using the same layout and the same symbols and color schemes, this time for the DIG component.
%(difference output/input) from the mean values for DIG, this case is illustrated in figure \ref{fig:dig}, where parameters variation is same as for figures \ref{fig:rrr} and \ref{fig:hii}.
In general, the recovery of the considered parameters for the DIG component show clear trends, unlike the case of the \hii\ regions for some particular simulations. Thus, it is evident in which cases the diffuse is better (or worse) recovered.

%some simulations cases of \hii\ regions line ratios, this is a good indicator since we suggest that the characterization of decontamination of both components is appropriate giving us information as "clean" as possible when separating these ionization sources. 

Going deeper into the content of Figure \ref{fig:dig}, in the case of $\Delta$N2 (top panels) the values range between -0.12 and 0.03 dex, although in most of the simulations (in particular all those with $f_{\rm leak}$ $>$20\%) its value is restricted to a much narrower range of $\pm$0.03 dex around the zero value. 
The first set of simulations (varying $b/a$ and \lkf) is the one that shows the smallest dispersion, with values between zero and 0.025 dex, showing a bias towards values slightly larger ($\sim$0.01 dex) and a clear trend to larger values as $b/a$ increases, a trend that it is more clear for larger values of \lkf\ (being almost absent for \lkf$\sim$0). The second set of simulations (varying the number of arms and the $b/a$ ratio) present a slightly larger dispersion, with a clear trend towards lower values of $\Delta$N2 as the number of arms increases and the galaxy becomes more inclined. On the contrary, $\Delta$N2 increases as the number of arms decreases and the galaxy becomes less inclined. The worst results are derived for the third set of simulations, those in which \lkf\ and the number of arms are varied for an edge-on galaxy. In this case, $\Delta$N2 may present a value as low as -0.1 dex (strongest negative bias) for the lowest values of \lkf\ and the larger number of spiral arms. There is a clear trend to improve the bias, increasing $\Delta$N2 as the \lkf\ increases and the number of spiral arms decreases. Indeed, for 3-4 spiral arms the recovered value is essentially the same as the simulated one. Curiously, for the case of the lowest spiral arms, there is a positive bias of $\sim$0.02 dex for almost any value of \lkf. 

The distributions described for O3, shown in the middle panel are remarkably similar, with just a few minor differences. $\Delta$O3 ranges between -0.09 and 0.02 dex for the three sets of simulations, describing similar trends with $b/a$, number of arms and \lkf. If any difference is apparent, it may be that for the second simulation the trend with $b/a$ seems to get inverted (from a positive to a negative trend) between high and low inclinations. The most remarkable result from the exploration of both $\Delta$N2 and $\Delta$O3 is the un-ability to recover the right line ratios as the number of arms increases, in particular for the largest number of arms and the lowest value of \lkf. This result is completely natural since in this case the field is crowded with \hii\ regions (thus, it is more difficult to sample the diffuse regions) and the line ratios are more different than those of the nearby \hii\ regions (therefore, the contribution of the wings of the \hii\ regions pollute the diffuse with a very different emission).

%and the variation of the \lkf\ as function of arms number (right panel) the ones that has the greatest dispersion (-0.11 to 0.03). We show that in the variation b/a as function of \lkf\ values (left panel) a greater recovery is obtain at low \ values with low b/a values, but with high b/a values there is a tendency to overestimate N2 output values, opposite to that we retrieve from \hii\ regions case. For the variation of the arms number as function of b/a (middle panel), N2 recovery increases at low values of b/a while arms number is also low, however, when increases, the trend is reverse, and the better recovery is at high values of b/a. Finally, for the \lkf\ variation as function of arms number (right panel), the best recovery occurs at lower number of arms.

%For the middle panels, we show $\Delta$O3, where the total dynamic range is between -0.9 and 0.012, with trends similar to those corresponding to the recovery N2, being the variation of b/a ratio as function of \lkf\ with less dispersion and the variation of \lkf\ as function of arms number with greater dispersion. From this we find that the best recovery occurs at a lower arms number when b/a ratio is varied simultaneously with low values, and in general \lkf\ factor is the parameter that least affect.

Finally, the recovery of the EW(\Ha) for the diffuse emission is shown in the lower panels. For the three sets of simulations, the range of \rEW\ covers a dynamical range between 0 and 7, although most of the range is restricted to values below 0.5. This is the case of all simulations of the second set, and those with \lkf$>10$\% of the first and third set. Like in case of the line ratios the worst recovery of the EW(\Ha) of the DIG component is found when the number of spiral arms is high and in particular when the \lkf\ is low. This result completely agrees with those found for the line ratios in agreement with the proposed scenario: as the field becomes more crowded and the line ratios more different between the DIG from those of the adjacent \hii\ regions, the ability to recover their properties is worse.

\subsection{Comparison with other codes}
\label{sec:comp_other_codes}

Once the ability of \pyHII\ to recover the properties of both the \hii\ regions and the DIG for an idealized set of simulations has been characterised, we now establish how our code compares with a subset of codes of similar purposes: {\sc pyhiiexplorer}, {\sc SourceExtractor}, {\sc HIIphot} and {\sc astrodendro}. We choose a heterogeneous set of detecting codes, trying to cover different methodologies ({\sc SourceExtractor}, {\sc HIIphot}, {\sc astrodendro}) and the code from which the current one evolved ({\sc pyhiiexplorer}). These algorithms are indeed the most frequently used in the recent literature \citep{huwanglin2018,schinnererhughesleroy2019, zhangzavagnolopez2021}. However, the comparison can be easily extended to other codes like ProFound \citep{profound}, MTObjects \citep{teeninga2015}, NoiseChisel \citep{noisechisel}, DAOPHOT \citep{daophot}, and many others, but we consider that the current selection is sufficient to place the strengths (or weaknesses) of our code with respect to others in the literature.

\subsubsection{Description of the adopted codes}

Here we present a brief description of the different algorithms to be compared with \pyHII, highlighting their similarities and differences with our code. To review particularities please consult the quoted articles in which each code is presented.

%We present a brief description of the different adopted codes, highlighting their similarities and differences with our code, below, we will briefly describe each one of them (to review particularities please consult the quoted articles in which each code was presented).

\noindent \underline{{\it PYHIIEXPLORER}}: This code is written in {\sc Python} and it is based on a previous version called {\sc hiiexplorer} \citep[originally written in {\sc Perl}][]{sanchez12b}, but with all the advantages for distribution of the code, installation and compatibility among operative systems offered by {\sc Python}. The main assumptions of this code for detecting candidate \hii\ regions are essentially the same as those adopted by {\sc hiiexplorer} (and our code). Thus, it relies on (i) a typical size for the \Hii\ regions (a few hundred parsecs, which correspond to a few arcsec at the distances of the used galaxy sample) and (ii) the high contrast of the emission line intensity maps.

The procedure to detect clumpy structures comprises three basic steps: ($i$) the identification of local maxima (i.e., peaks), ($ii$) aggregation of pixels adjacent to each peak to construct a segmentation image (this is done iteratively until no new local maximum is found) and ($iii$) final iteration over the segmentation map to redistribute pixels to the nearest peak when there are two \Hii\ regions that overlap. The procedure is controlled by selecting a set of thresholds that defined a lowest flux intensity threshold, the lowest intensity to define a peak, the largest relative difference in flux intensity between a peak and any pixel to be aggregated to the same \hii\ region, and finally, the largest size allowed for those regions. To extract the fluxes of each emission line for each \hii\ region, the code uses the generated segmentation map and calculates an \Ha\ luminosity-weighted flux. In addition, it extracts the properties of the underlying stellar populations following a similar scheme. For a more detailed description of the code see \cite{espi20}.

\noindent \underline{{\it SExtractor}} or Source Extractor: This code is written in C and it was designed to process large digital images for large amounts of survey data to detect and segregate individual sources. It analyzes an image in six steps: ($i$) estimation and subtraction of the background; ($ii$) peak finding above a defined threshold; ($iii$) source deblending based on an isophotal analysis; ($iv$) filtering of the detections to avoid spurious sources; ($v$) photometry extraction of each source; and ($vi$) separation between galaxies and stars. Despite the fact that this code was developed to detect, segregate and derive the photometry of deep field images, with the ultimate goal to explore galaxies, it has been used to detect \Hii\ regions in emission line images \citep{rousseauneptonrobertmartin2018}. For a more detailed description of the code see \cite{bert96} and \cite{holwerda2005}.

\noindent \underline{{\it HIIphot:}} {\sc HIIphot} is an automated method for photometric characterization of \Hii\ regions, written in IDL. The processing of an image consists on the following four steps: ($i$) initial detection of the \Hii\ region based on a shape matching procedure (using a pre-defined set of shapes); ($ii$) generation of a seed catalog of regions cleaning multiple and spurious detections; ($iii$) aggregation of pixels to each detected region until a certain limiting brightness is reached; and ($iv$) correction by the background (or DIG), based on those pixels not assigned to any region. For a more detailed description of each parameter and a more detailed description, see \cite{thilk00}.

%GO TO THESIS.. all the processing of an image consists of four steps: i) initial detection of sources making a match between six different morphologies (covering from Gaussians to rings), where for each morphology it is predict the surface brightness model and it is determinate a tentative match of sky coordinates for each model with local maxima, convolving the data with appropriately sized circular Gaussians, thus leaving only objects those that have a minimum value of noise-corrected version of Pearson's linear correlation, $\rho$; ii) building the footprint removing multiple detections from the same source taking into consideration the "claimed" centroid and the ratio signal-noise (critic) with a cut-off defined by 50\% median data value ("seeds"); iii) iterative growth of detections, making the seed pixels add adjacent pixels in each iteration carefully, limiting the growth of a region until the observed surface brightness profile is flattened or no more pixels can be reached; iv) correction of background (or DIG) which is defined as unclaimed pixels at a projected distance of 250 pc from an HII region, with the posterior selection of some "control" points to represent the background and finally compute the median value of all background and a an image of DIG with those values. The output parameters are images (continuum-subtracted, background surface or DIG image, footprint, seed and grown region maps) and a catalog of the detected regions (that includes an ID number, sky coordinates, pixel position, number of pixels by regions, and others). 

\noindent \underline{{\it astrodendro:}} This package is written in {\sc Python}, being developed to derive a 2D dendrogram structure from any astronomical image (or cube). A dendrogram is a tree structure where the data are hierarchical. It has has two components: branches (structures that are divided into sub-structures, either on branches or leaves) and leaves (structures without sub-structures). The branches and leaves are all united to a single trunk, which is a structure without any parent structure. The association is based on the level of flux intensity or surface brightness in the image. In summary, the code assigns each pixel to a leaf, that is associated with several branches, and a trunk, generating different segmentation maps for each level of flux intensity. Thus, a dendrogram can be thought of as a segregation based on an isophotal analysis.

%GO TO THESIS...In this algorithm there are three main parameters: min\_value, min\_delta and min\_npix, which are the minimum pixel value to be considered, the minimum height for a sheet to be defined as independent and the minimum number of pixels for a sheet to be defined as independent, respectively. 
%Subsequently, what is relevant is the statistics that can be performed with the astrodendro.analysis module, either structure by structure or as a catalog. Some of the output parameters of the statistics are: flux, major\_sigma, minor\_sigma, position angle, radius, area\_exact, area\_ellipse, x\_cen, and others.

Once the dendrogram has been calculated,  it is possible to access to the entire tree, or at each level: trunk, branches and leaves. In our particular study case, the latter can be associated with individual \Hii\ regions \citep[e.g.][]{zaragozabeckmanfont2015, rodriguezbaumefeinstein2019}. Therefore, it is possible to create a catalog and extract the properties of those regions either using a segmentation map or a weighted extraction procedure. For a more detailed description of each parameter and the code, see \cite{astrodendro}.

%Once the additional codes that are used to compare and characterize our code have been introduced, below, we will describe second set of tests: implementation of the simulations in pyHIIextractor and the other 4 codes.

%%%%%%%%%%%%%%%%%%%%%%%%%%%%%%%%%%%%%%%%%%%%%%%%%%%%%%%%%%%%%%%%%%%%%%%%%%
%  Rad vs. Flux: Other codes
\begin{figure}
    \minipage{0.99\textwidth}
    %\centering
    \includegraphics[scale=0.38]{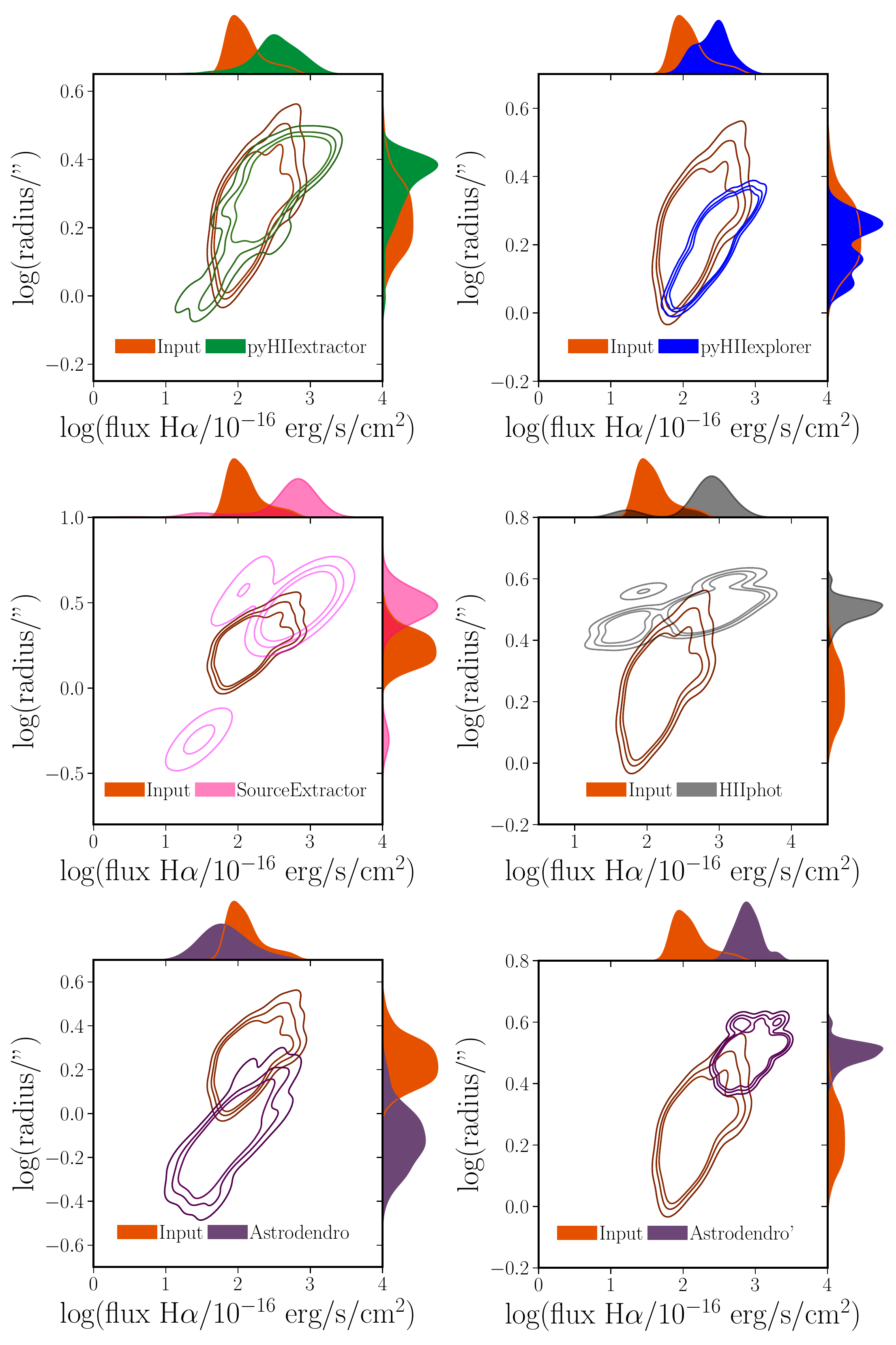}
    \endminipage
    \caption{Distribution of the recovered sizes along the \Ha\ fluxes estimated for the simulated \Hii\ regions on the archetypal galaxy adopted to compare the different explored codes, together with the input/simulated distribution. Each panel shows the results for a different code, as a set of contours enclosing a 15\%, 50\% and 99\% of the values, respectively. Similar contours are provided for the input values. In addition, each panel shows the projected distributions for each value.}
    \label{fig:frad_all_sim}
\end{figure}
%%%%%%%%%%%%%%%%%%%%%%%%%%%%%%%%%%%%%%%%%%%%%%%%%%%%%%%%%%%%%%%%%%%%%%%%%%%

\subsubsection{Comparison based on simulations}

To characterize how our code compares with the ones described above, we decided to first apply all of them to the same simulated galaxy. From the set of simulations described in Sec. \ref{sec:sim} and Sec. \ref{sec:ref_sim}, we choose as an archetypal galaxy a face-on one, with two spiral arms and a \lkf\ of 60\%. This is the case with the best recovery rate for any of the explored parameters, as described in the previous sections. Then, we apply the full suite of codes (including \pyHII) described previously to this simulated galaxy. For doing so we first need to optimize and homogenize the different input parameters required by each code (that have been designed for different purposes as indicated before). Therefore, we tune those parameters with the goal of maximizing the recovery of \hii\ regions, obtaining a distribution of sources, fluxes, and radii that follows as much as possible the input one. As we are not expert users of the comparison codes, it is possible that our selection of parameters could be improved, however, we tried to find those parameters as if we had chosen those codes for a common science goal (i.e., to explore the \hii\ regions in a sample of galaxies observed with MUSE at $z\sim$0.02).

Once each code was applied to the chosen simulated galaxy we need to homogenize their outputs and recover from each one the main parameters recovered by {\sc pyHiiextractor} for an emission line image, i.e., the coordinates, radii, and fluxes of the \hii\ regions. This way it is possible to make a homogeneous comparison among the results provided by the different codes, and between those results and the input/simulated parameters. Radii and coordinates for \pyHII\ and {\sc SourceExtractor} are derived directly by these codes.
On the other hand, {\sc pyhiiexplorer} and {\sc HIIphot} produce segmentation maps as output parameters, whose transformation to radii and coordinates involves additional processing. Based on the segmentation map, for each region, it is possible to estimate the barycenter and the area (as the sum of the number of pixels). Then, from these areas we derive approximate radii as $r=\sqrt{\frac{A}{\pi}}$, i.e., assuming an almost circular shape. Finally, {\sc astrodendro} provides its own estimation of the centroids and radii of the leaves. However, we noticed that these radii are in general smaller than the apparent size of the leaves. Therefore, for this code, we repeat the calculation outlined before and provide an alternative estimation of the coordinates and radii, based on the generated segmentation map. Once we obtain the radii and the coordinates, the extraction of the fluxes is carried out with the algorithms included in \pyHII, except {\sc pyhiiexplorer} that already includes in its code flux derivations.

%In the particular case of {\sc astrodendro}, for the derivation of the radii with the alternate methodology to its code, we use the calculated leaves of the dendrogram to calculate the barycenter, area and radius.
%SFS 23.10.21

%for two reasons: ($i$) to building a luminosity function and ($ii$) to building an image derived from each one to qualitatively compare.

%For doing so
%All the discussed codes have different input parameters because, as previously summarized, they are used for different purposes, however, in this study we optimize each case to have the highest qualitative recovery of \hii\ regions with the best structural recovery. Our aim in this section is to homogenize the outputs of codes and recover from each one, coordinates, radii, and fluxes of the \hii\ regions for two reasons: ($i$) to building a luminosity function and ($ii$) to building an image derived from each one to qualitatively compare.
%In the order of importance of obtaining, we first seek from each code to obtain the radii and the coordinates, since from them we can derive the fluxes from said sources.

%%%%%%%%%%%%%%%%%%%%%%%%%%%%%%%%%%%%%%%%%%%%%%%%%%%%%%%%%%%%%
% Simulated images by codes
\begin{figure}
    \minipage{0.49\textwidth}
    \centering
    \includegraphics[scale=0.32]{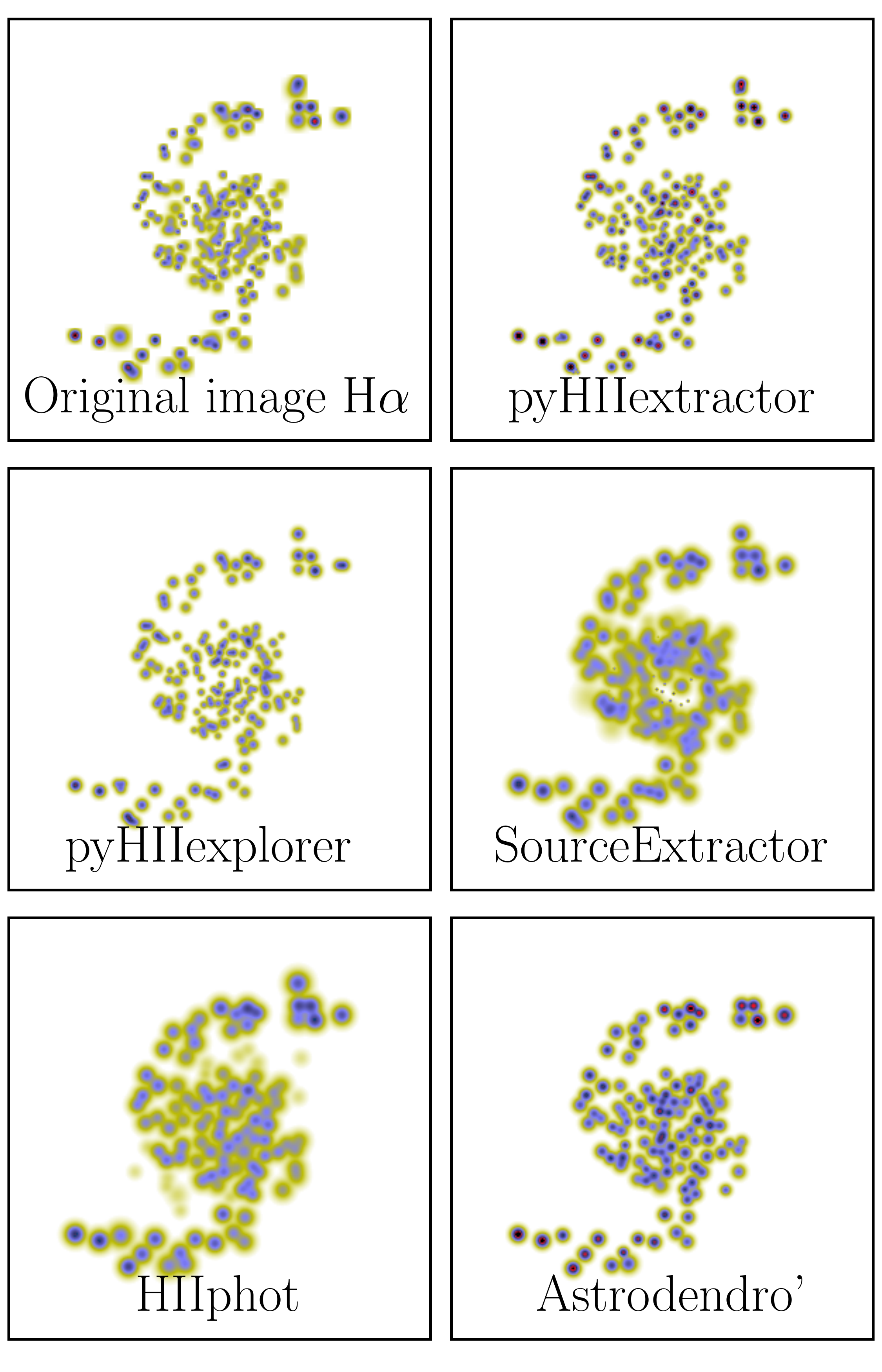}
    \endminipage
    \caption{Simulated \Ha\ intensity distribution for an archetypal spiral galaxy (top-left panel), together with the corresponding models created based on the distribution of recovered \Hii\ regions by the different explored codes:
    \pyHII\ (top-right), {\sc pyhiiexplorer} (middle-left), {\sc SourceExtractor} (middle-right), {\sc HIIphot} (bottom-left) and {\sc astrodendro'} (bottom-right)} %for the simulation of a spiral galaxy type with 2 arms. }
    \label{fig:sim_oc}
\end{figure}
%%%%%%%%%%%%%%%%%%%%%%%%%%%%%%%%%%%%%%%%%%%%%%%%%%%%%%%%%%%%%

Figure \ref{fig:frad_all_sim} shows the distribution of the recovered radii as a function of the recovered fluxes for the different explored codes (one for each panel), together with the same distribution for the simulated data (shown in all panels).
% SFS: This is more for caption than for the text.
%The summary of the obtaining of the radii, coordinates and fluxes is illustrated in the figure \ref{fig:frad_all_sim}, where each panel represents the comparison between output and input of each program for the size distribution of \hii\ regions and fluxes. Contour levels are 15\%, 50\% and 99\%, orange contours represent the input: size distribution of the \hii\ regions simulated along the fluxes (which is the same for all panels), and green, blue, pink, gray and purple contours levels represent different outputs programs: pyHIIextractor, {\sc pyhiiexplorer}, {\sc SourceExtractor}, {\sc HIIphot} and {\sc astrodendro}, respectively. 
In the case of {\sc astrodendro'} we show the two derivations of the parameter indicated before, with the estimation based on the segmentation map labelled as {\sc astrodendro'}.

%case and being the only one from which we can derive radii in two different methodologies: using the same code or deriving them as described for the segmentation maps, we decide to show both distributions; for practical purposes the distribution whose radii are derived with segmentation maps type methodology we define it as: {\sc astrodendro}'.
Based on the results shown in Fig. \ref{fig:frad_all_sim} we find that all the codes present clear differences between the recovered and simulated distributions. The code that reproduces better the trend/correlation observed between the two parameters is {\sc pyHIIextractor} (upper left panel). However, our code overestimates the number of \Hii\ regions at high fluxes (large radii), underestimating this number at low fluxes (small radii). We attribute this effect to the inability of the code to segregate/recover the smaller (fainter) regions, that are aggregated to larger (brighter) ones. On the contrary, {\sc pyhiiexplorer} (upper right panel), the code that was the basis for the current development, systematically underestimates the recovered radii, overestimating both the fluxes and the number of regions at high fluxes. Although there is a trend between both recovered parameters, it presents an offset with respect to the simulated one (as a consequence of the overestimation indicated before). 

For the remaining comparison codes we find larger discrepancies. For instance, the distribution recovered by {\sc SourceExtractor} (middle left panel) shows a set of discontinuities, with different peaks in the space of parameters. Most of the values are concentrated above the observed ones, showing a clear overestimation in both the fluxes and radii and with a much narrower dynamical range (in particular the radii). In addition, there are a few regions in which recovered fluxes and radii are much lower than the observed ones.
%a distribution divided, so to speak, into two large distributions (the contour levels corresponding to largest sizes distribution is broader so that systematically larger \hii\ regions are detected), one whose estimate of radii is high with high derivation of fluxes (overestimated), and another distribution of small radii with flux values that are also small (underestimated), but both large distributions are not connected to each other. 
As a consequence, {\sc SourceExtractor} does not return a distribution compatible with the simulated one. In the case of {\sc HIIphot} and {\sc astrodendro'} we find similar results, with a clear overestimation of both parameters, and values of the radii concentrated within a much narrow range than the input ones. The main difference with respect to the results found using {\sc SourceExtractor} is the lack of secondary peaks in the distribution (i.e., there is no regions with underestimated radii and fluxes, in average). 
%(middle right panel) shows level contours only for \hii\ input regions with large sizes and an estimate of fluxes that are also large, however, we could say that this program derives radii of the same size from each other but overestimating and underestimating the fluxes. The contours between input and output only coincide for 50\% and 99\%, the input contour of 15\% does not coincide with the output contour for this case. 
Finally, in the case {\sc astrodendro'} (bottom left panel), both the radii and fluxes are clearly underestimated. This implies that the leaves derived by the program as the ionization sources corresponds to fragments or just the peaks of larger \hii\ regions. On the other hand, this code is able to reproduce the trend between radii and fluxes, although with an offset in both parameters.
%As for the contours, these do not coincide more than at the edges for the values of 99\% and 50\%, leaving aside the coincidence of the corresponding contour to 15\%. And {\sc astrodendro} case 2 (bottom right panel) the radii and fluxes are overestimated, but there is a better coincidence of the level contours, and with a smaller dispersion between the values.
Hereafter we will use the values for fluxes and radii derived using the estimation based on the segmentation map, i.e., {\sc Astrodendro}', that recover larger values for both parameters.

%The latter implies that then {\sc astrodendro} 2 case, where we derive the radii without using those estimated by the same {\sc astrodendro}, is the most appropriate radii estimate to continue with the comparisons and evaluations between codes.

%\noindent { pyHIIextractor:} We use the image \Ha\ from the simulation as the main input parameter. This map pass through a gaussian filter with a sigma value=2 to create a continuum map which is use for detection, some other parameters necessary are FWHM, spax\_scale, MUSE\_1sig, MUSE\_1sig\_V, refined, num\_sigma, DIG\_lim and max\_size which values are 1.0, 0.2, 1, -1, 3, 300, 1, 1.75, respectively. We obtain a image with the best model of candidates to \hii\ regions and a catalog of the \hii\ regions with the positions and radii and later we derive fluxes.

%\noindent { PYHIIEXPLORER:} We use the simulation cube and other parameters such as max\_dist, frac\_peak, F\_min, dist, min\_flux, PSF, XC and YC which values are 3, 0.1, 100, 1, 0, 0.5 and 160, 160, respectively, to obtain a segmentation map. Additionally, we process the segmentation map outside of the code content. For each segmented region we calculate barycenter, area and radius, to construct the image of the candidates to \hii\ regions with a function of pyHIIextractor.

%Subsequently, we construct the image that only contains candidates to \hii\ regions detected, with radii and coordinates as input parameters of each program. All the images of \hii\ regions models are illustrated in the figure \ref{fig:sim_oc}, and are performed based on a pyHIIextractor function. 

Based on the fluxes, radii and spatial distribution of the detected \hii\ regions by each code, we construct the corresponding model image, following the procedure described in Sec. \ref{sec:sim_Hii}. Figure \ref{fig:sim_oc} shows those images for the different codes explored here, together with the one corresponding to the input simulation. It is appreciated that qualitatively all codes recover to a certain extent the simulated distribution of \hii\ regions. 
%As a consequence of the biases in radii and fluxes described before, 
In the case of {\sc pyHIIextractor} (upper right panel) the code recovers well the regions with smaller radii at the center of the simulated galaxy (upper left panel). While it presents a bias, described before, towards aggregating small/faint regions into larger/brighter ones, it is the code that best reproduces the observed distribution in the central and inter-arm  regions, allowing proper recovery of the galaxy's structure. 
This code is the one that presents the highest recovery rate of the ones explored here, with a \frec$\sim$83\%.
{\sc pyhiiexplorer} (middle left panel) recovers equally well the structure of the simulated distribution, although its deficiency to detect fainter regions is more notable in the central region of the galaxy. Its \frec\ is also high, $\sim$81\%, similar to the one derived using {\sc pyHIIextractor}.

{\sc SourceExtractor} (middle right panel, Figure \ref{fig:sim_oc}), qualitatively recovers systematically larger regions. Therefore it clearly aggregates the small regions into larger ones to a larger extent compared to two previous codes. This is translated into a lower \frec\ of $\sim$70.5\%.
Although it does not introduce spurious detections in the inter-arm regions, as the derived regions are larger than the simulated ones they are effectively overlapped and mixed, reducing the spaces between them. Another case in which the detected regions are systematically larger is {\sc HIIphot} (bottom left panel). However, this is not its only drawback because {\sc HIIphot} introduces spurious/ghost \hii\ regions in inter-arm areas that are not found in the original \Ha\ map. Despite these false detections, it presents the lowest \frec\ of all the explored codes ($\sim$56\%).
Finally, {\sc astrodendro}' (bottom right panel) present deficiencies in the recuperation of simulated regions in both the center and the spiral arms, affecting not only faint/small regions but bright ones too (e.g., the bright region in the upper-right edge of the upper spiral arm).  As a consequence, it presents a low detection rate too (\frec$\sim$56.5\%). We note that the use of the radii intrinsically derived by {\sc astrodendro'} instead of the ones derived using the segmentation maps is irrelevant regarding the effect in the detection rate.
%the simulation and in the spiral arms, without being able to recover the regions with smaller radii.
%Although it does not introduce detections in the inter-arm regions, its recovery percentage is only 56.5\%.

% SFS 25.10.2021

%%%%%%%%%%%%%%%%%%%%%%%%%%%%%%%%%%%%%%%%%%%%%%%%%%%%%%%%%%%%
% Size-Flux for NGC4030 by different codes
\begin{figure}
    \minipage{0.49\textwidth}
    \centering
    \includegraphics[scale=0.5]{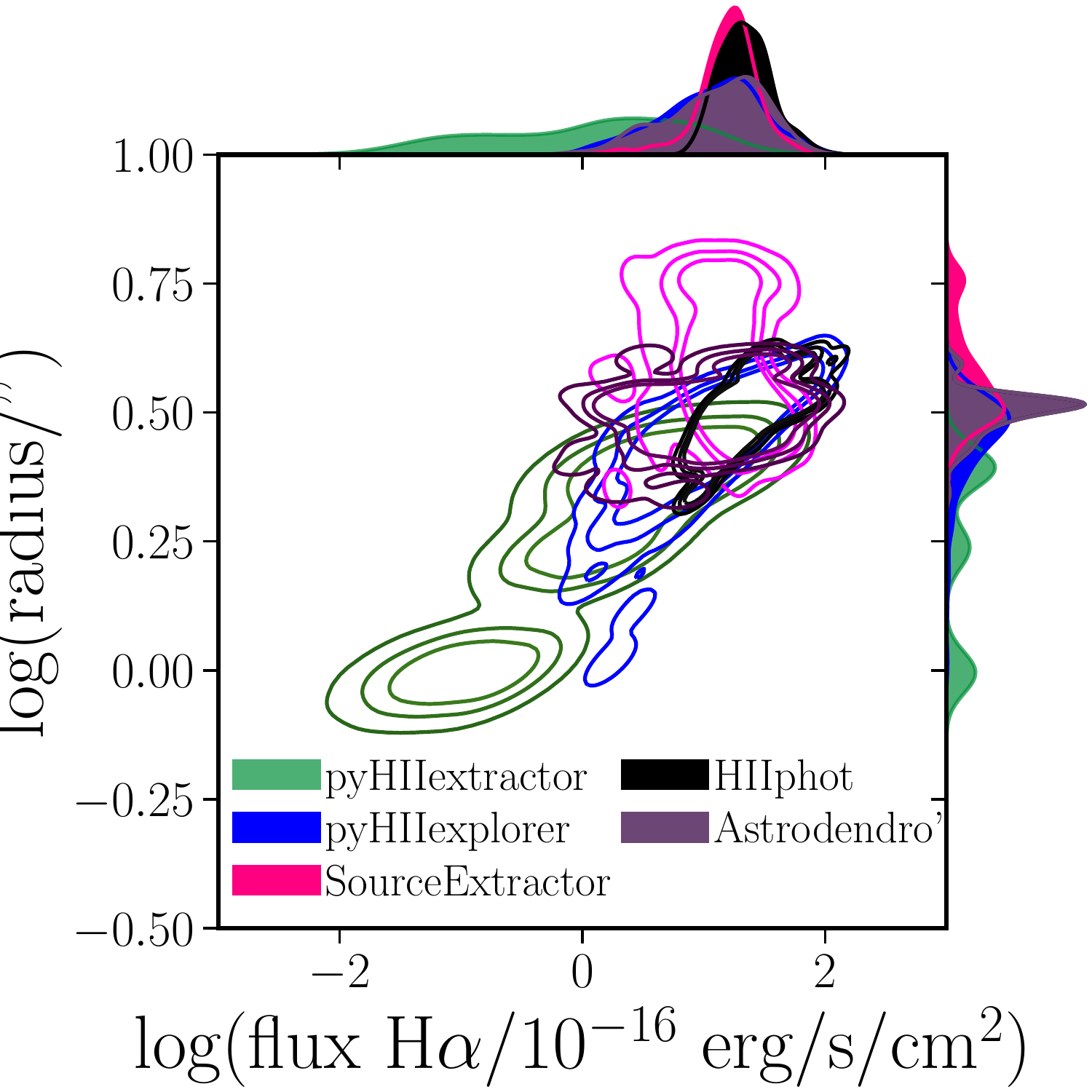}
    \endminipage
    \caption{Distribution of the size along the fluxes in \Ha\ estimated for the candidate \Hii\ regions in the explored NGC4030 MUSE data as derived for the different codes adopted in this comparison (indicated in the legend). Each contour corresponding to the area that encircles a 99\%, 50\%, and 15\% of the regions, respectively}
    \label{fig:frad_all}
\end{figure}
%%%%%%%%%%%%%%%%%%%%%%%%%%%%%%%%%%%%%%%%%%%%%%%%%%%%%%%%%%%%

%the simulated galaxy had been explored, we chose a galaxy with MUSE data, in order to test the detection of candidates for \hii\ regions with the codes to compare. The galaxy for case study is NGC4030.
%at a distance of 29 Mpc

\subsubsection{Comparison based on real data}
%Candidates to \Hii\ regions for NGC 4030}

After demonstrating the different outputs of the explored codes when applied to a simulated galaxy in the previous subsection, we now apply them to real data. In this case, we cannot control the shape, size, fluxes, and distributions of the regions, and therefore, we cannot compare with the input/simulated values. On the contrary, this exploration allows us to determine which code reproduces better the observed intensity distributions. Furthermore, it allows one to explore the recovery of  well known relations (e.g., flux-size) and properties (e.g., distribution in diagnostic diagrams) of both the \Hii\ regions and the DIG.

For this experiment, we chose the MUSE data of the NGC 4030 galaxy. This is an Sbc spiral galaxy at $z\sim$0.004, with multiple arms not following a classical Gran Design pattern. It is a star-forming galaxy with the presence of a young stellar population and a large number of \Hii\ regions. The data were observed under the umbrella of the MUSE Atlas of Disks \citep[MAD,][]{erroz19, denbrokcarollo2020}, being included in the AMUSING compilation \citep[AMUSING,][]{galbanyandersonrosales2016}, as it hosted the SN2007aa \citep{sn2007aafolatelli}. These data have been used to explore the anomalous presence of N2 regions \citep{cidfernandescarvalho2021}, being part of the exploration of the frequency of outflows done by \citet{lopezcobasanchez2020} too. It was also included in previous IFS studies, such as the SAURON survey \citep{ganda:2006p3135}, using data of lower spatial resolution (covering a smaller FoV). 

%For all these reasons this object and data 
%, MUSE Atlas of Disks (MAD) survey \citep{erroz19, denbrokcarollo2020} and the AMUSING++ Nearby Galaxy Compilation \citep{lopezcobasanchez2020}.
%, being ideal for testing the set of selected codes. 

%GO TO THESIS ... PROPERTIES OF NGC4030... characterizing and estimating some parameters such as its inclination (47°), effective radius (4.4 kpc), stellar mass ($10^{11.2}$M$\odot$), gas phase metallicity (12+log(O/H)$\sim$0.9), star formation rate (11 M$_{\odot}$ $yr^{-1}$), and the null presence of outflows. 
%In particular this galaxy is interesting due to the existence of unresolved sourced with enhanced forbidden line emission, that although some may be planetary nebulae (PNe), supernova remnants (SNRs), \hii\ regions, or variable blue stars (SN impostors) as described in \cite{cidfernandescarvalho2021}, its nature is not fully known from such unresolved sources. 

%NGC 4030 was observed with the Multi Unit Spectroscopic Explorer (MUSE) at ESO's Very Large Telescope as part of MUSE Atlas of Disks (MAD) survey and later added to AMUSING++ Nearby Galaxy Compilation, the data using in this analysis were taken, being downloaded from ESO archive. 

%%%%%%%%%%%%%%%%%%%%%%%%%%%%%%%%%%%%%%%%%%%%%%%%%%%%%%%%%%%%%%%%%%%%%%%%%%%%%%%%
% Ha and BPT comparison among the different codes
\begin{figure*}
    \minipage{0.99\textwidth}
    \includegraphics[trim=0 0 0 0, clip, width=\textwidth]{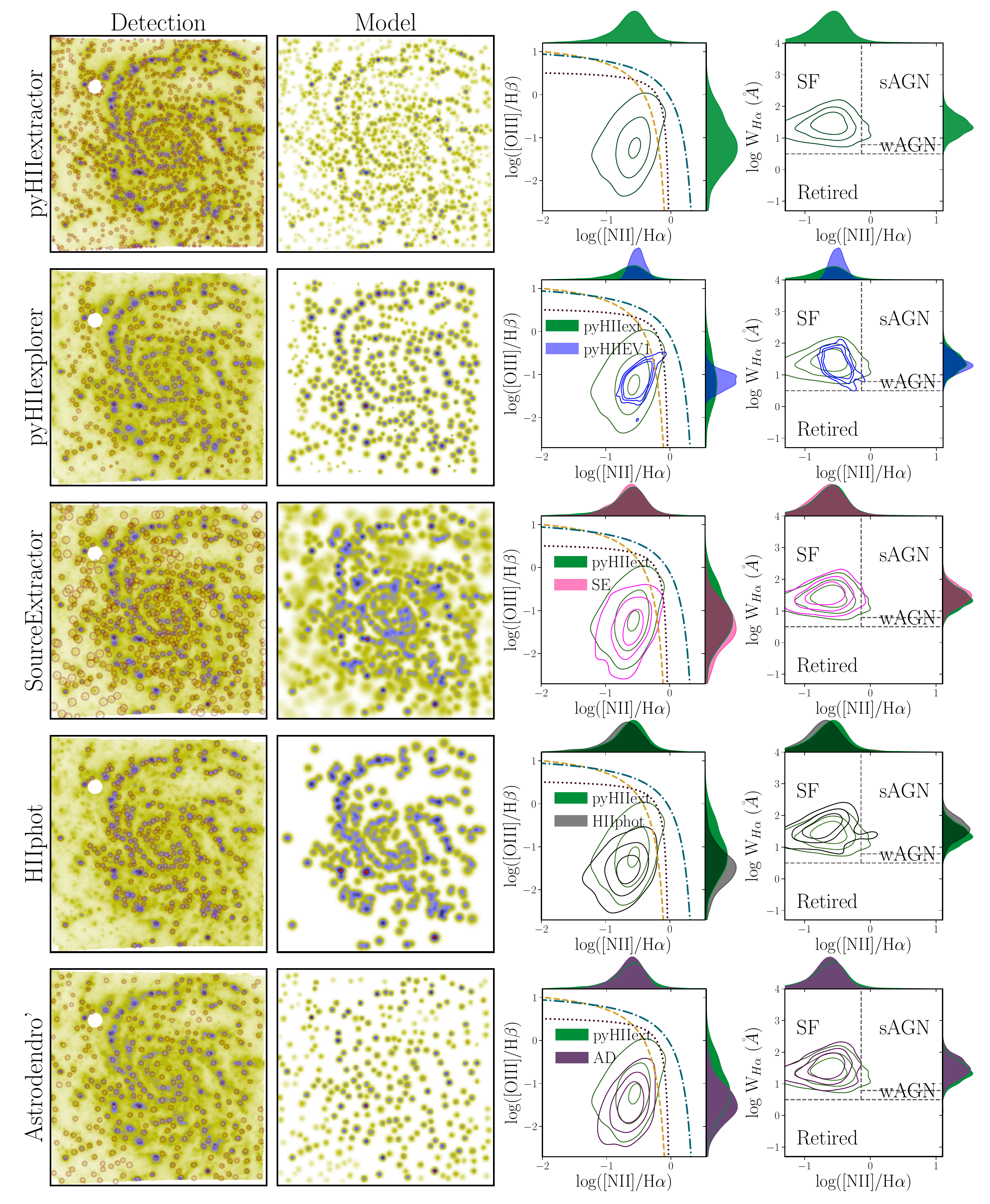}
    \endminipage
    \caption{Comparison of the recovered properties of the \hii\ regions using different codes. From left to right each panel shows, for each of the explored codes: (i) the original \Ha\ image provided by {\sc Pipe3D}, together with the detected \hii\ regions (red circles, with the size corresponding to the recovered radii); (ii) \Ha\ model-image generated based on the radii, fluxes and spatial distribution of recovered \Hii\ regions, assuming a Gaussian shape, as described in Sec. \ref{sec:sim_Hii}; (iii) BPT and (iv) WHAN diagnostic diagrams showing the density distribution of \hii\ regions as contours (encircling 90\%, 50\%, and 15\%, from the outside-in respectively). In each of these diagrams, the lines correspond to the same demarcation lines and regions described in Fig. \ref{fig:simula}. Each row corresponds to a different code (from top to bottom): \pyHII\ (pyHIIext), {\sc pyhiiexplorer} (pyHIIEV1), {\sc SourceExtractor} (SE), {\sc HIIphot} (HIIphot) and {\sc astrodendro}' (AD), respectively.}
    \label{fig:codes}
\end{figure*}
%%%%%%%%%%%%%%%%%%%%%%%%%%%%%%%%%%%%%%%%%%%%%%%%%%%%%%%%%%%%%%%%%%%%%%%%%%%%%%%%

% SFS 25.10.21
The MUSE data downloaded from the ESO archive cover a field of view (FoV) of 1$\arcmin\times$1$\arcmin$  with a spatial sampling of 0.2"/spaxel and a spatial resolution limited by the seeing (FWHM $\sim 1"$, i.e., $\sim$0.1 kpc at the redshift of the object). The spectral data cover the wavelength range between 4750-9300 \AA, with a spectral sampling of 1.25 \AA\ and an FWHM resolution of ~2.4 \AA. As part of the AMUSING++ compilation, the MUSE datacube was analyzed using {\sc Pipe3D}, as described in Sec. \ref{sec:data}. From this analysis we make use of the emission line dataproducts, which comprise, as indicated before, the intensity and EW maps of a set of emission lines. In particular, it includes all the lines involved to explore the BPT and WHAN diagrams (i.e., the N2 and O3 line ratios and the EW(\Ha), in a similar way achieved for the simulations discussed in Sec. \ref{sec:sim_rat}. We consider that these data are suitable to test our code and compare its behavior with that of our set of comparison codes.

%In summary therefore the original datacube have a wide field of view (FoV) of 1' $\times$ 1', a spatial sampling of 0.2" $\times$ 0.2" per spaxel with a spatial resolution is seeing limited, covers the whole optical range from 4750 \AA\ to 9300 \AA\ with a spectral sampling of 1.25 \AA\ and a FWHM$\sim$2.4 \AA\ \cite{lopezcobasanchez2020}. 

% GO TO THESIS ...Regarding the processing of the data, raw data reduction is done with {\sc REFLEX} \cite{freudlingromaniello2013} using version 0.18.5 of the MUSE pipeline \cite{weilbacherstreicher2014}, the NGC 4030's datacube was downloaded from the ESO archive, and subsequently for compilation 

%an adjustment routine to analyze IFS data using package FIT3D \citep{sanchez+2016a}. The adjustment procedure is completely summarized in \cite{lopezcobasanchez2020} and the extended version in \cite{pipe3d_ii}.

We apply \pyHII\ and the literature codes to detect candidate \hii\ regions in the processed MUSE data of NGC 4030, extracting their properties. As indicated before each code makes different assumptions and requires different adjustments of the input parameters to perform the optimal detection and extraction. Following the same scheme adopted for simulations, we perform several experiments for each code trying to maximize the number of detected candidate \hii\ regions and obtain the best representation of the original flux distribution. Then, we first derive the radii, positions, and flux intensities in \Ha\ obtained by each code for the detected candidates.

%As already mentioned previously, in this distribution of sizes along the fluxes, we chose use {\sc astrodendro}' case, due to that this modality of the sizes with our post-processing not fragments regions as the {\sc astrodendro} original processing to derive leaf radii and obtained sizes are more appropriate according to the results of the simulations (see figure \ref{fig:frad_all_sim}).

Figure \ref{fig:frad_all} shows the equivalent to Fig. \ref{fig:frad_all_sim}, but this time using real data. It shows the distribution of sizes along fluxes in \Ha\ estimated for all candidate \hii\ regions for each code. It is seen that the sizes and fluxes derived by {\sc pyhiiexplorer}, {\sc SourceExtractor}, {\sc HIIphot} and {\sc astrodendro'} tend to be larger than those estimated by \pyHII, in general. This is in agreement with what was obtained in the simulations, except for the case of {\sc pyhiiexplorer}. For this later code, based on the simulations, we expected that the sizes cover a similar range of values as the ones covered by {\sc pyHIIextractor}. This is the largest difference found between simulations and real data. We consider that this discrepancy is most probably due to the fact that {\sc pyhiiexplorer} was designed for lower spatial resolution data than the one achieved by the real data analyzed here. On the contrary, our simulation is tuned to replicate a galaxy at a slightly larger redshift (the average one of the AMUSING++ compilation). Thus, the simulated data have a resolution three times worse than the one of the adopted real data, and therefore it is expected that {\sc pyhiiexplorer} provides a worse result in this second case.

As a consequence of selecting smaller (and fainter) regions, covering a wider dynamical range in both parameters than any of the other codes, \pyHII\ recovers the largest number of candidates, with 1787 detections. On the contrary the rest of the codes recovers a much lower number of regions: 365 by {\sc pyhiiexplorer}, 564 by {\sc SExtractor}, 283 by {\sc HIIphot} and finally 287 by {\sc astrodendro'}. In general these codes aggregate several regions segregated by our code into a single one (e.g., {\sc pyhiiexplorer}, {\sc SExtractor} and {\sc astrodendro'}), or they are unable to detect the fainter (smaller) regions (e.g., {\sc pyhiiexplorer} and {\sc HIIphot}).

%worse than {\sc pyhiiextractor}.
%. Regarding, \Ha\ fluxes intensities, all the codes overestimate the values derived by pyHIIextractor, recovering candidates to large \hii\ regions with also large flux intensities.
%of the process also we generate for each code an \Ha\ model image of the candidates to \hii\ regions using our own code, making use of the coordinates, flux intensities and sizes derives (similar to simulations). 

The comparison between the recovered radii and fluxes among the different codes provides us with only relative information. Using real data it is not possible to know a priori which distribution is the closest to the intrinsic one. We generate a model image using the distribution, sizes, and \Ha\ fluxes of the regions recovered by each code (as described in Sec. \ref{sec:sim_Hii}). Figure \ref{fig:codes} shows how these \Ha\ model images (middle-left panels) from the different codes compare with the original image provided by {\sc Pipe3D}. Detected candidates are overlaid on the observed \Ha\ image (left panels).
%This \Ha\ model image can be easily compared with the original image provided by Pipe3D for each of the codes. This qualitative comparison is shown in Figure \ref{fig:codes}, where it is shown the original \Ha\ image, with the \Hii\ regions detected by each code overlaid.

It is appreciated that {\sc pyHIIextractor} provides a model image whose structure is richer than that of any of the tested codes, and qualitatively more similar to the observed one. Other codes, such as {\sc SExtractor} are able to reproduce most of the observed patterns in the original image, or at least the patterns of the brightest regions, like {\sc HIIphot}, although with a lower detail than {\sc pyHIIextractor}. Finally, the global structure of the original image is less clearly identified in the model recovered by {\sc pyhiiexplorer} or {\sc astrodendro'}. This is in agreement with the exploration of the distributions shown in Fig. \ref{fig:frad_all}, and the differences in the number of recovered regions. Our code recovers smaller and fainter regions, being able to reproduce a richer structure, while the remaining codes create larger and brighter regions, being unable to reproduce to the same extent the observed intensity distribution. { %In a similar way that our fiducial mock-galaxy (face-on, two arms, and ~60\% of leaking), in this case, 
In addition, like in the case of the fiducial simulation, we estimate the contribution to the total flux in \Ha\ of both the \hii\ regions candidates ($\sim$44\%) and the DIG ($\sim$63\%) recovered by our code (that produced a $\sim$3\%\ of mismatch).
Those percentages are again similar to those previously reported in the literature \citep[e.g.][]{rela12}}. 

% SFS: Too much text, Alejandra!!!!
%e derive that for NGC4030, the total flux in the \Ha\ emision of the map containing only the candidates is ~44\%, while the diffuse ionized gas is ~63\% (with a 3\% mismatch from the estimated and total recovered flux). All percentages are calculated with respect to the sum of the total flux of emission in \Ha\ from the original map. This simple test is not possible with all codes because not all codes provide an estimate of the ionized diffuse gas.}
%{ From this we highlight that the percentages obtained from the \Ha\ emission either for the case of NGC4030 and the mock galaxy, are in accordance with what the literature indicates, as we cited in the introduction.}

% SFS: 26.10.21
%Once detected the regions and recovered their coordinates, fluxes and radii, we extract the same information for the set of 30 emission lines prodi, but we focus in the ratios \nii/\Ha, \oiii/\Hb\ and EW(\Ha), to construct corresponding BPT and WHAN diagrams to information retrieved by each code. 

%As indicated before, once detected and recovered the structural parameters of the distribution of candidates to \hii\ regions we extract their properties (fluxes and EWs) for the set of lines analyzed by Pipe3D. 

Finally, we use the fluxes and EWs extracted from the {\sc Pipe3D} datacubes for each region detected by each code to explore and compare the distributions along the BPT and WHAN diagram. The results of this exploration have been included in the right-hand panels of Fig. \ref{fig:codes}. We first highlight that despite the number of detected regions and their differences in sizes and fluxes, in all cases the vast majority of the regions are located in the areas of both diagrams usually associated with ionization due to young OB stars (i.e., the loci of SF/\Hii\ regions). In more detail, the distributions recovered by {\sc pyHIIextractor} are the ones that cover a wider range of parameters. In the particular case of the BPT diagram, it is the code that recovers more regions in the so-called intermediate region between the K03 and K01 demarcation lines, despite the fact that those are a marginal number (between the 50\% and 90\% density contour). Consequently, it also presents a tail in the WHAN diagram towards the sAGN and wAGN zone. On the other hand, the distributions recovered by {\sc pyhiiexplorer} are more compact in both diagrams, being nicely concentrated around the peaks of the distributions observed by {\sc pyHIIextractor}. This is a direct consequence of the bias introduced by this former code, which detects a lower number of larger regions, aggregating within each of them several regions resolved by the later code. The concentration of values in the diagnostic diagrams is a pure consequence of averaging values within a certain aperture, as already demonstrated in different experiments \citep[e.g.][]{mast14}. Furthermore, the distributions recovered by {\sc pyhiiexplorer} present an elongation towards the intermediate region (BPT) and the Retired region (WHAN), which could be a consequence of the contamination for the diffuse component, since this code does not perform any correction of the DIG contribution.

%All this is summarized in figure \ref{fig:codes}.
%compared to original \Ha\ map. The density contours in BPT diagram are below the boundary lines, except for the contour edge corresponding to 99\%. As for the contours of the WHAN diagram, they are in the star formation region, except for the border of the contour corresponding to 99\%. In general, our code recovers candidates for \hii\ regions with smaller radii, thus obtaining 1787 detections.
%In the case of {\sc pyhiiexplorer}, the candidates detected are visually and systematically larger, this causes more large regions to be recovered and very few small regions, this implies that structure is lost, mainly of the arms of NGC4030, reflecting on a low number of candidates detected by this code: only 365. 
%egarding the BPT and WHAN diagrams, contours of {\sc pyhiiexplorer} are closer to each other, and the contours corresponding to 15\% match acceptably with pyHIIextractor contours. The contour edge corresponding to 99\% {\sc pyhiiexplorer} in the BPT diagram tends in a similar way as pyHIIextractor. In the WHAN diagram {\sc pyhiiexplorer} contours show a trend towards retired regions. The lowest N2 values are not estimated by {\sc pyhiiexplorer}, while the highest and lowest values are not derived from the O3 values as pyHIIextractor does.

The distributions recovered by {\sc SourceExtractor} are the most similar ones to those recovered by {\sc pyHIIextractor} among the compared ones. In general, there is a broad agreement between both distributions, with a mild shift towards lower values of O3 and N2, and maybe slightly large EW(\Ha). This code is the one that detects the second largest number of regions, recovering fainter regions than {\sc pyhiiexplorer}, in particular in the inter-arm and outer regions of the galaxy. Therefore, although it recovers larger regions and therefore it presents a mixing effect, this effect is clearly weaker than in the case of {\sc pyhiiexplorer}. Furthermore, since this code comprises a correction of the background (i.e., the DIG), the distributions do not present the tail towards the retired/intermediate locations in the diagnostic diagrams shown in the values recovered by {\sc pyhiiexplorer}.

A different case is offered by the distributions recovered by {\sc HIIphot}. This is the code for which we have recovered the lowest number of regions, being all of them large with a very small range of recovered radii. Furthermore, we were unable to recover most of the inner-arms and outer/faintest regions, clearly detected by {\sc pyHIIextractor} and {\sc SExtractor}, and even those recovered by {\sc pyhiiexplorer}. As a consequence, the distributions in the BPT and WHAN diagram are affected by the mixing effect, but even more clear, they are affected by this detection bias. Therefore, they are limited to those regions with high EW(\Ha) (the brightest ones that follow the spiral-arms), with the distribution in the WHAN clearly shifted towards these values (i.e., to the upper envelope of the distribution described by {\sc pyHIIextractor}). Regarding the BPT diagram, there is a clear shift towards lower values of N2 and O3, which is less clearly explained by the mixing and detection biases. Considering that the non-detected regions are those with larger values of both parameters (upper-right envelope of the distributions described by {\sc pyHIIextractor} and {\sc SExtractor}), we consider that this shift is a consequence of an overcorrection of the background (DIG). In other words, the background is estimated using locations corresponding to regions detected by other codes, and its subtraction affects the line ratios of the remaining detected regions.

%On the other hand, {\sc HIIphot} performs detections of the brightest and largest candidates, leaving aside those candidates from the outermost part of NGC 4030 and detecting only 283 candidates. In the central part of NGC 4030 it is observed that {\sc HIIphot} detects more candidates but these are systematically large as in the case of {\sc SourceExtractor}. With regard to N2, O3 and EW(\Ha) are derived by {\sc HIIphot} underestimating, underestimating and overestimating respectively, compared to the values derived by pyHIIextractor. The latter is also seen in the coincidence of the contours of the BPT and WHAN diagrams, the contours of {\sc HIIphot} in the BPT diagram are systematically lower than the contours of pyHIIextractor, while the contours of {\sc HIIphot} in the WHAN diagram are systematically higher than the contours of pyHIIextractor. In WHAN diagram there is a small contour corresponding to 99\% in sAGN region.

Finally, {\sc astrodendro'} presents similar results as the one produced by {\sc HIIphot}, regarding the low number and the large size of the recovered regions. However, in contrast to this later code, it is able to recover outer/faint regions, although it is unable to recover many of the inter-arms/faint ones. As a consequence, the distributions present mixed patterns. For instance, the distribution in the WHAN is very similar to the one derived using {\sc SExtractor} (and {\sc pyHIIextractor} to a lower extent). On the other hand, the distribution in the BPT diagram is in between the one observed for {\sc SExtractor} and the one for {\sc HIIphot}. In summary, although this code presents an improvement over {\sc HIIphot}, it does not recover the observed distributions as well as {\sc SExtractor} or \pyHII.

We should stress that the main aim of the comparison with other tools available in the literature is not to make a ranking of which is the best code for detecting and extracting the properties of candidate \Hii\ regions using the considered datasets and simulations. We tried to demonstrate that the presented code, \pyHII, provides a reliable representation of the light distribution in the considered emission line maps, a fair detection and extraction of the properties of the ionized regions, and distribution of the properties (locations, fluxes, radii, and line ratios) as good as the one provided by existing codes. Despite our best efforts to tune the input parameters of the existing codes to provide with the best detection and extractions of the properties of \Hii\ regions, we acknowledge that users with more expertise could provide better results. This is most probably true for the case of {\sc HIIphot} since we did not find the last version of this code publicly available, and we find the documentation scarce regarding the required input parameters. Furthermore, we also acknowledge that some of the adopted codes were not created for this specific purpose. This is the case of {\sc SExtractor}, whose main aim was the segregation of extended sources (galaxies) in deep fields, and {\sc astrodendro'}, which main aim is to segregate structures of different extension by their flux density (isophotes), and the connection between them (tree structures).

\section{Summary and conclusions}
\label{sec:summary}

We have developed a new code designated {\sc pyHIIextractor}, an open-source code written in {\sc Python}. Its main goal is to recover all possible information about emission lines and underlying stellar populations of candidate \hii\ regions with high-resolution IFS data, in particular MUSE data for galaxies at $z\sim$0.015. Additionally, the code provides a model of the DIG emission based on a minimum number of assumptions. This model is used to decontaminate the properties derived for the candidate \hii\ regions. This is a novelty with respect to other codes and procedures in the literature, that do not perform such decontamination, or propose a selection and decontamination a posteriori based on assumed physical properties of the DIG (e.g. through high values \sii/\Ha, \citealt{della20}). An additional important advantage of our code is its versatility since it can detect candidate \hii\ regions either from an IFS datacube, the dataproducts provided by IFS analysis pipelines, or just an emission line image (e.g, an \Ha\ image). Furthermore, the code can be used as a single script running the full analysis process, or the user may select which specific functions are more suitable for the considered analysis.

%either be used in its entirety or only use those functions the user needs to complement their results. 

The entire \pyHII\ procedure is divided into two steps: detection and extraction. The detection implements the routines to select the candidate \Hii\ regions in an emission line image. It provides their positions, sizes, and fluxes, together with two model images corresponding to the emission of the detected regions and the DIG. The extraction assigns to each \Hii\ region the properties of the emission lines and the underlying stellar population provided by the {\sc Pipe3D} datacube, including flux intensities, equivalent widths, ages, metallicities, among other properties, each one with their respective uncertainties. We note that this second procedure can be easily adapted to the dataproducts of other analysis pipelines or even to extract the spectra corresponding to each region from an IFS datacube (or similar data).

%has positions and size as the output parameters for DIG and candidates to \hii\ regions, with size being the only parameter for the latter. Subsequently, their maps, flux intensities in \Ha\ can be recovered, as well as segmentation maps if needed. Once the spatial information of candidates to \hii\ regions have been obtained, the code and a Pipe3D datacube of the galaxy can extract other flux intensities, equivalent widths, ages, metallicities, among other properties, each one with their respective uncertainties.

We characterize \pyHII\ exploring its behavior using both simulations and real data. First, we use simulations of an idealized spiral galaxy with three components (i) \hii\ regions, (ii) DIG due to old stars (HOLMES), and { (iii)} leaking of photons from \hii\ regions. In this simulation, we vary the inclination of the galaxy ($b/a$ ratio), number of arms, and the leaking factor. We perform approximately $\sim$200 simulations and we conclude that: ($i$) the parameter that least affects the fraction of recovered regions is the leaking factor compared to the inclination and the number of arms; ($ii$) the procedure provides better results regarding the recovery of simulated regions and their fluxes when the system is less inclined (face-on) and the number of spiral arms is lower; ($iii$) regarding the derived radii, our code introduces a systematic bias, obtaining values $\sim$20\%\ larger than the simulated ones; ($iv$) the N2 and O3 line ratios are slightly underestimated and overestimated respectively for the candidate \hii\ regions (with $\sim$0.1 dex of the simulated values), while the EW(\Ha) is well recovered for most of the cases. However, there are particular cases (edge-on, large number of \hii\ regions, and low \lkf values), where the EW(\Ha) is erroneously recovered (by a factor of two); and ($v$) N2 and O3 recovered for DIG tend to be slightly overestimated and underestimated, respectively (within $\sim$0.2 dex of the simulated values). Regarding the EW(\Ha), in most of the cases it is recovered appropriately, within 50\% of the original value. However, like in the case of the candidate \hii\ regions for an edge-on galaxy with a large number of candidates and a low \lkf, it presents errors up to a factor of six of the simulated value. In summary, we consider that the code may not be adequate to explore flocculent, highly inclined, and early-type spirals.

In addition we compare the performance of \pyHII\ with that of four different codes most frequently used in the literature with the same purpose: {\sc pyhiiexplorer}, {\sc SourceExtractor}, {\sc HIIphot} and {\sc astrodendro}. For doing so we apply all the codes to (i) the same simulated galaxy (a face-on, Grand Design galaxy, with a large \lkf), and (ii) the dataproducts derived by {\sc Pipe3D} for the MUSE data on the NGC 4030 galaxy. Both experiments show that \pyHII\ is a highly competitive code to recover the properties of the \Hii\ regions, performing as well or better than the other codes. In particular, we found that: ($i$) our code is the one that recovers the largest number of regions in both the simulation and the real data; ($ii$) it is also the one that best reproduces the observed distribution of sizes and fluxes (based on the simulation); ($iii$) most of the codes tend to increase the number of large/bright regions at the expense of small/faint ones. Thus, they all aggregate the later ones to existing former ones or create unrealistically large regions by aggregation. This bias is stronger in the comparison codes, except for {\sc astrodendro} since if the original sizes for the candidates are adopted in this case the sizes and fluxes are far too small/low. This is observed in the simulations, where both the sizes and fluxes are biased towards larger values than the input ones (for all codes, except for the case of {\sc pyhiiexplorer}), and in real data, where sizes and fluxes are condensed towards the larger values reported by {\sc pyHIIextractor}; ($iv$) all codes recover well the distribution, sizes, and fluxes, of the brightest/largest \Hii\ regions, tracing their location in the original image in both the simulation and the observations. However, the bi-dimensional intensity distribution is better recovered by {\sc pyHIIextractor} with respect to any of the other codes, either because it recovers fainter/smaller regions (not recovered by the remaining codes) or because those regions are erroneously aggregated into larger ones by other codes; finally, ($v$) the recovered distributions of the line ratios and EW(\Ha) of the recovered candidates are located for all the codes in the expected location for \hii\ regions and star formation areas (for the observations). Nevertheless, the distribution is wider in the case of {\sc pyHIIextractor}, covering a more ample range of values than any of the other codes. {\sc SExtractor} covers almost the same range, while {\sc HIIphot} and {\sc astrodendro'} present a much stronger bias. Finally {\sc pyhiiextractor} concentrates the values towards the peak of the distribution provided by {\sc pyHIIextractor}, as a consequence of the aggregation of ionized regions into larger ones. In summary, our results suggest that our code is as good or better than any of the existing codes. Even in the case that our implementation of the comparison codes could be improved (by experts users), our code has the advantage that it does not require any adjustment. All the required parameters can be self-defined by a simple exploration of the input data (such as the PSF FWHWM or max\_size values), or are directly estimated by the code itself (like the adopted thresholds for the detection of regions).

Our purpose for future work is to implement our tool for the IFS MUSE data included in the  AMUSING ++ compilation, already processed by {\sc Pipe3D} \citep{lopezcobasanchez2020}, to provide a refined catalog of physical properties of \hii\ regions. Our estimations, based on the analysis of NGC 4030 presented here suggest that we should recover of the order of $\sim$100,000 regions for the already compiled set of $\sim$500 galaxies.

%pyHIIextractor:1787
%%SE:564
%pyHIIexplorer:365
%HIIphot:283
%AD:330

\appendix

%%%%%%%%%%%%%%%%%%%%%%%%%%%%%%%%%%%%%%%%%%%%%%%%%%%%%%%%%%%%%%%%%%%
\begin{figure*}
    \minipage{0.99\textwidth}
    \includegraphics[trim=0 0 0 0, clip, width=\textwidth]{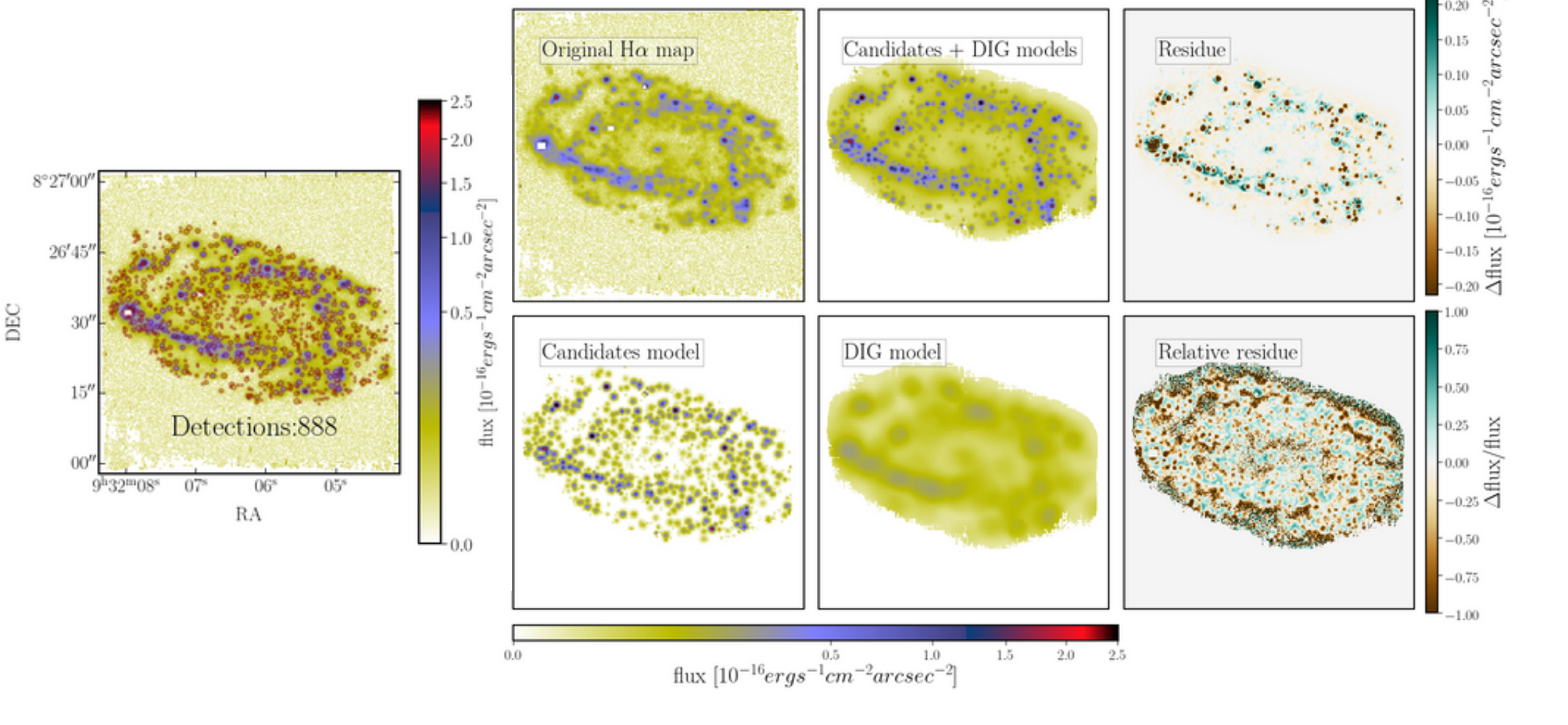}
    \endminipage
    \caption{Example of the use of \pyHII\ for the AMUSING++ data of the galaxy NGC 2906. Left panel shows the original \Ha\ map (intensity image) together with the detected candidate \Hii\ regions, as red circles. The radius of the circles is proportional to the size of the candidates. The number of detected regions is indicated in the label. Right panels shows, from the upper-left to the bottom-right, (i) the original \Ha\ image, (ii) the best model provided by the code (candidates + diffuse), (iii) the residual image created by subtracting the models to the original image, (iv) the model of the candidates to \hii\ regions, (v) the model of the DIG, and finally, (vi) the relative residual image (i.e., the residual image divided by the original \Ha\ image)}
    \label{fig:ngc2906_muse}
\end{figure*}

\begin{figure*}
    \minipage{0.99\textwidth}
    \includegraphics[trim=0 0 0 0, clip, width=\textwidth]{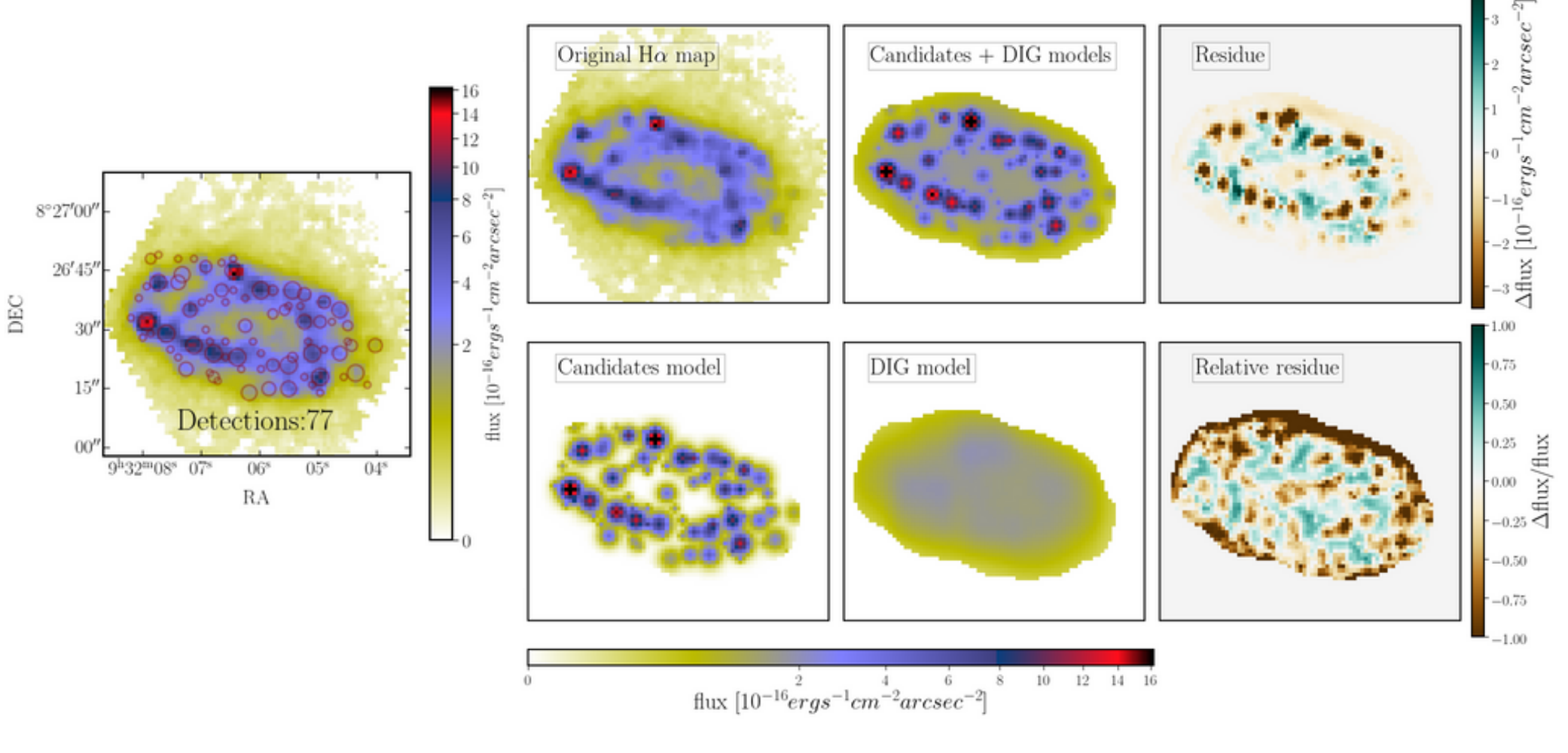}
    \endminipage
    \caption{Similar figure as Fig. \ref{fig:ngc2906_muse}, this time for the CALIFA data of NGC 2906.}
    \label{fig:ngc2906_califa}
\end{figure*}

\begin{figure*}
    \minipage{0.99\textwidth}
    \includegraphics[trim=0 0 0 0, clip, width=\textwidth]{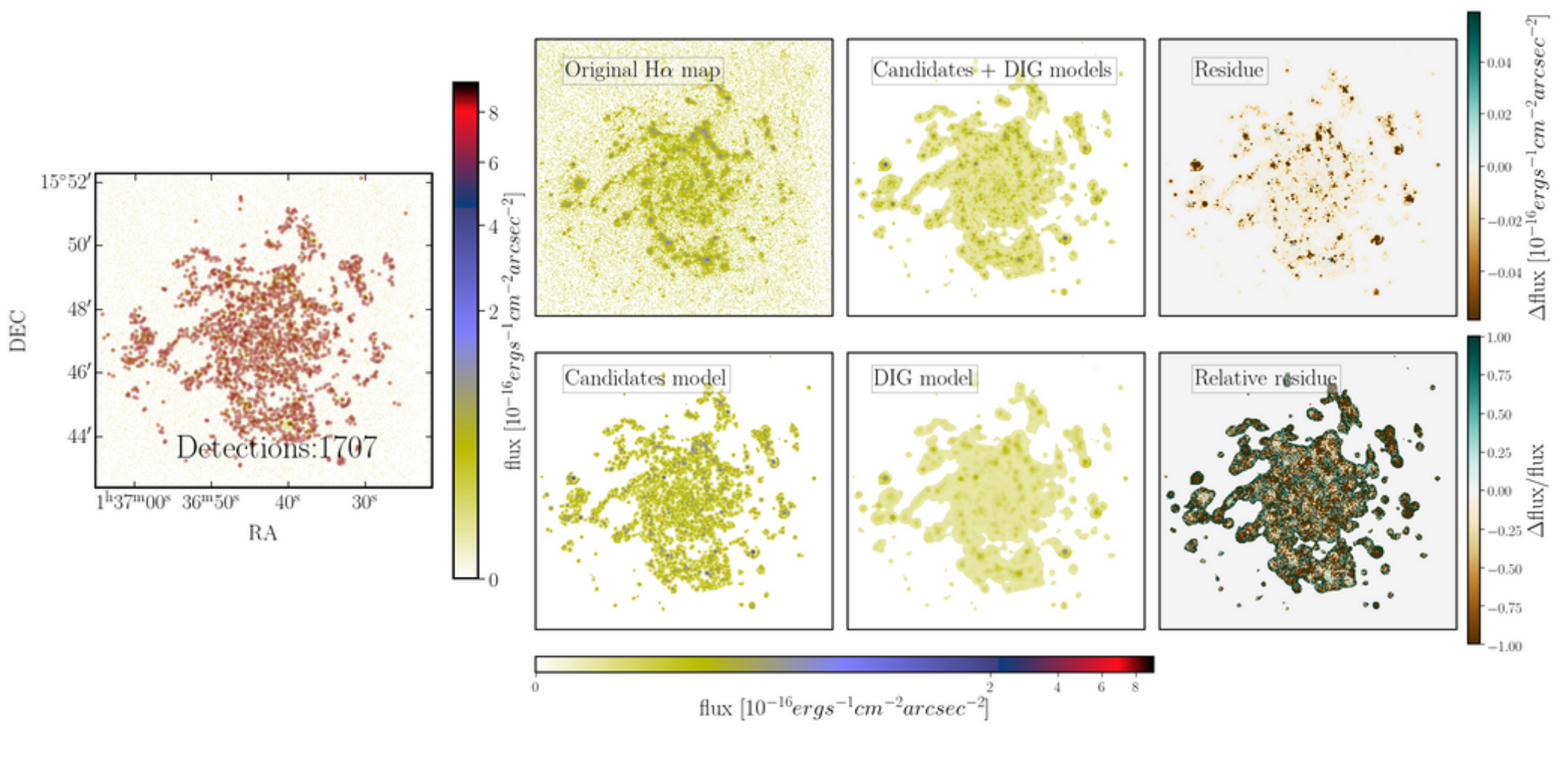}
    \endminipage
    \caption{Similar figure as Fig. \ref{fig:ngc2906_muse}, this time for the \Ha\ SINGS data of NGC 628.}
    \label{fig:ngc628_sings}
\end{figure*}

\begin{figure*}
    \minipage{0.99\textwidth}
    \includegraphics[trim=0 0 0 0, clip, width=\textwidth]{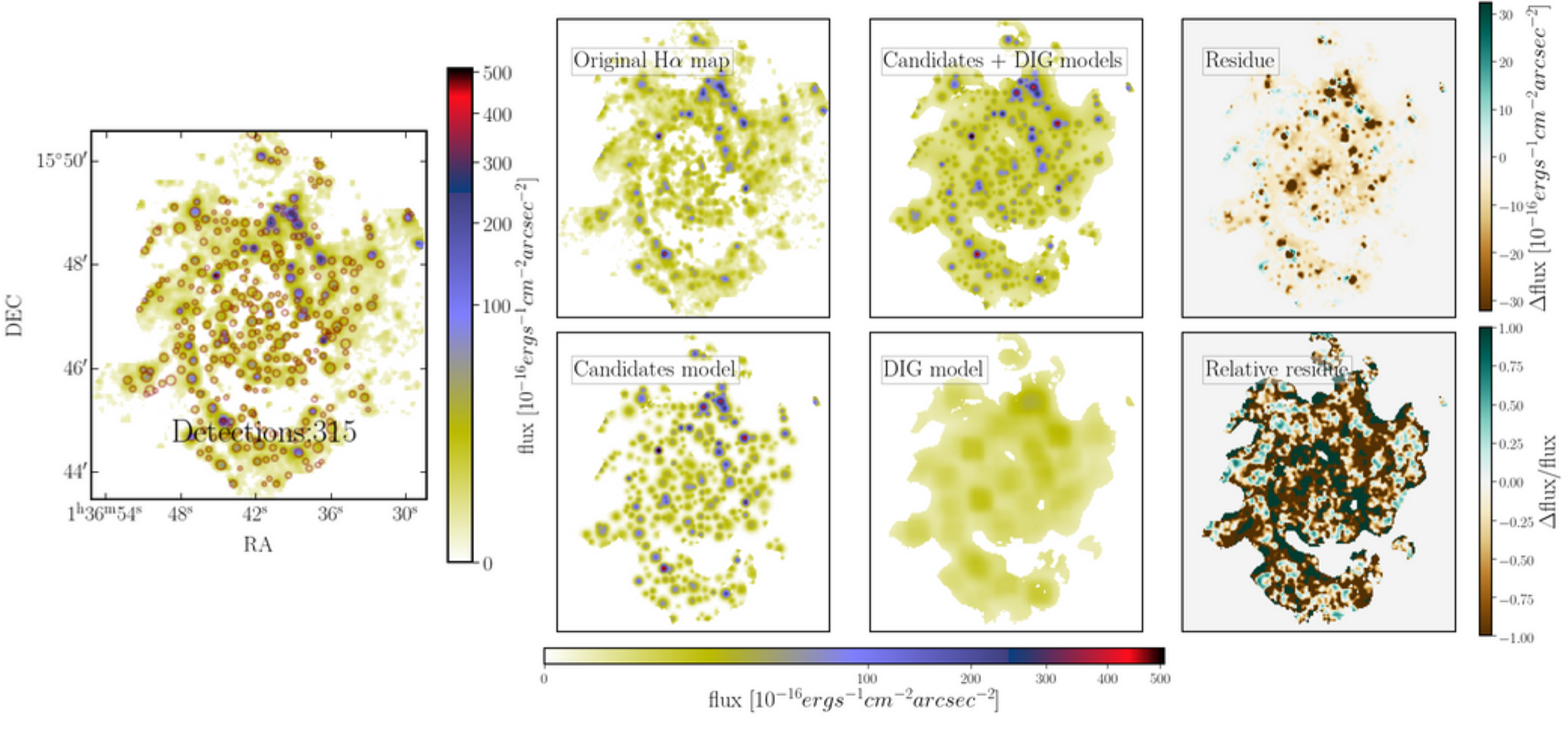}
    \endminipage
    \caption{Similar figure as Fig. \ref{fig:ngc2906_muse}, this time for the PINGS data of NGC 628.}
    \label{fig:ngc628_ppak}
\end{figure*}
%%%%%%%%%%%%%%%%%%%%%%%%%%%%%%%%%%%%%%%%%%%%%%%%%%%%%%%%%%%%%%%%%%%%

\section{Accessing and using \pyHII}
\label{sec:code}
%\Com{How to access and use the code}

{\sc pyHIIextractor} is a public and freely accessible code written in {\sc Python}. It can be downloaded from the following GitHub repository:  \url{https://github.com/sfsanchez72/pyHIIExplorer.git}. It requires the following versions of the language and packages: Python 3.6.5, Numpy 1.19.5, Astropy 4.1, Scipy 1.1.0, Matplotlib 2.2.2, Skimage 0.17.2, Argparse 1.1, Photutils 1.0.2, Astrodendro 0.2.0, extinction 0.4.6 and PIL 5.1.0. We recommend installing all in a virtual environment to avoid dependency problems. 

The GitHub repository comprises five different folders: (i) simulations, including the code required to replicate the simulations presented along this study; (ii) bin, comprising the scripts required to run the code as a command line from a terminal; (iii) data, that contains two datasets useful to test the code, a {\sc Pipe3D} datacube corresponding to the analysis performed for the CALIFA data of galaxy NGC 2906 and a table with the emission lines properties corresponding to the candidate \hii\ regions detected by our code on the MUSE data of the NGC 6975 galaxy; (iv) notebooks, including some useful tools to read and represent the extracted data; and (v) {\sc pyHIIExplorer}, comprising all scripts involved in the different steps of \pyHII. 

% An example for the user is in the file: bin/pyHIIdet\_ext\_pipe3d.py this program is a simple implementation to run all pyHIIextractor code (detection and extraction) in a CALIFA datacube: 

In order to test \pyHII, the user can run the {\sc pyHIIdet\_ext\_pipe3d.py} script, which implements both the detection and extraction of properties from a {\sc Pipe3D} datacube (e.g. NGC2906.Pipe3D.cube.fits.gz). It can be run with the command-line: {\sc python3 pyHIIdet\_ext\_pipe3d.py NGC2906 NGC2906.Pipe3D.cube.fits.gz 3 1 2 4 45 245 2.3 1.0 0 0 1 3 1 1 2 outpout\_files\_DIR}. For other scripts and stand-alone codes we refer the reader to the README.md file.
%The example data is a datacube of CALIFA survey by storage purposes.

%For any questions, doubts or comments please write to emails: \url{sfsanchez@astro.unam.mx} \& \url{alugo@astro.unam.mx}.

\section{Examples of use of \pyHII}
\label{sec:examples}

As an example of the use of our code we apply it to four different \Ha\ emission line maps of very different nature: (i) a map extracted from the IFS data of the galaxy NGC 2906 observed by MUSE (included in the AMUSING++ compilation); (ii) a similar map for the same galaxy extracted from the IFS data obtained by the CALIFA survey; (iii) an \Ha\ narrow-band image obtained for the galaxy NGC 628 as part of the {\it Spitzer} Infrared Nearby Galaxies Survey sample (SINGS) ancillary data \citep{SINGSkennicutt2003}; and (iv) an \Ha\ map extracted from the PPAK IFS Nearby Galaxy Survey (PINGS) observations \citep{rosales-ortega10} on the same galaxy.
The first two \Ha\ maps, corresponding to NGC 2906, were obtained using the {\sc Pipe3D} pipeline, following the procedures outlined in Sec. \ref{sec:data}. On the other hand, to derive the \Ha\ maps corresponding to NGC 628 the classical approach of obtaining a narrow-band image centered in the emission line of interest was adopted, to sample the emission by the ionized gas, and a broad-band image at the same wavelength, to sample the underlying continuum. Then, the final \Ha\ map was constructed by scaling and subtracting the broad-band image to the narrow-band one. The only difference among the two maps is that for the SINGS data both narrow- and broad-band images correspond to direct observations, while for the PINGS data they were synthesized from the IFS datacube. 

We should highlight that the four images correspond to very different apparent and physical resolutions. The resolution of the AMUSING++ and SINGS data is driven by the atmospheric seeing, with a FWHM$\sim$1$\arcsec$. On the other hand, both the CALIFA and PINGS data have a resolution of FWHM$\sim$2.5-2.7$\arcsec$, driven mostly by the size of the PPAK fibers and the observing strategy (i.e., the adopting of a dithering scheme or not). Finally, NGC 628 is three times nearer than NGC 2906, which implies that the physical scales are very different for the same apparent resolution. This way, the physical resolution of the NGC 2906 AMUSING++ data is FWHM$\sim$160 pc, while for the CALIFA data is FWHM$\sim$400 pc. In the case of NGC 628, the SINGS data have a physical resolution of FWHM$\sim$46 pc, while for the PINGS data is FWHM$\sim$124 pc. Thus, the physical resolution of the PINGS data for NGC 628 is similar to the one provided by AMUSING++ for NGC 2906.

For each image, we optimize the required input parameters ($\sigma_{\rm eline}$, $\sigma_{\rm cont}$, and max\_size) with the aim of obtaining the best compromise between (i) deriving the largest number of recovered candidate \Hii\ regions, (ii) minimizing the value of the $\chi^{2}$ (see Sec. \ref{subsec:adoptedrange}) and (iii) recovering the same spatial distribution of \Hii\ regions in the model than the one observed in the original data. Figures \ref{fig:ngc2906_muse}, \ref{fig:ngc2906_califa}, \ref{fig:ngc628_sings} and \ref{fig:ngc628_ppak} illustrate the results of this experiment, showing the default plots provided by \pyHII. Each figure comprises a panel showing the detected regions on top of the input emission line map (left panel), and a set of panels comparing this map with the best provided model, the models for the \hii\ regions and the DIG, and the corresponding residuals (right panels). As expected the image with best physical spatial resolution, covering the largest optical extension of the corresponding galaxy (i.e., the \Ha\ SINGS data for NGC 628), is the one in which the code recovers the largest number of candidate \HII\ regions. As the resolution is degraded the number of candidate \hii\ regions and the quality of the recovered model is lower (i.e., it is less similar to the original dataset). It is worth noticing that at a resolution below 100pc (i.e., \Ha\ SINGS data of NGC 628), the basic premise required by our code is that ionized regions are visually similar to clumpy/peaky almost circular regions is not fulfilled. For this reason, a detailed exploration of the best model provided by the code shows that in some cases, near large candidates with clear resolved structures the code over-segregates the regions creating artificial structures. On the other hand, for the dataset with the worst physical resolution (CALIFA data of NGC 2906), the code is unable to recover most of the detailed structure and the extracted candidates are in most cases the result of aggregating several regions well segregated in the data of a better spatial resolution. It is worth noticing that the rate of aggregation (between 3-10) was already predicted in the simulations performed by \citet{mast14}, on the very same topic. This is important for the interpretation of the physical quantities associated with those ionized regions \citep[e.g.][]{espi20}.

In summary, this experiment demonstrates that \pyHII, despite being optimized to work with data that adopts the format used by {\sc Pipe3D}, can be used in a wide variety of data, using as input the emission line maps generated using different techniques. Based on this experiment we consider that the optimal physical resolution on which the code is more reliable is FWHM$\sim$100 pc.

{ 
\section{Available information in {\sc Pipe3D} dataproducts}
\label{sec:info_pipe3d}

{\sc Pipe3D} pipeline derives at least four cubes of dataproducts (FLUX\_ELINES, SFH, SSP, and INDICES), in which each channel is a map a physical property or parameter of either the emission lines or the underlying stellar populations. We present here a brief summary of the content of those dataproducts \citep[for a more detailed information we refer to][]{sanchez06a}.
%Below, we briefly present the most relevant information that can be extracted from each dataproducts with its respective table to illustrate the order of information within the datacube \citep{sanchez06a}. 

i) FLUX\_ELINES: contains the main parameters of the 30 emission lines (for a MUSE cube) listed in the table \ref{tab:lines}, such as flux intensity, velocity, velocity dispersion and  equivalent width, with their corresponding errors. %flux error, velocity error, velocity dispersion error and equivalent width error.

ii) SFH: each channel includes the spatial distribution of the light fraction of each single stellar population within the adopted SSP-library. The currently SSP-library for the standard analysis performed for the AMUSING++ data comprises 156 SSPs, with 39 ages and 4 metallicities.

iii) SSP: each slice corresponds to the average properties of the stellar populations derived from the multi-SSP fitting procedure performed by {\sc Pipe3D}.

iv) INDICES: each channel contains the spatial distribution of a set of stellar indices obtained from the emission-line subtracted spectra.

\begin{table}
\begin{center}
\caption{List of emission lines includes in the FLUX\_ELINES cube.}
\begin{tabular}{| c | c | c | c | c | c |}
\hline\hline
\# & $\lambda$ (\AA) & ID & \# & $\lambda$ (\AA) & ID\\ \hline
0 & 4922.2 & He I & 15 & 5875.6 & He I \\
1 & 5006.8 & [O III] & 16 & 6300.3 & [O I] \\ 
2 & 4958.9 & [O III] & 17 & 6312.4 & [S III] \\ 
3 & 4861.3 & H$\beta$ & 18 & 6347.3 & [S II] \\
4 & 4889.6 & [Fe II] & 19 & 6363.8 & [O I] \\
5 & 4905.3 & [Fe II] & 20 & 6562.7 & H$\alpha$ \\
6 & 5111.6 & [Fe II] & 21 & 6548.1 & [N II] \\
7 & 5158.8 & [Fe II] & 22 & 6583.4 & [N II] \\
8 & 5199.6 & [N I] & 23 &  6678.0 & He I \\ 
9 & 5261.6 & [Fe II] & 24 & 6716.4 & [S II] \\ 
10 & 5517.7 & [Cl III] & 25 & 6730.7 & [S II] \\
11 & 5537.6 & [Cl III] & 26 & 7136.0 & [Ar III] \\ 
12 & 5554.9 & [O I] & 27 & 7325.0 & [O II] \\ 
13 & 5577.3 & [O I] & 28 & 7751.0 & [Ar III] \\
14 & 5754.5 & [N II] & 29 & 9068.6 & [S III] \\ \hline\hline
\end{tabular}
\label{tab:lines}
\end{center}
where {\tt \#} is the index of the corresponding emission line, $\lambda$ is the wavelength, and ID is the label to each emission line. 
\end{table}

\begin{table}
\begin{center}
\caption{Description of the FLUX\_ELINES extensions.}
\begin{tabular}{lll}\hline\hline
Channel	& Units	& Description of the map\\
\hline
{\tt \#}	    & 10$^{-16}$ erg/s/cm$^2$	& Flux intensity\\
{\tt \#}+N	    &km/s	            & Velocity \\
{\tt \#}+2N	&\AA	                & Velocity dispersion$^a$\\
{\tt \#}+3N	&\AA	                & Equivalent width$^b$\\
{\tt \#}+4N	&10$^{-16}$ erg/s/cm$^2$	& Flux error\\
{\tt \#}+5N	&km/s	            & Velocity error\\
{\tt \#}+6N	&\AA	            & Velocity dispersion error \\
{\tt \#}+7N	&\AA	                & Equivalent width error\\
\hline\hline
\end{tabular}\label{tab:fe}
\end{center}
{\tt \#} is the index of corresponding emission line listed in table \ref{tab:lines} and $N=$30 is the number of analysed lines. 
\end{table}

\begin{table}
\begin{center}
\caption{Description of the SFH extensions.}
\begin{tabular}{ll}\hline\hline
Channel	&  Description of the map\\
\hline
0 &  $w_{\star,L}$ for ($\mathcal{A}_\star$,$Z_\star$)=(0.001 Gyr, 0.0037) \\
... &  ... \\
155 &  $w_{\star,L}$ for ($\mathcal{A}_\star$,$Z_\star$)=(14.1254 Gyr, 0.0315) \\
\hline  
156 &  $w_{\star,L}$ for $\mathcal{A}_\star$=0.0010 Gyr\\
  ... &  ... \\
194 &  $w_{\star,L}$ for $\mathcal{A}_\star$=14.1254 Gyr\\
\hline  
195 &  $w_{\star,L}$ for $Z_\star$=0.0037 \\
... &  ... \\
198 &  $w_{\star,L}$ for $Z_\star$=0.0315 \\
\hline\hline
\end{tabular}\label{tab:sfh}
\end{center}
Channel column indicates the length on the Z-axis of the datacube, $w_{\star,L}$ is the fraction weight of light corresponding to an SSP, which can be of 3 types: a) to an age-metallicity ($\mathcal{A}_\star$ $Z_\star$), b) to an specific age ($\mathcal{A}_\star$) and c) to an specific metallicity ($Z_\star$).  
\end{table}

\begin{table*}
\begin{center}
\caption{Description of the SSP extension.}
\begin{tabular}{lll}\hline\hline
Channel	& Units	& Stellar index map\\
\hline
0	&	10$^{-16}$ erg/s/cm$^2$	& Unbinned flux intensity at $\sim$5500\AA, f$_{V}$ \\
1	&	none	                & Continuum segmentation index, $CS$\\
2	&	none	                & Dezonification parameter, $DZ$\\
3	&	10$^{-16}$ erg/s/cm$^2$	& Binned flux intensity at $\sim$5500\AA, f$_{V,CS}$\\
4	&	10$^{-16}$ erg/s/cm$^2$	& StdDev of the flux at $\sim$5500\AA, ef$_{V,CS}$ \\
5	&	Gyr	                & Lum. Weighted age, $\mathcal{A}_{\star,L}$\\
6	&	Gyr	                & Mass Weighted age, $\mathcal{A}_{\star,M}$\\
7	&	Gyr	                & Error of both $\mathcal{A}_{\star}$\\
8	&	dex	                & Lum. Weighted metallicity, $Z_{\star,L}$ in\\
    &                       & logarithm scale, normalized to the solar value\\
9	&	dex	                & Lum. Weighted metallicity, $Z_{\star,M}$ in\\
    &                       & logarithm scale, normalized to the solar value\\
10	&	dex	                & Error of both $Z_{\star}$\\
11	&	mag	                & Dust attenuation of the st. pop., A$_{V,\star}$\\
12	&	mag	                & Error of A$_{V,\star}$, e$_{{\rm A}_V}$\\
13	&	km/s	                & Velocity of the st. pop., vel$_{\star}$\\
14	&	km/s	                & Error of the velocity, $e_{\rm vel}$ \\
15	&	km/s	                & Velocity dispersion of the st. pop., $\sigma_{\star}$\\
16	&	km/s	                & Error of $\sigma_{\star}$, $e_\sigma$\\
17	&	log10(M$_\odot$/L$_\odot$)	& Mass-to-light ratio of the st. pop., $\Upsilon_{\star}$\\
18	&	log10(M$_\odot$/sp$^2$)	& Stellar Mass density per spaxel., $\Sigma_{\star}$\\
19	&	log10(M$_\odot$/sp$^2$)	& Dust corrected $\Sigma_{\star}$, $\Sigma_{\star,dust}$\\
20	&	log10(M$_\odot$/sp$^2$)	& error of $\Sigma_{\star}$\\
\hline\hline
\end{tabular}\label{tab:ssp}
\end{center}
Channel column indicates the length on the Z-axis of the datacube, units column indicates the units for each map, and stellar index map column indicates the labels of each property.
\end{table*}

\begin{table*}
\begin{center}
\caption{Description of the INDICES extension.}
\begin{tabular}{llllccc}\hline\hline
ID & Channel	& Units	& Channel content & Index $\lambda$ range (\AA) & blue $\lambda$ range (\AA) & red $\lambda$ range (\AA) \\
\hline
Hd     & 0/9	&	\AA     &H$\delta$ index/error& 4083.500-4122.250 &4041.600-4079.750 &4128.500-4161.000\\
H$\beta$    & 1/10	&	\AA	&H$\beta$ index/error& 4847.875-4876.625 &4827.875-4847.875 &4876.625-4891.625\\
Mgb    & 2/11	&	\AA	&Mg$b$ index/error  & 5160.125-5192.625 &5142.625-5161.375 &5191.375-5206.375\\ 
Fe5270 &3/12	&	\AA	&Fe5270 index/error  & 5245.650-5285.650 &5233.150-5248.150 &5285.650-5318.150\\
Fe5335 &4/13	&	\AA	&Fe5335 index/error & 5312.125-5352.125 &5304.625-5315.875 &5353.375-5363.375\\
D4000  & 5/14	&	\AA	&D4000 index/error  & 4050.000-4250.000 &3750.000-3950.000 & \\
Hdmod  &6/15	&	\AA	&H$\delta_{\rm mod}$/error & 4083.500-4122.250 &4079.000-4083.000 &4128.500-4161.000\\
Hg     &7/16	&	\AA	&H$\Upsilon$/error  & 4319.750-4363.500  &4283.500-4319.750  &4367.250-4419.750\\
Flux   &8/17	&	10$^{-16}$ erg/s/cm$^2$&median flux/standard deviation$^{1}$ & & & \\
\hline\hline
\end{tabular}\label{tab:index}
\end{center}
ID is the label to each index, channel column indicates the length on the Z-axis of the datacube, channel content column indicates the property . 
\end{table*}

}

\section*{Acknowledgements}

We thank the referees for their useful commennts and suggestions that have improved the quality of this manuscript.

We are grateful for the support of a CONACYT grant CB-285080 and FC-2016-01-1916, and funding from the PAPIIT-DGAPA-IN100519 (UNAM) project. 

J. K. B-B acknowledges support from the grant IA-100420 (DGAPA-PAPIIT ,UNAM) and funding from the CONACYT grant CF19-39578. DC acknowledges support from \emph{Deut\-sche For\-schungs\-ge\-mein\-schaft, DFG\/} project number SFB956A. T.W., Y.C., and Y.L. acknowledge support.

L.G. acknowledges financial support from the Spanish Ministerio de Ciencia e Innovaci\'on (MCIN), the Agencia Estatal de Investigaci\'on (AEI) 10.13039/501100011033, and by the European Social Fund (ESF) "Investing in your future" under the 2019 Ram\'on y Cajal program RYC2019-027683-I and the PID2020-115253GA-I00 HOSTFLOWS project.

This research made use of
Astropy,\footnote{http://www.astropy.org} a community-developed core {\sc Python} package for Astronomy \citep{astropy:2013, astropy:2018}.

\section*{Data Availability}
\label{sec:data_ava}

The MUSE data are available in ESO archive (\url{https://archive.eso.org/wdb/wdb/adp/phase3_main/form}), but MUSE data processed by {\sc Pipe3D} pipeline underlying this article will be shared on reasonable request to the corresponding author \citep{lopezcobasanchez2020}. The map of the NGC 2906 galaxy obtained by the CALIFA survey as part of Data Release 2 and Data Release 3 (DR2 and DR3) is available in CALIFA SURVEY website (\url{https://califaserv.caha.es/CALIFA_WEB/public_html/?q=content/califa-explorer-v01&califaid=275}). 

The \Ha\ narrow-band image obtained for the NGC 628 galaxy as part of SINGS sample is available in NASA/IPAC Infrared Science Archive (\url{https://irsa.ipac.caltech.edu/data/SPITZER/SINGS/}), while the \Ha\ map for the NGC 628 galaxy as part of PPAK IFS Nearby Galaxy Survey is available in PINGS project website (\url{https://calmecac.inaoep.mx/~frosales/pings/html/public/ngc628-1/index.html}).

%\section*{Data Availability}
%TBW

\bibliographystyle{mnras}
\bibliography{my_bib} %#,carlos,extra}

\end{document}